%% file: main.tex
\documentclass[sigconf, nonacm]{acmart}

\newcommand\vldbdoi{XX.XX/XXX.XX}
\newcommand\vldbpages{XXX-XXX}
\newcommand\vldbvolume{14}
\newcommand\vldbissue{1}
\newcommand\vldbyear{2020}
\newcommand\vldbauthors{\authors}
\newcommand\vldbtitle{\shorttitle} 
\newcommand\vldbavailabilityurl{}
\newcommand\vldbpagestyle{plain}

\usepackage{caption}
\usepackage{subcaption}
\usepackage{algorithm}
\usepackage[noend]{algpseudocode}
\usepackage{graphicx}
\usepackage{tabularx}
\usepackage{enumitem}
\usepackage{xcolor,listings}
\usepackage{textcomp}
\usepackage{tcolorbox}
\usepackage{colortbl}
\usepackage{soul}
\usepackage{makecell}
\usepackage{float}
\usepackage{bbm}
\usepackage{bm}
\usepackage{amsmath}
\usepackage{balance}
\usepackage{cleveref}
\usepackage{array,multirow}
\usepackage{thmtools}
\usepackage{thm-restate}
\usepackage[scaled=0.85]{beramono}
\usepackage{nccmath}
\graphicspath{ {./figures/} }

\newif\iffullversion
\fullversiontrue  

\newcommand{\eat}[1]{}

\makeatletter
\newcounter{phase}[algorithm]
\newlength{\phaserulewidth}

\makeatother

\algnewcommand\algorithmicintialize{\textbf{Initialize:}}
\algnewcommand\Initialize{\item[\algorithmicintialize]}

\algnewcommand\algorithmicprivate{\textbf{Private:}}
\algnewcommand\Private{\item[\algorithmicprivate]}
\algnewcommand\algorithmicpublic{\textbf{Public:}}
\algnewcommand\Public{\item[\algorithmicpublic]}
\algnewcommand\algorithmicuser{\textbf{User:}}
\algnewcommand\User{\item[\algorithmicuser]}

\AtBeginDocument{%
  \providecommand\BibTeX{{%
    \normalfont B\kern-0.5em{\scshape i\kern-0.25em b}\kern-0.8em\TeX}}}

\newtheorem{theorem}{Theorem}[section]
\newtheorem{observation}{Observation}[section]
\newtheorem{problem}{Problem}
\newcommand{\paratitle}[1]{\paragraph{\bf #1}}
\newcommand{\mechanism}{\ensuremath{\mathcal{M}}}
\newcommand{\predspace}{\ensuremath{\mathcal{P}}}

\newcommand{\remindera}[1]{\textcolor{brown}{[[[#1]]]}}

\newcommand{\yuchao}[1]{\remindera{\bf (Yuchao)~#1}{\typeout{#1}}}

\newcommand{\red}[1]{{ \color{red} #1 }}
\newcommand{\blue}[1]{{ \color{blue} #1 }}

\newcommand{\rev}[1]{{#1 }}

\newcommand{\cut}[1]{}
\newcommand{\dataset}[1]{{\texttt{#1}}}

\providecommand{\abs}[1]{\lvert#1\rvert}

\DeclareMathOperator\erf{erf}

\DeclareMathOperator\M{\mathcal{M}}
\DeclareMathOperator\imp{\textsc{Inf}}
\DeclareMathOperator\score{s}
\DeclareMathOperator\I{\mathcal{I}}

\DeclareMathOperator\rank{rank}

\newcommand{\Aagg}{A_{agg}}
\newcommand{\Agb}{A_{gb}}
\newcommand{\attrset}{\mathbb{A}}
\newcommand{\dom}{{\tt dom}}
\newcommand{\dommax}[1]{#1^{max}}

\newcommand{\neighbor}{\approx}
\newcommand{\rhoQuery}{{\rho_{q}}} 

\newcommand{\rhoTopk}{{\rho_{Topk}}}
\newcommand{\rhoInflu}{{\rho_{Influ}}}
\newcommand{\rhoRank}{{\rho_{Rank}}}
\newcommand{\userquery}{{\ensuremath{Q}}}

\newcommand{\uq}{uq}
\newcommand{\influ}{influ}
\newcommand{\rnk}{rank}
\newcommand{\relimp}{\widetilde{\imp}}
\newcommand{\noisyimp}{\widehat{\imp}}
\newcommand{\relinflu}{relinflu}

\newcommand{\exptab}{Explanation Table}
\newcommand{\privans}{Priv-answer}
\newcommand{\trueans}{True-answer}

\newcommand{\oursys}{{\sc DPXPlain}}
\newcommand{\subo}{s}

\newcommand{\EFFORT}[1]{\cellcolor{black!#1}}
\newcolumntype{P}[1]{>{\centering\arraybackslash}p{#1}}

\newcommand{\squishlist}{
	\begin{list}{$\bullet$}
		{
			\setlength{\itemsep}{0pt}
			\setlength{\parsep}{3pt}
			\setlength{\topsep}{3pt}
			\setlength{\partopsep}{0pt}
			\setlength{\leftmargin}{1.5em}
			\setlength{\labelwidth}{1em}
			\setlength{\labelsep}{0.5em} } }

\newcommand{\squishend}{
\end{list}  }

\begin{document}
\title{DPXPlain: Privately Explaining Aggregate Query Answers}
\author{Yuchao Tao, Amir Gilad, Ashwin Machanavajjhala, Sudeepa Roy\\ Duke University, USA}




\begin{abstract}
Differential privacy (DP) is the state-of-the-art and rigorous notion of privacy for answering aggregate database queries while preserving the privacy of sensitive information in the data. In today's era of data analysis, however, it poses new challenges for users to understand the trends and anomalies observed in the query results: Is the unexpected answer due to the data itself, or is it due to the extra noise that must be added to preserve DP? In the second case, even the observation made by the users on query results may be wrong. In the first case, can we still mine interesting explanations from the sensitive data while protecting its privacy? To address these challenges, we present a three-phase framework \oursys, which is the first system to the best of our knowledge for explaining group-by aggregate query answers with DP. In its three phases, \oursys\ (a) answers a group-by aggregate query with DP, (b) allows users to compare aggregate values of two groups and with high probability assesses whether this comparison holds or is flipped by the DP noise, and (c) eventually provides an explanation table containing the approximately `top-k' explanation predicates along with their relative influences and ranks in the form of confidence intervals, while guaranteeing DP in all steps. We perform an extensive experimental analysis of \oursys\ with multiple use-cases on real and synthetic data showing that \oursys\ efficiently provides insightful explanations with good accuracy and utility.
\end{abstract}

\maketitle

\iffullversion
\settopmatter{printfolios=true} 
\fi

\input{introduction_new}
\input{preliminary}
\input{model}
\input{question_ci}
\input{influence}

\input{topk_new}

\input{influence_ci}

\input{rank_ci_new}
\input{evaluation}
\input{related_work}
\input{future}

\begin{acks}
This work was supported by the NSF awards IIS-2016393, IIS-1552538, IIS-2008107, IIS-2147061, and IIS-1703431.
\end{acks}

\clearpage

\clearpage
\balance
\bibliographystyle{abbrv}
\bibliography{bibtex}

\iffullversion
\clearpage
\input{appendix}

\else
\fi

\clearpage
\end{document}

%% file: introduction_new.tex
\section{Introduction}\label{sec:intro}
\emph{Differential privacy (DP)} \cite{dwork2006calibrating, dwork2014algorithmic, dwork2016concentrated, bun2016concentrated} is the gold standard for protecting privacy in query processing and is critically important for sensitive data analysis. It has been widely adopted by organizations like the U.S. Census Bureau \cite{abowd2018us, dwork2019differential, ruggles2019differential, kenny2021use} and companies like Google \cite{erlingsson2014rappor, wilson2019differentially}, Microsoft \cite{ding2017collecting}, and Apple \cite{tang2017privacy}. 
The core idea behind DP is that a query answer on the original database cannot be distinguished from the same query answer on a slightly different database. This is usually achieved by adding random noise to the query answer to create a small distortion in the answer.
Recent works have made significant advances in the usability of DP, allowing for complex query support \cite{kotsogiannis2019privatesql, wilson2019differentially, kotsogiannis2019architecting, johnson2018towards, DBLP:journals/pvldb/McKennaMHM18, tao2020computing, dong2022r2t}, and employing DP in different settings \cite{dong2022r2t, tao2020computing, ferrando2020general, qiao2021oneshot, DBLP:conf/icml/GillenwaterJK21, yan2020private}. 
These works assist in bridging the gaps between the functionality of non-DP databases and databases that employ DP. 

Automatically generating meaningful \emph{explanations} for query answers in response to questions asked by users is an important step in data analysis that can significantly reduce human efforts and assist users. 
Explanations help users validate query results, understand trends and anomalies, and make decisions about next steps regarding data processing and analysis, thereby facilitating data-driven decision making. 
Several approaches for explaining aggregate and non-aggregate query answers have been proposed in  database research, including intervention \cite{wu2013scorpion, RoyS14, RoyOS15}, Shapley values \cite{LivshitsBKS20}, counterbalance \cite{miao2019going}, (augmented) provenance \cite{amsterdamer2011provenance,li2021putting}, 
responsibility \cite{MeliouGNS11, MeliouGMS11}, 
and entropy \cite{Gebaly+2014-expltable} (discussed in Section~\ref{sec:related}). 
\par
One major gap that remains wide open is 
to provide 
explanations for analyzing query answers from sensitive data under DP.
Several new challenges arise from this need.
First, in DP, 
the (aggregate) query answers shown to users are distorted due to the noise that must be added for preserving privacy, 
so 
the explanations need to separate the contributions of the noise from the data. Second, even after removing the effect of noise, new techniques have to be developed to provide explanations based on the sensitive data and measure their effects. 
For instance, standard explanations methods in non-DP settings are typically deterministic, while it is known that DP methods must be randomized. Therefore, no deterministic explanations can be provided, 
even no deterministic scores or ranks of explanations can be displayed in response to user questions if we want to guarantee DP in the explanation system. 
Third, the system needs to ensure that the returned explanations, scores, and ranks still have high accuracy while being private.  

In this paper, we 
propose \oursys, a novel three-phase framework 
that generates explanations \footnote{\rev{The explanations we provided should not be considered as causal explanations.
}} under DP for aggregate queries based on the notion of \emph{intervention} \cite{wu2013scorpion, RoyS14} 
\footnote{
\rev{A graphical user interface for \oursys\ is an onging work.}
}
. 
\oursys\ surmounts the 
aforementioned challenges
and is the first system combining DP and explanations to the best of our knowledge.
We illustrate \oursys\ through an example.



\begin{figure*}[t]
    \centering
    \begin{subfigure}[b]{0.49\linewidth}
        {\footnotesize
        \setlength\tabcolsep{2pt}
        \begin{center}
        \ttfamily
        \begin{tabular}{|l|l|l|l|l|}
            \hline
            \rowcolor{gray!40}
            \textbf{marital-status} & \textbf{occupation} & $\ldots$ & \textbf{education} & \textbf{high-income} \\
            \hline
            Never-married & Machine-op-inspct & $\ldots$ & 11th & 0 \\ 
            Married-civ-spouse & Farming-fishing & $\ldots$ & HS-grad & 0 \\ 
            Married-civ-spouse & Machine-op-inspct & $\ldots$ & Some-college & 1 \\
            $\ldots$ & $\ldots$ & $\ldots$ & $\ldots$ & $\ldots$ \\ \hline
        \end{tabular}
        \end{center}
        }
        \caption{Example of the \dataset{Adult} dataset.}
        \label{fig:intro_example_data_adult}
        \begin{tcolorbox}[colback=white,left=1.5pt,right=1.5pt,top=0pt,bottom=0pt]
        {\bf \small Question-Phase-1:}\\[-2pt]
        {\small
        SELECT {\tt marital-status}, AVG({\tt high-income}) as {\tt avg-high-income}\\[-3pt]
        FROM Adult GROUP BY {\tt marital-status};\\
        }
        
        \vspace{-6pt}
        \begin{minipage}{\linewidth}
        \begin{minipage}{0.2\linewidth}
        {\bf \small Answer-Phase-1:}
        \end{minipage}
        \hfill
        \begin{minipage}{0.79\linewidth}
            \footnotesize
            \setlength\tabcolsep{1.5pt}
            \
            \begin{tabular}{r|c | c} 
            
                \hline
                \textbf{\tt group} & \textbf{\privans} & \EFFORT{25}\textbf{\trueans}  \\
                \tt marital-status & \tt avg-high-income & \EFFORT{25}(hidden) \\ \hline
                    \tt Never-married  & 0.045511 & \EFFORT{25} 0.045480 \\
                    \tt Separated  & 0.064712 & \EFFORT{25}0.064706 \\
                    \tt Widowed  & 0.082854 & \EFFORT{25}0.084321 \\
                    \tt Married-spouse-absent  & 0.089988 & \EFFORT{25}0.092357 \\
                    \tt Divorced  & 0.101578 & \EFFORT{25}0.101161 \\
                    \tt Married-AF-spouse  & 0.463193 & \EFFORT{25}0.378378 \\
                    \tt Married-civ-spouse  & 0.446021 & \EFFORT{25}0.446133 \\
                 \bottomrule
            \end{tabular}
        \end{minipage}
        \end{minipage}
        \end{tcolorbox}
    \caption{Phase-1 of \oursys: Run a query and receive noisy answers by DP. \trueans s are not visible to the user and for illustration only.}
    \label{fig:phase1}
    \end{subfigure}
    \hfill
    \begin{subfigure}[b]{0.49\linewidth}
        \begin{tcolorbox}[colback=white,left=1.5pt,right=1.5pt,top=0pt,bottom=0pt]
        {\bf \small Question-Phase-2:}
        {\small Why {\tt avg-high-income} of group {\bf \ttfamily "Married-civ-spouse"} $>$ that of group {\bf \ttfamily "Never-married"}?}\\
        
        \vspace{-6pt}
        
        {\bf \small  Answer-Phase-2:}
        {\small 
        The 95\% confidence interval of group difference is $(0.399, 0.402)$, 
        hence the noise in the query is possibly not the reason.
        }
        \end{tcolorbox}
        \caption{Phase-2 of \oursys: Ask a comparison question and receive a confidence interval of the comparison.}
        \label{fig:phase2-B}
        \begin{tcolorbox}[colback=white,left=1.5pt,right=1.5pt,top=0pt,bottom=0pt]
            {\bf \small Answer-Phase-3:}
            \normalsize
            \setlength\tabcolsep{1.5pt}
            \begin{center}
            {\footnotesize
            \begin{tabular}{@{} >{\raggedright}p{4.2cm} c c P{0.8cm} P{0.8cm} @{}}
                \toprule
                \multirow{2}{*}{explanation predicate} &
                \multicolumn{2}{c}{Rel Influ 95\%-CI} &
                \multicolumn{2}{c}{Rank 95\%-CI} \\
                \cmidrule(lr){2-3}
                \cmidrule(lr){4-5}
                & L & U & L & U \\
                \midrule
{\tt occupation = "Exec-managerial"} & 3.25\% & 10.12\%  & 1  & 9 \\
{\tt        education = "Bachelors"} & 2.93\% & 9.80\%  & 1  & 8 \\
{\tt               age = "(40, 50]"} & 2.76\% & 9.63\%  & 1  & 8 \\
{\tt  occupation = "Prof-specialty"} & 0.94\% & 7.81\%  & 1 & 18 \\
{\tt     relationship = "Own-child"} & -0.49\% & 6.38\%  & 1 & 96 \\
                \bottomrule
            \end{tabular} 
            }
            \end{center}
        \end{tcolorbox}  
        \caption{Phase-3 of \oursys: Receive an explanation table from data for the previous question that passed Phase-2.}
        \label{fig:phase3}
    \end{subfigure}
    \caption{Database instance and the three phases of the \oursys\ framework.}
    \vspace{-3mm}
    \label{fig:framework}
\end{figure*}

\sloppy
\begin{example}\label{ex:intro}
Consider the \dataset{Adult} (a subset of Census) dataset \cite{Dua:2019} with 48,842 tuples.
We consider the following attributes:
{\tt age,}
{\tt workclass,}
{\tt 
education,}
{\tt 
marital-status,}
{\tt 
occupation,}
{\tt 
relationship,}
{\tt 
race,}
{\tt 
sex,}
{\tt 
native-country,}
{\tt 
and high-income},
where {\tt high-income} is a binary attribute indicating whether the income of a person is above 50K or not; 
some relevant columns are illustrated in \Cref{fig:intro_example_data_adult}. 
\end{example}

\sloppy
In the {\bf first phase (Phase-1)} of \oursys, the user submits a query and gets the results as shown in Figure~\ref{fig:phase1}.
This query is asking the fraction of people with high income in each {\tt marital-status} group. As \Cref{fig:phase1} shows, the framework returns the answer with two columns: {\tt group} and {\tt \privans}. 
Here {\tt group} corresponds to the group-by attribute  {\tt marital-status}. However, since the data is private, instead of seeing the actual aggregate values {\tt avg-high-income}, the user sees a perturbed answer {\tt \privans} for each group as output by some differentially private mechanism with a given privacy budget (here computed by the Gaussian mechanism with privacy budget $\rho = 0.1$ \cite{bun2016concentrated}). 
The third column {\tt \trueans}\ shown in grey (hidden for users) in \Cref{fig:phase1} shows the {\bf true aggregated output} for each group. 



    \cut{
    \begin{subfigure}[b]{0.28\linewidth}
        \centering
        ``\texttt{SELECT marital-status, AVG(high\_income) FROM R GROUP BY marital-status}''
        \caption{Query}
        \label{fig:demo_query}
    \end{subfigure}

    }

In the \textbf{second phase  (Phase-2)} of \oursys, the user selects two groups to compare their aggregate values and asks for explanations. However, unlike standard explanation frameworks \cite{wu2013scorpion, RoyS14, Gebaly+2014-expltable, miao2019going, li2021putting} where the answers of a query are correct and hence the question asked by the user is also correct, in the DP setting, the answers that the users see are perturbed. Therefore, the user question and the direction of comparison may not be valid. 
Hence our system first tests the validity of the question. If the question is valid, our system provides a data dependent explanation of the user question. We explain this below with the running example. 

\begin{figure}[h]
\begin{tcolorbox}[colback=white,left=1.5pt,right=1.5pt,top=0pt,bottom=0pt]
{\bf \small Question-Phase-2:}
{\small Why avg-high-income of group {\bf \ttfamily "Married-AF-spouse"} $>$  that of group {\bf \ttfamily "Married-civ-spouse"}?}\\

\vspace{-6pt}

{\bf \small Answer-Phase-2:}
{\small The 95\% confidence interval of group difference is $(-0.259, 0.460)$, 
hence the noise in the query is possibly the reason.}
\end{tcolorbox}
\caption{A user question 
explained by high noise. 
}
\label{fig:phase2-A}
\end{figure}

First consider the question in \Cref{fig:phase2-A} comparing the last two groups in \Cref{fig:phase1} (spouse in armed forces vs. a civilian). 
In this example, even though the noisy {\tt avg-high-income} for {\tt "Married-AF-spouse"} is larger than the noisy value for {\tt "Married-civ-spouse"}, this might not be true in the real data (as is the case in the {\tt \trueans} column). Hence, our system tests whether the user question could potentially be explained just using the noise introduced by DP rather than from the data itself. To do this, our system tests the validity of the user question by computing a confidence interval around the difference between these two outputs. In this case, the confidence interval is $(-0.259, 0.460)$. Since it includes 0 and negative values, we cannot conclude with high probability that {\tt "Married-AF-support"} > {\tt "Married-civ-spouse"} is true in the original data.
{\bf Since the validity of the user question is uncertain, we know that any further explanation might not be meaningful and the user may choose to stop here
.} In other words, the explanation for the comparison in the user question is primarily attributed to the added noise by the DP mechanism. 
\rev{If the user chooses to proceed to the next phase for further explanations from the data, they might not be meaningful.}







%
%
Now consider the comparison between two other groups 
{\tt "Never-married"} and {\tt "Married-civ-spouse"}, 
in \Cref{fig:phase2-B}.
In this case, the confidence interval about the difference 
does not include zero and is tight around a positive number 0.4, which indicates that the 
user question is correct with high probability. 
\rev{Notice that it is still possible for a valid question to have a confidence interval that includes zero given sufficiently large noise.}
Since the question is valid, the user may continue to the next phase. 


In the {\bf third phase (Phase-3)} of \oursys, for the questions that are likely to be valid, 
\oursys\ can provide a further detailed data dependent explanation for the question.
To achieve this again with DP, our framework reports an ``{\bf \exptab}''\footnote{We note that our notion of explanation table is unrelated to that described by Gebaly et al. \cite{Gebaly+2014-expltable} for summarizing dimension attributes to explain a binary outcome attribute.} to the user as \Cref{fig:phase3} shows, which includes the top-5 {\em explanation predicates}. The explanation predicates explain the user question using the notion of \emph{intervention} as done in previous work \cite{wu2013scorpion, RoyS14} for explaining aggregate queries in non-DP setting. Intuitively, if we intervene on the database by (hypothetically) removing tuples that satisfy the predicate, and re-evaluate the query, then the difference in the aggregate values of the two groups mentioned in the question will reduce. In the simplest form, explanation predicates are singleton predicates of the form ``{\tt attribute = <value>}'', while in general, our framework supports more complex predicates involving conjunction, disjunction, and comparison ($>, \geq$ etc.).
In \Cref{fig:phase3}, the top-5 simple explanation predicates, as computed by \oursys, are shown out of 103 singleton predicates, according to their influences to the question but perturbed by noises to satisfy DP.
The amount of noise is proportional to the sensitivity of the influence function, the maximum possible change of the influence of any explanation predicate when adding or removing a single tuple from the database.
Once the top-5 predicates are selected, the explanation table also shows both their {\em relative influence} (intuitively, how much they affect the difference of the group aggregates in the question) and their {\em ranks} in the form of confidence interval (upper and lower bounds) to preserve DP. 
\rev{Since the selected top-k explanations are not guaranteed to be the true top-k, showing a confidence interval of rank can help give an indication on whether they are close to the true top-k. }
\cut{
The privacy budget in this step includes the mechanisms for privately selecting top-k explanation predicates, and privately computing confidence intervals of influence and 
ranks of explanations. 
}

From this table, {\tt occupation = "Exec-managerial"} is returned as the top explanation predicate, indicating that the people with this job contribute more to the average high income of the married group compared to the never-married group. In other words, managers tend to earn more if they are married than those who are single, which probably can be attributed to the intuition that married people might be older and have more seniority, which is consistent with the third explanation {\tt age = "(40, 50]"} in Figure~\ref{fig:phase3} as well.
Although these explanations are chosen at random, 
\iffullversion
(see \Cref{fig:another_phase3} for another random example)
\fi
we observe that the first three explanations are almost constantly included.
This is consistent with the narrow confidence interval of rank for the first three explanation predicates, which are all around $[1, 8]$.
Looking at the confidence intervals of the relative influence and ranks in the explanation table, the user also knows that the first three explanations are likely to have some effect on the difference between the married and unmarried groups. However, for the last two explanations,
the confidence intervals of influences are closer to 0 and the confidence intervals of ranks are wider, especially for the fifth one which includes negative influences in the interval and has a wide range of possible ranks (96 out of 103 simple explanation predicates in total).\\
\subsection*{Our Contributions}
\begin{itemize} [leftmargin=0.1in, topsep=0pt]
\item We develop \oursys, the first framework, to our knowledge, that generates explanations for query answers under DP adapting the notion of intervention \cite{wu2013scorpion, RoyS14}. It explains user questions comparing two group-by aggregate query answers (COUNT, SUM, or AVG) with DP in three phases: private query answering, private user question validation, and private explanation table. 
\rev{
\item We develop multiple novel techniques that allow DPXPlain to provide explanations under DP including (a) computing confidence intervals to check the validity of user questions, (b) choosing explanation predicates, and (c) computing confidence intervals around the influence and rank of the predicates.
\item We design a low sensitivity influence function inspired by previous work on non-private explanations \cite{wu2013scorpion}, which is the key to the accurate selection of the top-k explanation predicates.
\item We design an algorithm that uses a noisy binary search technique to find the confidence intervals of the explanation ranks. This algorithm is able to overcome the high sensitivity challenge of the rank function.
}
\item We have implemented a prototype of \oursys\ \cite{fullversion} 
to evaluate our approach. We include two case studies 
on a real and a synthetic dataset showing the entire process and the obtained explanations. We have further preformed a comprehensive accuracy and performance evaluation, showing that \oursys\ correctly indicates the validity of the question with 100\% accuracy for 8 out of 10 questions, selects at least 80\% of the true top-5 explanation predicates correctly for 8 out of 10 questions, and generates 
descriptions about their influences and ranks with high accuracy. 
\end{itemize}

%% file: preliminary.tex
\section{Preliminaries}\label{sec:preliminary}
We now give the necessary background for our model.
\cut{\yuchao{Remove this sentence if we don't show the notation table.}
The notations used in the paper (from Sections~\ref{sec:preliminaries} and \ref{sec:model}) are summarized in Table \ref{tbl:notations}.}
%
The \oursys\ framework supports single-block SELECT - FROM - WHERE - GROUP BY queries with aggregates (\Cref{fig:query}) on single tables\footnote{Unlike some standard explanation framework \cite{wu2013scorpion}, in DP, we cannot consider materialization of join-result for multiple tables, since the privacy guarantee depends on {\em sensitivity}, and removing one tuple from a table may change the join and query result significantly. We leave it as an interesting future work.}. Hence the database  schema $\attrset = (A_1, \ldots, A_m)$ is a vector of attributes of a single relational table. Each attribute $A_i$ is associated with a domain $\dom(A_i)$, which can be continuous or categorical.
A database (instance) $D$ over a schema $\attrset$ is a bag of tuples (duplicate tuples are allowed) $t_i = (a_1, \ldots, a_m)$, where $a_i \in \dom(A_i)$ for all $i$. The domain of a tuple is denoted as $\dom(\attrset) = \dom(A_1) \times \dom(A_2) \times \ldots \times \dom(A_m)$.
We denote $\dommax{A_i} = \max\{\abs{a} \mid a \in \dom(A_i)\}$ as the maximum absolute value of 
$A_i$. 
The value of the attribute $A_i$ of tuple $t$ is denoted by $t.A_i$.

\begin{figure}[h]
\begin{tcolorbox}[colback=white,left=1.5pt,right=1.5pt,top=1.5pt,bottom=1.5pt]
\begin{equation*}
\small
q = \texttt{SELECT $\Agb$, agg($\Aagg$) FROM D WHERE $\phi$ GROUP BY $\Agb$;}
\end{equation*}
\end{tcolorbox}
\caption{Group-by query with aggregates supported by \oursys. The true results are denoted by $(\alpha_i, o_i)$ and the noisy results released by a DP mechanism are denoted by $(\alpha_i, \hat{o}_i)$ where $\alpha_i$ is the value of  
$\Agb$ and $o_i, \hat{o}_i$ are aggregate values.}
\label{fig:query}
\end{figure}

\sloppy
In this paper, we consider group-by aggregate queries $q$ of the form shown in Figure~\ref{fig:query}.
Here $\Agb$ is the group-by attribute and $\Aagg$ is the aggregate attribute, $\phi$ is a predicate without subqueries, and $\texttt{agg} \in$ $\{COUNT,$ $SUM, AVG\}$ is the aggregate function. 
When query $q$ is evaluated on database $D$, its result is a set of tuples $(\alpha_i, o_i)$, where $\alpha_i \in \dom(\Agb)$ and 
$o_i = agg(\{t.\Aagg ~|~ t \in D, \phi(t) = true, t.\Agb = \alpha_i\})$. For brevity, we will use $\phi'(D)$ to denote $\{t ~| ~ \phi'(t) = true\}$ for any predicate $\phi'$, and $agg(\Aagg, D')$, or simply $agg(D')$ when it is clear from context, to denote $agg(\{t.\Aagg ~|~ t \in D'\})$ for any $D' \subseteq D$. Hence, $o_i = agg(\Aagg, g_i(D))$, where $g_i = \phi \land (\Agb=\alpha_i)$.

\begin{example}
\begin{sloppypar}
Consider Example \ref{ex:intro}. The schema is $\attrset$ =  ({\tt marital-status}, {\tt  occupation}, {\tt  age}, {\tt  relationship}, {\tt  race}, {\tt  workclass}, {\tt  sex}, {\tt  native-country}, {\tt  education}, {\tt  high-income)}. All the attributes are categorical attributes and the domain of {\tt high-income} is $\{0, 1\}$.  The query is shown in \Cref{fig:phase1} and the true result for each group is shown in the {\tt \trueans} column. Here $\Agb = {\tt marital-status}$, $\Aagg = {\tt high-income}$, and $\texttt{agg} = AVG$.
\end{sloppypar}

\end{example}

\paratitle{Differential Privacy}
In this work, we consider 
query-answering and providing explanations using 
\emph{differential privacy (DP)} \cite{dwork2014algorithmic} to protect  private information in the data.
In standard databases, a query result can give an adversary the option to find the presence or absence of an individual in the database, compromising their privacy. 
DP allows users to query the database without compromising the privacy by guaranteeing that the query result will not change too much (defined in the sequel) even if it is evaluated on any two different but \emph{neighboring} databases defined below.

\begin{definition}[Neighboring Database]
Two databases $D$ and $D'$ are neighboring (denoted by $D \neighbor D'$) if $D'$ can be transformed from $D$ by adding or removing
\footnote{There are two variants of 
neighboring databases. 
The definition by addition/deletion of tuples is called ``unbounded DP'', and by updating tuples is called ``bounded DP'', since the size of data is fixed. In this work, we assume the unbounded version, while \oursys\ can be adapted also for the bounded version by adapting the noise scale.}
a tuple in $D$. 
\end{definition}


To achieve DP, it is necessary to randomize the query result such that given any two neighboring databases, it is highly possible that the answers are the same. 
Informally, the more similar the two random distributions are, the harder it is to distinguish which database is the actual database, therefore the privacy is better protected.

\sloppy
In this paper, we consider a relaxation of DP called {\bf $\rho$-zero-concentrated differential privacy (zCDP)} \cite{bun2016concentrated, dwork2016concentrated} for several reasons. First, we use Gaussian noise to perturb query answers and derive confidence intervals, which does not satisfy pure $\epsilon$-DP \cite{dwork2014algorithmic} but satisfies approximate $(\epsilon, \delta)$-DP \cite{dwork2014algorithmic} and $\rho$-zCDP. Second, $\rho$-zCDP only has one parameter $\rho$, comparing to $(\epsilon, \delta)$-DP  which has two parameters, so it is easier to understand and control. Third, $\rho$-zCDP allows for tighter analyses for tracking the privacy loss over multiple private releases, which is the case for this framework.  The parameter $\rho$ is also called the {\em privacy budget} of the mechanism. A lower $\rho$ value implies a lower privacy loss.

\begin{definition}[Zero-Concentrated Differential Privacy (zCDP) \cite{bun2016concentrated}]
A mechanism $\M$ is said to be \emph{$\rho$-zero-concentrated differential private}, or $\rho$-zCDP for short, if for any neighboring datasets $D$ and $D'$ and all $\alpha \in (1, \infty)$ it holds that
\begin{align*}
    D_\alpha(\M(D) \| \M(D')) \leq \rho \alpha
\end{align*}
where $D_\alpha(\M(D) \| \M(D'))$ denotes the R\'enyi divergence of the distribution $\M(D)$ from the distribution $\M(D')$ at order $\alpha$ \cite{mironov2017renyi}.
\end{definition}


{\em Unless otherwise stated, from now on, we will refer to zero-concentrated differential privacy simply as DP. }





A popular approach for providing zCDP 
to a query result 
is to add Gaussian noise to the result before releasing it to user. 
This approach is called \emph{Gaussian mechanism} \cite{dwork2014algorithmic, bun2016concentrated}. 

\begin{definition}[Gaussian Mechanism]
Given a query $q$ and a noise scale $\sigma$,  Gaussian mechanism $\mechanism^G$ is given as:
\begin{align*}
    \mechanism^G(D; q, \sigma) = q(D) + N(0, \sigma^2)
\end{align*}
where $N(0, \sigma^2)$ is a random variable from a normal distribution\footnote{The probability density function of a normal distribution 
$N(\mu, \sigma^2)$ is given as  
$exp(-((x-\mu)/\sigma)^2 / 2) / (\sigma\sqrt{2\pi})$.
}
with mean zero and variance $\sigma^2$.
\end{definition}

\begin{example}\label{eg:gaussian}
Suppose there is a database $D$ with 100 tuples.
Consider a query $q$ = ``\texttt{SELECT COUNT(*) FROM D}'', which counts the total number of tuples in a database $D$. Here $q(D) = 100$.  Now we use Gaussian mechanism to release $q(D)$, which is to randomly sample a noise $z$ from distribution $N(0, \sigma^2)$. Here we assume $\sigma = 1$. Finally, we got a noisy result $\hat{q}(D) = 102.32$, which we may round to an integer in postprocessing without sacrificing the privacy guarantee (\Cref{prop:composition} below).
\end{example}

The 
privacy guarantee 
from the Gaussian mechanism depends on both the noise scale it uses and the sensitivity of the query. Query sensitivity reflects how sensitive the query is to the change of the input. More noise is needed for a more sensitive query to achieve the same level of privacy protection.

\begin{definition}[Sensitivity]
Given a scalar query $q$ that outputs a single number, its sensitivity is defined as:
\begin{align*}
    \Delta_q = \sup_{D \neighbor D'} |q(D) - q(D')|
\end{align*}

\end{definition}

\begin{example}
Continuing \Cref{eg:gaussian}, since the query $q$ returns the database size, for any two neighboring databases, their sizes always differ by 1, so the sensitivity of $q$ is 1.
\end{example}

The next theorem provides the bound on DP guaranteed by a Gaussian mechanism.

\begin{theorem}[Gaussian Mechanism \cite{bun2016concentrated}]
\label{thm:gaussian_mechanism}
Given a query $q$ with sensitivity $\Delta_q$ and a noise scale $\sigma$, its Gaussian mechanism $\mechanism^G$ satisfies $(\Delta_q^2 / 2\sigma^2)$-zCDP. Equivalently, given a privacy budget $\rho$, choosing  $\sigma = \Delta_q / \sqrt{2\rho}$ in Gaussian mechanism satisfies $\rho$-zCDP.
\end{theorem}

\paratitle{Composition Rules}
In our analysis, we will use the following standard composition rules and other known results from the literature of DP \cite{mcsherry2009privacy} (in particular, zCDP \cite{bun2016concentrated}) frequently:
\begin{proposition}\label{prop:composition}
The following holds for zCDP \cite{mcsherry2009privacy, bun2016concentrated}:
\begin{itemize}[leftmargin=0.1in, topsep=0pt]
    \item {\bf Parallel composition:} if two mechanisms take disjoint data as input, the total privacy loss is the maximum privacy loss from each. 
    \item {\bf Sequential composition:} if we run two mechanisms in a sequence on overlapping inputs, the total privacy loss is the sum of each privacy loss. 
    \item {\bf Postprocessing:} if we run a mechanism and postprocess the result without accessing the data, the total privacy loss is only the privacy loss from the mechanism.
\end{itemize}
\end{proposition}

\paratitle{Private Query Answering}
Recall that we have group-by aggregation query of the form $q = $ \texttt{SELECT $\Agb$, agg($\Aagg$) FROM D WHERE $\phi$ GROUP BY $\Agb$}, and it returns a list of tuples $(\alpha_i, o_i)$ where $\alpha_i \in \dom(\Agb)$ and $o_i$ is the corresponding aggregate value. Since no single tuple can exist in more than one group, adding or removing a single tuple can at most change the result of a single group. As mentioned earlier, Phase-1 returns noisy aggregate values $\hat{o}_i$ for each $\alpha_i$ instead of $o_i$.
The following holds:
\begin{observation}
According to the \emph{parallel composition rule} (\Cref{prop:composition}),
if for each $\alpha_i$, 
its (noisy) aggregate value $\hat{o}_i$ is released under $\rhoQuery$-zCDP, the entire release of results including all groups $\{\alpha_i, \hat{o}_i ~:~ \alpha_i \in \dom(\Agb)\}$ satisfies $\rhoQuery$-zCDP.
\end{observation}
For a $COUNT$ or $SUM$ query, we use the Gaussian mechanism for each group $\alpha_i$: $\hat{o}_i = o_i + N(0, \sigma^2)$, where the noise scale $\sigma = \Delta_q/\sqrt{2\rhoQuery}$ to satisfy $\rhoQuery$-zCDP by \Cref{thm:gaussian_mechanism}. 
The sensitivity term $\Delta_q$ is 1 for $COUNT$ and $\dommax{\Aagg}$ for $SUM$, the maximum absolute value of the aggregation attribute in its domain. 
For an $AVG$ query, since $AVG = SUM / COUNT$, we decompose the query into a $SUM$ query and a $COUNT$ query, privately answer each of them by half of the privacy budget $\rhoQuery/2$, get $\hat{o}^{S}_i$ and $\hat{o}^{C}_i$ for each group $\alpha_i$ for the $SUM$ query and $COUNT$ query separately
\footnote{The intermediate releases $\hat{o}^{S}_i$ and $\hat{o}^{C}_i$ are stored in the framework for computing the confidence interval of question, which will be discussed in \Cref{sec:private_question_ci}.}, and release $\hat{o}_i = \hat{o}^{S}_i / \hat{o}^{C}_i$ as a post-processing step.
The noisy query answers of the group-by query with AVG satisfies $\rhoQuery$-zCDP by the sequential composition rule (\Cref{prop:composition}), since each of $SUM$ and $COUNT$ queries satisfies $\rhoQuery/2$-zCDP. 



\paratitle{Confidence Level and Interval}
Confidence intervals are commonly used to determine the error margin in uncertain computations and are used in various fields from estimating the error in predictions by machine learning models \cite{JiangZC08} to providing query results with added noise due to DP \cite{ferrando2021parametric}. 
In our context, we use confidence intervals to measure the uncertainty in the user question and our explanations.

\begin{definition}[Confidence Level and Interval \cite{wasserman2004all}]\label{def:conf-interval}
Given a confidence level $\gamma$ and an unknown but fixed parameter $\theta$, a random interval $\I=(\I^L, \I^U)$ is said to be its confidence interval, or CI, with confidence level $\gamma$ if the following holds:
\begin{align*}
    Pr[\I^L \leq \theta \leq \I^U] \geq \gamma
\end{align*}
\end{definition}

Notice that $\theta$ is a fixed quantity and $\I^L, \I^U$ are random variables. One interpretation of a confidence interval is that with probability at least $\gamma$, a random draw of the pair $(\I^L, \I^U)$ as an interval will contain the unknown parameter $\theta$. 
Two bounds are sampled together unless they are independent.

\begin{example}
Let $\theta = 0$. Suppose with probability 50\% we have $I^L = -1$ and $I^U = 1$, and with another probability 50\% we have $I^L = 1$ and $I^U = 2$. Therefore, $Pr[\I^L \leq \theta \leq \I^U] = 50\%$, 
and we can conclude that the random interval $\I=(\I^L, \I^U)$ is a 50\% level confidence interval for $\theta$.
\end{example}

%% file: model.tex
        

\section{Private Explanations in \oursys}\label{sec:model}


In this section we provide the model for private explanations of query results in \oursys, outline the key technical problems addressed by \oursys, and highlight the difference from existing database explanation frameworks.

\paratitle{User Question and Standard Explanation Framework}
In Phase-2 of \oursys, given the noisy results of a group-by aggregation query from Phase-1, 
users can ask questions comparing the aggregate values of two groups\footnote{\rev{Our framework can handle more general user questions involving single group or more than two groups; details are deferred to 
\iffullversion
\Cref{sec:ext_general_question}.
\fi
}}:

\begin{definition}[User Question]
Given a database $D$, a group-by aggregate query $q$ as shown in \Cref{fig:query}, a DP mechanism $\mechanism$, and  two noisy answer tuples $(\alpha_i, \hat{o}_i), (\alpha_j, \hat{o}_j) \in \mechanism(D; q)$ where $\hat{o}_i > \hat{o}_j$, a \emph{user question} has the form \emph{``why is the (noisy) aggregate value $\hat{o}_i$ of group $\alpha_i$ larger than the aggregate value $\hat{o}_j$ of group $\alpha_j$?'')}, which is denoted by ``why $(\alpha_i, \alpha_j, >)$?''.
\end{definition}

\begin{example}\label{ex:user-question}
The question from \Cref{fig:phase2-B} is denoted as \emph{``why ({\tt `Married-civ-spouse', `Never-married'}, $>$)?''}.
\end{example}


To explain a user question, several previous approaches return top-k predicates that have the most influences to the group difference in the question as explanations \cite{wu2013scorpion, RoyS14, Gebaly+2014-expltable,  li2021putting}. We follow this paradigm and define explanation predicates.

\begin{definition}[Explanation Predicate]
\label{def:explanation_predicate}
Given a database $D$ with a set of attributes $\attrset$, a group-by aggregation query $q$ (\Cref{fig:query}) with group-by attribute $\Agb$ and aggregate attribute $\Aagg$ and a predicate size $l$, an explanation predicate $p$ is a Boolean expression of the form $p = \varphi_1 \land ... \land \varphi_l$, where each $\varphi_i$ has the form $A_i = a_i$ such that $A_i \in \attrset \setminus \{\Agb, \Aagg\}$ is an attribute, and $a_i \in \dom(A_i)$ is its value.
\end{definition}

\rev{We assume $\dom(A_i)$ is discrete, finite and data-independent.
We focus here on conjunction of equality predicates. However, our framework can also handle predicates that contain disjunctions and inequalities of the form $A_i \circ a_i$ where $\circ \in \{>, <, \geq, \leq, \neq \}$ when the constant $a_i$ is from a finite and data-independent set.}

\cut{
We assume that a set of explanation predicates $\mathcal{P}$ is given (e.g., all singleton predicates of the form $A = a_i$).
These explanation predicates are ranked by an influence function \cite{wu2013scorpion,RoyS14,RoyOS15} and the top-k predicates are reported to the user as the explanations for the question. 
}

\paratitle{New challenges for explanations with DP}  Unlike standard explanation framework on aggregate queries \cite{wu2013scorpion, RoyS14, li2021putting}, the existing frameworks are not sufficient to support DP and need to be adapted: (i) the question itself might not be valid due to the noise injected into the queries, (ii) the selection of top-k explanation predicates needs to satisfy DP, which further requires the influence function to have low sensitivity so that the selection is less perturbed, and  (iii) 
since the selected explanation predicates are not guaranteed to be the true top-k, it is also necessary to output extra descriptions under DP for each selected explanation predicate about their actual influences and ranks.
We detail the adjustments as follows.

\paratitle{Question Validation with DP (Phase-2)}
While the user is asking ``why is $\hat{o}_i > \hat{o}_j$?'', 
in reality, it may be the case that the true results satisfy $o_i \leq o_j$, i.e., they have opposite relationship than the one observed by the user. This indicates that $\hat{o}_i > \hat{o}_j$ is the result of the noise being added to the results. 
In this scenario, one option to explain the user's observation of $\hat{o}_i > \hat{o}_j$ will be releasing the true values (equivalently, the added exact noise values), which will violate DP. Instead, to provide an explanation in such scenarios, we generate a confidence interval for the  difference of two (hidden) aggregate values $o_i - o_j$, which can include negative values 
(discussed in detail in Section \ref{sec:private_question_ci}). 
This leads to the first problem we need to solve in the \oursys\ framework:

\cut{
\paratitle{First Problem of \oursys: Private confidence interval of user question}
Given a user question w.r.t two query results $\userquery = (\alpha_i, >, \alpha_j)$, the solution to the first problem determines with high probability whether the user question holds in the true results or is just the effect of the noise added to them. Specifically, we want to find a confidence interval for $o_i - o_j$.  }

\begin{problem}[Private Confidence Interval of Question]
\label{pro:private_question_ci}
Given a dataset $D$, a query $q$, a DP mechanism $\mechanism$, a privacy budget $\rhoQuery$, a confidence level $\gamma$, and a user question 
$(\alpha_i, \alpha_j, >)$ on the noisy query answers output by $\mechanism$ satisfying 
$\rhoQuery$-zCDP, find a confidence interval (see Definition~\ref{def:conf-interval}) for the user question $\I_{\uq} = (\I_{\uq}^L, \I_{\uq}^U)$
for $o_i - o_j$ at confidence level $\gamma$ without extra privacy cost. 
\end{problem}

In Phase-2, the framework returns a confidence interval of $o_i - o_j$ to the user. If it includes zero or negative numbers, it is possible that $o_i \leq o_j$, and the user's observation of $\hat{o}_i > \hat{o}_j$ is the result of the noise added by the DP mechanism. 
In such cases, the user 
may stop at Phase-2. 
If the user is satisfied with the confidence interval for the validity of the question, she can proceed to Phase-3.

\paratitle{Influence Function (Phase-3)}
When considering DP, the order of the explanation predicates is perturbed by the noise we add to the influences according to the sensitivity of the influence function (discussed in detail in \Cref{sec:private_topk}). To provide useful explanations, this sensitivity needs to be low, which means the influence does not change too much by adding or removing a tuple from the database. 
For example, a counting query that outputs the database size $n$ has sensitivity 1, since its result can only change by 1 for any neighboring databases. 
Following this concept, we propose the second and a core problem 
for the \oursys\ framework, which is also critical to the subsequent problems defined below.

\begin{problem}[Influence Function with Low Sensitivity]
\label{pro:influence}
Find an influence function $\imp: \mathcal{P} \rightarrow \mathcal{R}$ that maps an explanation predicate to a real number and has low sensitivity. 
\end{problem}

\paratitle{Private Top-$k$ Explanations (Phase-3)}
In \oursys, to satisfy DP, in Phase-3 we output the top-$k$ explanation predicates ordered by the noisy influences, and release the influences and ranks of these predicates in the form of confidence intervals to describe the uncertainty. 
To achieve this goal, we tackle the following three sub-problems. 

\begin{problem}[Private Top-$k$ Explanation Predicates]
\label{pro:private_topk}
Given a set of explanation predicates $\predspace$, an integer $k$, and a privacy parameter $\rhoTopk$, find the top-k highest influencing predicates $p_1, p_2, \ldots, p_k$ from $\predspace$ while satisfying $\rhoTopk$-zCDP. 
\end{problem}


\begin{problem}[Private Confidence Interval of Influence]
\label{pro:private_influence_ci}
Given a confidence level $\gamma$, 
k explanation predicates $p_1, p_2, \ldots, p_k$, and 
a privacy parameter $\rhoInflu$,
find a confidence interval 
$\I_{\influ} = (\I_{\influ}^L, \I_{\influ}^U)$ for influence $\imp(p_u)$ at confidence level $\gamma$ for each $u \in \{1, \ldots, k\}$
satisfying $\rhoInflu$-zCDP (overall privacy budget). 
\end{problem}

\begin{problem}[Private Confidence Interval of Rank]
\label{pro:private_rank_ci}
Given a confidence level $\gamma$, 
k explanation predicates $p_1, p_2, \ldots, p_k$, and a privacy parameter $\rhoRank$, find a confidence interval $\I_{\rnk} = (\I_{\rnk}^L, \I_{\rnk}^U)$ for 
rank of $p_u$ at confidence level $\gamma$ for each $u \in \{1, \ldots, k\}$ satisfying $\rhoRank$-zCDP (overall privacy budget). 
\end{problem}

%% file: question_ci.tex
\section{Computing Explanations Under DP}
Next we provide solutions to problems \ref{pro:private_question_ci}, \ref{pro:influence}, \ref{pro:private_topk}, \ref{pro:private_influence_ci}, and \ref{pro:private_rank_ci} in Sections \ref{sec:private_question_ci}, \ref{sec:influence}, \ref{sec:private_topk}, \ref{sec:private_influence_ci}, and \ref{sec:private_rank_ci} respectively, and 
analyze their properties.
We summarize the entire \oursys\ framework in \Cref{sec:put_it_together}.

\subsection{Confidence Interval for a User Question} 
\label{sec:private_question_ci}

For {\bf \Cref{pro:private_question_ci}}, the goal is to find a confidence interval of $o_i - o_j$ for the user question at the confidence level $\gamma$ without extra privacy cost in Phase-2. We divide the solution into two cases. (1) When the aggregation is COUNT or SUM, the noisy difference $\hat{o}_i - \hat{o}_j$ follows Gaussian distribution, which leads to a natural confidence interval. (2) When the aggregation is AVG, the noisy difference does not follow Gaussian distribution, but we 
show that the confidence interval in this case can be derived through multiple partial confidence intervals. The solutions below only take the noisy query result as input, which does not incur extra privacy loss according to the post-processing property of DP
(\Cref{prop:composition}). 
The pseudo codes can be found in
\iffullversion.
\Cref{sec:ext_ci_question}
\fi

\paratitle{Confidence interval for COUNT and SUM}
For a COUNT or SUM query, recall from \Cref{sec:preliminary} that $\hat{o}_i$ and $\hat{o}_j$ are produced by adding Gaussian noises to $o_i$ and $o_j$ with some noise scale $\sigma$. Therefore, the difference between $\hat{o}_i$ and $\hat{o}_j$ also follows Gaussian distribution with mean $o_i - o_j$ and scale  $\sqrt{2}\sigma$ (since the variance is $2 \sigma^2$).
Following the standard properties of Gaussian distribution, the interval with center $c$ as $\hat{o}_i - \hat{o}_j$ and margin $m$ as $\sqrt{2}(\sqrt{2}\sigma)\erf^{-1}(\gamma)$ ~\footnote{$\erf^{-1}$ is the inverse function of the error function $\erf z= (2/\sqrt{\pi})\int_{0}^{z} e^{-t^2}dt$. 
}, or (c-m, c+m),  
is a $\gamma$ level confidence interval of $o_i - o_j$ \cite{wasserman2004all}. 

\paratitle{Confidence interval for AVG}
For an AVG query,
even the single noisy answer $\hat{o}_i$ does not follow Gaussian distribution, because it is a division between two Gaussian variables as described in \Cref{sec:preliminary}: $\hat{o}_i = \hat{o}^{S}_i / \hat{o}^{C}_i$. 
\rev{However, we can still infer a range for $\hat{o}_i$ based on the confidence intervals of $\hat{o}^{S}_i$ and $ \hat{o}^{C}_i$.}
\rev{More specifically,} we first derive partial confidence intervals for $o^{S}_i$ and $o^{C}_i$ as discussed above, denoted by $\I^{S}$ and $\I^{C}$,  individually at some confidence level $\beta$. Let $I^A = \I^{S} / \I^{C} \coloneqq \{x/y \mid x \in \I^{S}, y \in \I^{C}\}$ 
\footnote{In the algorithm, we only need the maximum and the minimum of the set to construct the interval, which can be solved by a numerical optimizer.} 
to be the set that includes all possible divisions between any numbers from $\I^S$ and $\I^C$.
Especially, if $I^{C}$ contains zero, we return a trivial confidence interval $(\infty, -\infty)$ that is always valid. Otherwise, 
$I^A$ is a $2 \beta - 1$ level confidence interval for the division, as stated in the following proposition. 


\begin{restatable}{lemma}{avgCI}
\label{lem:avg_ci}
Given $\I^S$ and $\I^C$ as two $\beta$ level confidence intervals of $o^{S}_i$ and $o^{C}_i$ separately, the derived interval $\I^A = \{x/y \mid x \in \I^S, y \in \I^C\}$ is a $2 \beta - 1$ level confidence interval of $o_i^{S} / o_i^{C}$.
\end{restatable}
\begin{proof}

The following holds:\\
{
\small
$
    Pr[o_i^{S} / o_i^{C} \in \I^{A}] 
    \geq Pr[o_i^{S} \in \I^{S} \land  o_i^{C} \in \I^{C}] 
    \geq 1 - (Pr[o_i^{S} \not\in \I^{S}] + Pr[o_i^{C} \not\in \I^{C}]) \geq 1 - ((1 - \beta) + (1 - \beta))  = 2 \beta - 1 
$
}%
The first inequality above is due to fact that the second event is sufficient for the first event\rev{: if two numbers are from $\I^{S}$ and $\I^{C}$, their division belongs to the set $\I^{A}$ by definition.} The next inequality holds by applying the union bound. \rev{The third inequality is by definition.}
\end{proof}

Furthermore, the difference $\hat{o}_i - \hat{o}_j$ is a subtraction between two ratios of two Gaussian variables, which can be expressed as an arithmetic combination of multiple Gaussian variables: $\hat{o}_i - \hat{o}_j = X_i / Y_i - X_j/Y_j$, where $X_t = N(o_t^S, \sigma_{S}^2)$ and $Y_t = N(o_t^C, \sigma_{C}^2)$ for $t \in \{i, j\}$. Similar to \Cref{lem:avg_ci}, we can derive the confidence interval for $\hat{o}_i - \hat{o}_j$  based on 4 partial confidence intervals of $o_i^S$, $o_i^C$, $o_j^S$, and $o_j^C$ instead of 2. 
The confidence level we set for each partial confidence interval is $\beta = 1-(1-\gamma)/4$ by applying union bound on the failure probability $1-\gamma$ that one of the four variables is outside its interval. After we have 4 partial confidence intervals $\I_i^{S}$, $\I_i^{C}$, $\I_j^{S}$, and $\I_j^{C}$ for $o_i^S$, $o_i^C$, $o_j^S$, and $o_j^C$ separately, similar to \Cref{lem:avg_ci}, we combine them together as $\I^A = \I_i^{S} / \I_i^{C} - \I_j^{S} / \I_j^{C} $ and derive the confidence interval for $o_i - o_j$ as $(\inf \I^A, \sup \I^A)$, which is guaranteed to be at confidence level $\gamma$. If 0 is included in either $\I_i^{C}$ or $\I_j^{C}$, we set the confidence interval to be $(\infty, -\infty)$ instead. 
\rev{
Although there is no theoretical guarantee of the interval width, in two case studies at \Cref{sec:case_studies}, we demonstrate narrow confidence intervals of AVG queries in practice, and observe no extreme case $(\infty, -\infty)$ in the experiments.
}

%% file: influence.tex
\subsection{Influence Function with Low Sensitivity}
\label{sec:influence}

For \textbf{\Cref{pro:influence}}, the goal is to design an influence function that has low sensitivity. Inspired by PrivBayes \cite{zhang2017privbayes}, we start by adapting a known influence function to our framework, and then adapt it to have a low sensitivity.

Our influence function of an explanation predicate with respect to a comparison user question is inspired by the Scorpion framework \cite{wu2013scorpion}, where the user questions seek explanations for outliers in the results of a group-by aggregate query.
The Scorpion framework identifies predicates on input data that cause the outliers to disappear from the output results.
Given the group-by aggregation query shown in \Cref{fig:query} and a group $\alpha_i \in \dom(\Agb)$, recall from Section~\ref{sec:preliminary} that 
the true aggregate value for $\alpha_i$ is $o_i = agg(\Aagg, g_i(D))$, where $g_i = \phi \land (\Agb=\alpha_i)$, i.e., $g_i(D)$ denotes the set of tuples that contribute to the group $\alpha_i$.

\cut{
$g_i = $
we denote by $g_i(D) \subseteq D$  the set of tuples = $\{t ~|~ \phi(t) = True \land t[\Agb]=\alpha_i\}$ as the tuples selected for group $\alpha_i$, and $q(g_i(D)) = agg(g_i(D)[\Aagg])$ as the aggregation result of group $\alpha_i$.
}Scorpion measures the influence of an explanation predicate $p$ to some group $\alpha_i$ as the ratio between the change of output aggregate value and the change of group size: 

\vspace{-1pt}
\begin{footnotesize}
\begin{equation}
    \frac{agg(g_i(D)) - agg(g_i(\neg p(D)))}{\abs{g_i(p(D))}}
    \label{eq:scorpion}
\end{equation}
\end{footnotesize}
Here $\neg p(D)$ denotes $D - p(D)$, i.e., the set of tuples in $D$ that do not satisfy the predicate $p$.
To adapt this influence function to \oursys, 
we make the following two changes.
\begin{itemize}[leftmargin=0.1in,nolistsep]
    \item First, it should measure the influence w.r.t. the comparison from the user question $(\alpha_i, \alpha_j, >)$ instead of a single group. 
    A natural extension is to change the target aggregate on $g_i$ in the numerator in (\ref{eq:scorpion}) 
    to the difference between the aggregate values of two groups $g_i, g_j$
    before and after applying the explanation predicate $p$, and change the denominator 
    as the maximum change in $g_i$ or $g_j$ when $p$ is applied, which gives the following influence function:  
    \begin{footnotesize}
    \begin{equation}
        \frac{\left(agg(g_i(D)) - agg(g_j(D))\right)
          - \left(agg(g_i(\neg p (D))) - agg(g_j(\neg p (D)))\right)}
            {\max (\abs{g_i(p(D))}, \abs{g_j(p(D))})}
            \label{eq:scorpion-adapt}
    \end{equation}
    \end{footnotesize}
    \noindent
    
    \item Second and more importantly, in \oursys, we need to preserve DP when we use influence function to sort and rank multiple explanation predicates, or to release the influence and rank of an explanation predicate. Therefore, {\bf we need to account for the sensitivity of the influence function}, which is determined by the worst-case change of influence when a tuple is added or removed from the database. If the predicate only selects a small number of tuples, the denominator in (\ref{eq:scorpion-adapt})
    is small and thus changing the denominator in (\ref{eq:scorpion-adapt}) by one (when a tuple is added or removed) can result in a big change in the influence as illustrated in the following example, making 
    (\ref{eq:scorpion-adapt}) unsuitable for \oursys.
\end{itemize}

\begin{example}[The Issue of the Influence Sensitivity]
\label{emp:high_sens_influ}
\sloppy
Suppose there are two groups $\alpha_i$ and $\alpha_j$ in $D$ with 1000 tuples in each, 
aggregate function $agg = SUM$ on attribute $\Aagg$ with domain $[0, 100]$, and the explanation predicate $p$ matches only 1 tuple from the group $\alpha_i$ with $\Aagg = 100$ and no tuple from $\alpha_j$. Suppose $agg(g_i(D)) = 20,000$, $agg(g_j(D)) = 10,000$, then $agg(g_i(\neg p(D))) = 19,900$ and $agg(g_j(\neg p(D))) = 10, 000$. Therefore, from \Cref{eq:scorpion-adapt}, the influence of $p$ is $((20,000 - 10,000) - (19,900 - 10,000)) / \max \{1, 0\} = 100$ on the original database $D$. However, suppose a new tuple that satisfies $p$ and belongs to group $\alpha_i$ is added with $\Aagg$ = 2. Now the influence in \Cref{eq:scorpion-adapt} becomes $((20,002 - 10,000) - (19,900 - 10,000)) / \max \{2, 0\} = 102/2 = 51$. Note that while we added a tuple contributing only 2 to the sum, it led to a change of 100-51 = 49 to the influence function because of the small denominator.

\end{example}


Therefore, we propose a new influence function that is inspired by \Cref{eq:scorpion-adapt} 
but has lower sensitivity. Note that the denominator in Scorpion's influence function in \Cref{eq:scorpion-adapt} acts as a normalizing factor, whose purpose is to penalize the explanation predicate that selects too many tuples, e.g., to prohibit removal of the entire database by a dummy predicate.
To have a similar normalizing factor with low sensitivity,
we multiply the numerator in \Cref{eq:scorpion-adapt} by $ \frac{
        \min(\abs{g_i(\neg p (D))}, \abs{g_j(\neg p (D)})
    }{
        \max(\abs{g_i(D)}, \abs{g_j(D)}) + 1
    }$.
From this new normalizing factor, the numerator captures the minimum of the number of tuples that are not removed from each group, and the denominator is a constant, 
which does not change for different explanation predicates and keeps the normalizing factor in the interval $[0, 1]$. Similar to Scorpion, if $p(D)$ constitutes a large fraction of $D$ (e.g., if $p(D) = D$), then the normalizing factor is small, reducing the value of the influence.
Also note that, unlike standard SQL query answering where only non-empty groups are shown in the results, in DP, all groups from the actual domain have to be considered, hence unlike \Cref{eq:scorpion}, $g_i(D), g_j(D)$ could be zero, hence  $1$ is added in the denominator to avoid division by zero. 
When $agg = AVG$, we remove the constant denominator 
to boost the signal of the influence and keep the sensitivity low, which will be discussed in the sensitivity analysis after \Cref{thm:influ_sens} and in \Cref{eg:avg-sens}.
We formally define the influence as follows.


\begin{definition}[Influence of Explanation Predicates]
\label{def:influence}
Given a database $D$, a query $q$ as shown in \Cref{fig:query}, and a user question $(\alpha_i,  \alpha_j, >)$, 
the influence of an explanation predicate $p$ is defined as $\imp(p; (\alpha_i,  \alpha_j, >), D)$, or simply $\imp(p)$ when it is clear from context:
\begin{footnotesize}
\begin{align}
    \imp(p) = & \left( 
    \left(agg(g_i(D)) - agg(g_j(D))\right)
     - \left(agg(g_i(\neg p (D))) - agg(g_j(\neg p (D)))\right)
    \right)
    \nonumber\\ 
    &\times
    \begin{cases}
        \frac{
            \min(\abs{g_i(\neg p (D))}, \abs{g_j(\neg p (D)})
        }{
            \max(\abs{g_i(D)}, \abs{g_j(D)}) + 1
        }
        & \textrm{ for } agg \in \{COUNT, SUM\} \\
        \min(\abs{g_i(\neg p (D))}, \abs{g_j(\neg p (D)})
        & \textrm{ for } agg = AVG
    \end{cases}
    \label{eq:our-influence}
    \end{align}
    \end{footnotesize}
\end{definition}

    

The proposition below guarantees the sensitivity of the influence function. 


\begin{restatable}{proposition}{influSens}[Influence Function Sensitivity]
\label{thm:influ_sens}
Given an explanation predicate $p$ and an user question with respect to a group-by query with aggregation $agg$, the following holds:
\begin{enumerate}
\itemsep0em
    \item If $agg = COUNT$, the sensitivity of $\imp(p)$ is 4.
    \item If $agg = SUM$, the sensitivity of $\imp(p)$ is $4~\dommax{\Aagg}$.
    \item If $agg = AVG$, the sensitivity of $\imp(p)$ is $16~\dommax{\Aagg}$. 
\end{enumerate}
\end{restatable}

\rev{
We give an intuitive proof as follows, where the formal proofs  
\iffullversion
can be found at \Cref{proof:influence}.
\fi
When $agg$ = COUNT, we combine two group differences
$\left(agg(g_i(D)) - agg(g_j(D))\right)
     - \left(agg(g_i(\neg p (D))) - agg(g_j(\neg p (D)))\right)$
into a single group difference as $agg(g_i(p(D)) - agg(g_j(p(D))$, which is considered as a subtraction between two counting queries. We prove that the sensitivity of a counting query after a multiplication with the normalizing factor will multiply its original sensitivity by 2. Since we have two counting queries, the final sensitivity is 4. When $agg = SUM$, the proof is similar except we need to multiply the final sensitivity by $\dommax{\Aagg}$ , the maximum absolute domain value of $\Aagg$.  For AVG, we view it as a summation of 4 AVG queries that times with $\min(\abs{g_i(\neg p (D))}, \abs{g_j(\neg p (D)})$. Intuitively, we change AVG to SUM, and, therefore, reduce to the case of SUM and bound the sensitivity. This sensitivity now becomes relatively small since we have amplified the influence. 
}

Intuitively, the sensitivity of $\imp(p)$ is low compared to its value. When $agg = COUNT$, $\imp(p)$ is $O(n)$ and $\Delta_{\imp}$ is $O(1)$, where $n$ is the size of database. When $agg \in \{SUM, AVG\}$, $\imp(p)$ is $O(n\dommax{\Aagg})$ and $\Delta_{\imp}$ is $O(\dommax{\Aagg})$. Therefore, the sensitivity of influence $\Delta_{\imp}$ is low comparing to the influence itself. However, as the example below shows, if we define the influence function for $AVG$ the same way as $COUNT$ or $SUM$, both $\imp(p)$ and $\Delta_{\imp}$ will become $O(\dommax{\Aagg})$, which makes the sensitivity (relatively) large. 

\begin{example}[The Issue with $AVG$ Influence.]\label{eg:avg-sens}
Consider an $AVG$ group-by query 
where the domain of the aggregate attribute is $[0, 100]$, and an explanation predicate $p$ such that for group $\alpha_i$ we have 2 tuples with $AVG(g_i(D)) = 100/2=50$, $AVG(g_i(\neg p(D))) = 0/1 = 0$, and for group $\alpha_j$ we
have two tuples with $AVG(g_j(D)) = 100/2=50$ and $AVG(g_j(\neg p(D))) = 100/2 = 50$. Suppose we define the influence function for $AVG$ the same way as $COUNT$ or $SUM$, therefore the influence of $p$ in \Cref{eq:our-influence} is $\imp(p) = ((50 - 50) - (0 - 50))(\min(1,2)/(\max(2,2)+1) = 50/3$. However, suppose we remove the single tuple from $g_i$, so  $|g_i(\neg p(D))|$ becomes 0, now the influence in \Cref{eq:our-influence} (for COUNT/SUM) becomes 0. Note that a single removal of a tuple completely changes the influence to 0, and this change is equal to the influence itself, which is relatively large and therefore is not a good choice for AVG. 
\end{example}

Note that the user question ``why $(\alpha_i, \alpha_j, >)$'' is asked based on the noisy results $\hat{o}_i > \hat{o_j}$, while the influence function uses the true results, i.e., even if 
$o_i \leq o_j$, we still consider $agg(g_i(D)) - agg(g_j(D))$ in $\imp(p)$. 
Hence $\imp(p)$ can be positive or negative and removing tuples satisfying $p$ can make the gap smaller or larger. In \iffullversion
\Cref{sec:influence_monotonicity}.
\fi
we show that 
$\imp(p)$ is not monotone with $p$-s.

%% file: topk_new.tex
\subsection{Private Top-k Explanations}

In this section, we discuss the computation of the top-k explanation predicates and the confidence intervals of influences and ranks.

\subsubsection{\bf Problem 3: Private Top-k Explanation Predicates}
\label{sec:private_topk}
The goal is to find  with DP the top-$k$ explanation predicates from a set of explanation predicates $\mathcal{P}$ in terms of their (true) influences $\imp(p)$, which is 
the first step in Phase-3 of \oursys\ (Figure \ref{fig:framework}). 
Note that simply choosing the \emph{true} top-$k$ explanation predicates in terms of their $\imp(p)$ is not differentially private.

In \oursys, we adopt the {\bf One-shot Top-k mechanism} \cite{DBLP:conf/nips/DurfeeR19, DPorg-one-shot-top-k} to privately select the top-$k$.
It works as follows. For each explanation predicate $p \in \mathcal{P}$, it adds a Gumbel noise
\footnote{For a Gumbel noise $Z \sim Gumbel(\sigma)$, its CDF is $Pr[Z \leq z] = \exp(-\exp(-z/\sigma))$.}
to its influence with scale $\sigma = 2\Delta_{\imp}\sqrt{k/(8\rhoTopk)}$, where $\Delta_{\imp}$ is the sensitivity of the influence function (discussed in \Cref{thm:influ_sens}), reorders all the explanation predicates in a descending order by their noisy influences, and outputs the first $k$ explanation predicates. It satisfies $\rhoTopk$-zCDP
\cite{DBLP:conf/nips/DurfeeR19, dong2020optimal, cesar2021bounding, DPorg-exponential-mechanism-bounded-range, DPorg-one-shot-top-k}, since it is equivalent to iteratively applying $k$ exponential mechanisms \cite{dwork2014algorithmic}, where each satisfies $\epsilon^2/8$-zCDP \cite{DBLP:conf/nips/DurfeeR19, dong2020optimal, cesar2021bounding, DPorg-exponential-mechanism-bounded-range} and $\epsilon = \sqrt{8\rhoTopk/k}$ \cite{DBLP:conf/nips/DurfeeR19, DPorg-one-shot-top-k}. Therefore, in total it satisfies $(k\epsilon^2/8)$-zCDP by the sequential composition property (\Cref{prop:composition}) which is also $\rhoTopk$-zCDP. The returned list of top-k predicates is close to that of the true top-k in terms of their influences; the proof is based on the utility proposition of exponential mechanism in Theorem 3.11 of \cite{dwork2014algorithmic}. 
\rev{Since this algorithm iterates over each explanation predicate, the time complexity is proportional to the size of the explanation predicate set $\mathcal{P}$. By \Cref{def:explanation_predicate}, this number is $O({m \choose l} N^l)$, where $N$ is the maximum domain size of an attribute, $l$ is the number of conjuncts in the explanation predicate and $m$ is the number of attributes. 
In our experiments (\Cref{sec:experiments}), we fix $l=1$ and use all the singleton predicates as the set $\mathcal{P}$, so its size is linear in the number of attribtues.} 
We summarize the properties of this approach in the following proposition and defer the pseudo codes and proofs to
\iffullversion
\Cref{sec:ext_ci_question} and 
\Cref{sec:private_explanation}.
\fi

\begin{restatable}{proposition}{PrivateTopk}
\label{thm:private_top_k}
Given 
an influence function $\imp$ with sensitivity $\Delta_{\imp}$, a set of explanation predicates $\mathcal{P}$, a privacy parameter $\rhoTopk$ and a size parameter $k$, the following holds:
\begin{enumerate}
    \item One-shot Top-k mechanism finds k explanation predicates while satisfying $\rhoTopk$-zCDP.
    \item Denote by $OPT^{(i)}$ the $i$-th highest (true) influence, and by $\mathcal{M}^{(i)}$ the $i$-th explanation predicate selected by the One-shot Top-k mechanism. For $\forall t$ and $\forall i \in \{1, 2, \ldots, k\}$, we have
    \begin{small}
    \begin{align}
        Pr\left[\imp(\mathcal{M}^{(i)}) \leq OPT^{(i)} - \frac{2\Delta_{\imp}}{\sqrt{8\rhoTopk/k}}(\ln(|\mathcal{P}|) + t)\right] \leq e^{-t}
        \label{eqn:topk_utility}
    \end{align}
    \end{small}
\end{enumerate} 
\end{restatable}

\begin{example}
Reconsider the user question in \Cref{fig:phase2-B}.
For this question, we have in total 103 explanation predicates as the set of explanation predicates. The privacy budget $\rhoTopk = 0.05$, the size parameter $k=5$, and the sensitivity $\Delta_{\imp} = 16$. For each of the explanation predicate, we add a Gumbel noise with scale $\sigma = 113$ to their influences. For example, for the predicates shown in \Cref{fig:phase3},
their noisy influences are 
990, 670, 645, 475, 440, 
which are the highest 5 among all the noisy influences. 
The true influences for these five ones are 
547, 501, 555, 434, 118.
To see how close it is to the true top-5, we compare their true influences with the true highest five influences:
555, 547, 501, 434, 252, which shows the corresponding differences in terms of influence are 8, 46, 54, 0, 134. 
By \Cref{eqn:topk_utility}, in theory 
the probability that such difference is beyond 864 is at most 5\% for each explanation predicate.
Finally, we sort explanation predicates by their noisy influences and report the top-k.  
These $k$ predicates will be reordered as discussed in \Cref{sec:put_it_together}.
\end{example}

%% file: influence_ci.tex
\subsubsection{\bf \Cref{pro:private_influence_ci}: Private Confidence Interval of Influence}
\label{sec:private_influence_ci}
The goal is to generate a confidence interval of influence $\imp(p)$ (\Cref{def:influence}) of each explanation predicate $\imp(p_1)$, $\imp(p_2)$, $\ldots$, $\imp(p_k)$ from the selected top-k (\Cref{sec:private_topk}).
For each $\imp(p_i)$, we apply the Gaussian mechanism (\Cref{thm:gaussian_mechanism}) with privacy budget $\rhoInflu / k$
to release a noisy influence $\noisyimp_i$ with noise scale $\sigma = \Delta_{\imp} / \sqrt{2 \rhoInflu/k}$. The sensitivity term $\Delta_{\imp}$ is determined by \Cref{thm:influ_sens}.
Following the standard properties of Gaussian distribution, for each $\imp(p_i)$, we set the confidence interval by a center $c$ as $\noisyimp_i$ and a margin $m$ as $\sqrt{2}\sigma\erf^{-1}(\gamma)$, or (c-m, c+m), as a  $\gamma$ level confidence interval of $\imp(p_i)$
\cite{wasserman2004all}. Together, it satisfies $\rhoInflu$-zCDP according to the composition property by  \Cref{prop:composition}.
Pseudo codes can be found in
\iffullversion
\Cref{sec:ext_influence_ci}.
\fi

%% file: rank_ci_new.tex
\subsubsection{\bf \Cref{pro:private_rank_ci}: Private Confidence Interval of Rank}
\label{sec:private_rank_ci}
The goal is to find the confidence interval of the rank of each explanation predicate
from the selected top-k (\Cref{sec:private_topk}). 
We denote $\rank(p)$ as the rank of $p \in \mathcal{P}$ by the natural ordering of the predicates imposed by their (true) influences according to the influence function $\imp$, and denote $\rank^{-1}(t)$ (for an integer $1\leq t \leq |\mathcal{P}|$) as the predicate ranked in the $t$-th place according to $\imp$. 
One trivial example of a confidence interval of rank is $[1, |\mathcal{P}|]$, which has no privacy loss and always includes the true rank. 

Unlike the sensitivity of the influence function, the sensitivity of $\rank(p)$ is high, since adding one tuple could possibly changing the highest influence to be the lowest and vice versa. Fortunately, we can employ a critical observation about rank and influence. 

\begin{restatable}{proposition}{rankInfluSens}
\label{thm:indicator_sens}
Given a set of explanation predicates $\mathcal{P}$, an influence function $\imp$ with global sensitivity $\Delta_{\imp}$, and an integer $1\leq t \leq |\mathcal{P}|$, $\imp(\rank^{-1}(t))$ has sensitivity $\Delta_{\imp}$.
\end{restatable}

The intuition behind this proof (details in \iffullversion
\Cref{sec:private_explanation})
\fi
 is that, fixing an explanation predicate $p = \rank^{-1}(t)$, for a neighboring database, if its influence is increased, its rank will be moved to the top which pushes down other explanation predicates with lower influences, so the influence at the rank $t$ in the neighboring database is still low. \cut{We defer the full proof to
\iffullversion
\Cref{sec:private_explanation}.
\fi
}
For a target explanation predicate $p$,
since both $\imp(p)$ and $\imp(\rank^{-1}(t))$ have low sensitivity as $\Delta_{\imp}$, intuitively we can check whether $t$ is close to the rank of $p$ by checking whether their influences $\imp(p)$ and $\imp(\rank^{-1}(t))$ are close by adding a little noise to satisfy DP.
Given this observation, we devise a binary-search based strategy to find the confidence interval of rank.

\paratitle{Noisy binary search mechanism} We decompose the problem into finding two bounds of the confidence interval separately by a subroutine $\Call{RankBound}{p, \rho, \beta, dir}$ that guarantees that it will find a lower ($dir = -1$) or upper ($dir = +1$) bound of rank with probability $\beta$ for the explanation predicate $p$ using privacy budget $\rho$. 
We \rev{divide the privacy budget $\rho$ into two parts by a parameter $\eta \in (0, 1)$ and return $(\Call{RankBound}{p_{u}, \eta\rho, \beta, -1}, \Call{RankBound}{p_{u}, (1-\eta)\rho, \beta, +1})$}
as the confidence interval of rank for each predicate $p_u$ for $u \in \{1, \ldots, k\}$, where $\rho = \rhoRank / k$ to divide the \rev{total} privacy budget equally, and $\beta = (\gamma+1)/2$ to ensure 
an overall confidence of $\gamma$. 

The subroutine $\Call{RankBound}{p, \rho, \beta, dir}$ works as follows. It is a noisy binary search with at most $N = \lceil\log_2\abs{\mathcal{P}}\rceil$ loops. We initialize the search pointers $t_{low} = 1$ and $t_{high} = \abs{\mathcal{P}}$ as the two ends of possible ranks. Within each loop, 
we check the difference of influences at $t = \lfloor(t_{high} + t_{low})/2\rfloor$ by adding a Gaussian noise: 
\begin{align}
{\small
    \hat{s} = \imp(p) - \imp(\rank^{-1}(t)) + \mathcal{N}(0, \sigma^2)
    }
\end{align}
The noise scale is set as $\sigma = (2 \Delta_{\imp}) / \sqrt{2(\rho/N)}$ to satisfy $\rho/N$-zCDP. Instead of comparing the noisy difference $\hat{s}$ with 0 to check whether $t$ is a close bound of $\rank(p)$, we compare it with the following slack constant $ \xi$ below so that $t$ is a true bound of $\rank(p)$ with high probability. 
\begin{align}
{\small
    \xi = \sigma\sqrt{2\ln(N / (1-\beta))} \times dir
    }
\end{align}
We update the binary search pointers by the comparison as follows: if $\hat{s} \geq \xi$, we set $t_{high} = \max\{t-1, 1\}$, otherwise $t_{low} = \min \{t+1, \abs{\mathcal{P}}\}$. The binary search stops when $t_{high} \leq t_{low}$ and returns $t_{high}$ as the rank bound.  \Cref{ex:rank_ci} gives an illustration. We defer the pseudo codes of the {\bf noisy binary search mechanism} to 
\iffullversion
\Cref{sec:appendix_private_rank_ci}.
\fi

\begin{figure}[h]
    \centering
    \includegraphics[width=\linewidth]{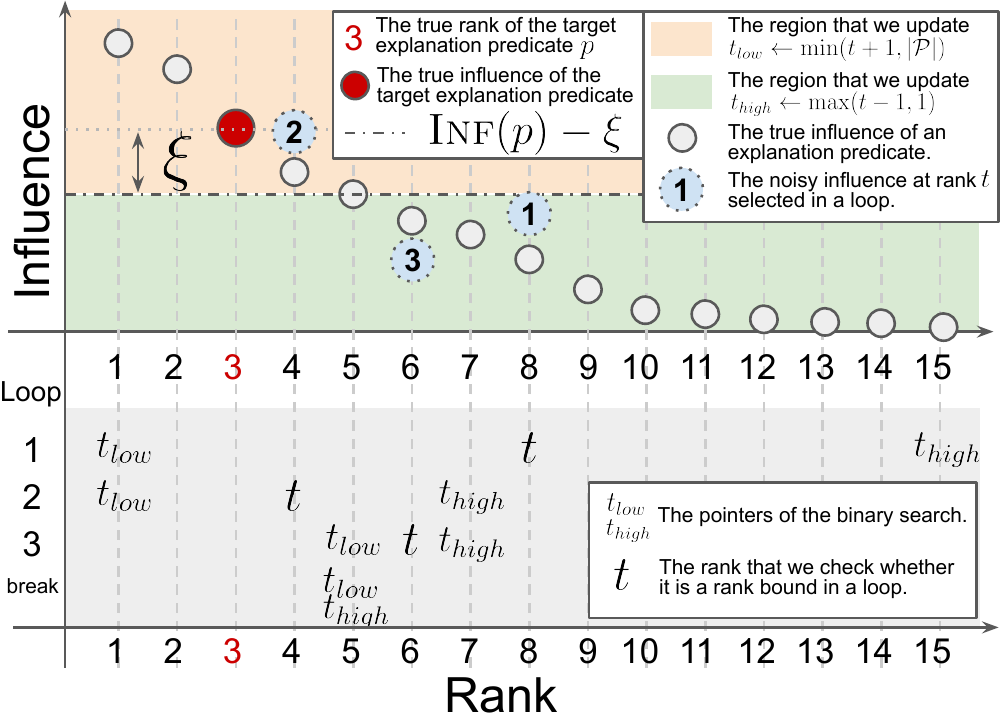}
    \caption{The execution of \textsc{RankBound} for finding the upper bound of the confidence interval of rank for the predicate $p$ (with true rank 3 shown in red) from a toy example. }
    \label{fig:diagram-ci-rank}
\end{figure}

\begin{example}
\label{ex:rank_ci}
\Cref{fig:diagram-ci-rank} shows an example of \textsc{RankBound} for finding the upper bound of the confidence interval for $\rank(p)$ for some explanation predicate $p$ (with true rank 3 shown in red). The upper part of the figure shows the influences of all the explanation predicates in a descending order, and the lower part shows the status of the binary search pointers in each loop. 
The search contains three loops starting from $t_{low} = 1$ and $t_{high} = 15$. Within each loop, to illustrate the idea, it is equivalent to adding a Gaussian noise to $\imp(\rank^{-1}(t))$, which is shown as a blue circle, compare it with $\imp(p)-\xi$, which is shown as a dashed line, and update the pointers accordingly.
For example, in loop 1, the blue circle 1 is in the green region, so the pointer $t_{high}$ is moved from 15 to 7 (shown in the lower part). Finally it breaks at $t_{low} = t_{high} = 5$. 
\end{example}

\eat{
\paratitle{Privacy budget allocation} 
\red{We allocate more to the rank upper bound since in the common case, only the first few explanation predicates have large influences and the gaps between them are also large, while the rest of explanation predicates have low and similar influences. 
As \Cref{fig:diagram-ci-rank} from \Cref{ex:rank_ci} shows, given a small privacy budget, the slack constant $\xi$ will be large. If, for example, it lowers the horizontal dash line below the smallest influence, this would make it easier to have the noisy influence at each step closer to the orange region, which always moves the pointer of rank to the larger range in the binary search, eventually resulting in a large upper bound. On the other hand, for finding the lower bound of rank, given large influence gaps within the first few explanation predicates, even with a large slack constant $\xi$, it is unlikely that the dashed line will come across other explanation predicates (which can considered as the dashed line above the red dot in \Cref{fig:diagram-ci-rank} instead of below it for this case), which makes this problem less severe.}
}

We now show that {\bf noisy binary search mechanism} satisfies the privacy requirement, and outputs valid confidence intervals.
In \Cref{sec:experiments}, we show that the interval width is empirically small. 

\begin{restatable}{theorem}{rankCI}
\label{thm:rank_ci}
Given a database $D$, a predicate space $\mathcal{P}$, an influence function $\imp$ with sensitivity $\Delta_{\imp}$, explanation predicates $p_{1}, p_{2}, \ldots, p_{k}$, a confidence level $\gamma$, and a privacy parameter $\rhoRank$,  noisy binary search mechanism returns confidence intervals $\I_1, \I_2, \ldots, \I_k$ such that
\begin{enumerate}[noitemsep,topsep=0pt]
    \item\label{itm:priv} { Noisy binary search mechanism} satisfies $\rhoRank$-zCDP.
    \item\label{itm:ci-conf} For $\forall u \in [1, k]$, $\I_u$ is a $\gamma$ level confidence interval of $\rank(p_u)$.
\end{enumerate}
\end{restatable}

The proof of \cref{itm:priv} follows from the composition theorem and the property of Gaussian mechanism \cite{bun2016concentrated}. The proof of \cref{itm:ci-conf}
is based on the property of the random binary search. We defer the formal proofs and a \rev{weak} 
utility bound to
\iffullversion
\Cref{sec:private_explanation}.
\fi


\subsection{Putting it All Together}
\label{sec:put_it_together}

After we selected top-$k$ explanation predicates (\Cref{sec:private_topk}), and constructed the confidence intervals of their influences (\Cref{sec:private_influence_ci}) and ranks (\Cref{sec:private_rank_ci}), the final step is to combine them together into a single explanation table. 

\paratitle{Relative Influence} Recall that the influence defined from \Cref{def:influence} is the difference of $(o_i - o_j)$ before and after removing the tuples related to an explanation predicate (first term), and multiplies with a normalizer to penalize trivial  predicates (second term).  Since the absolute value of influence is hard to interpret, to help user better understand the confidence interval of influence, we show the \emph{relative influence} compared to the original difference  $\abs{o_i - o_j}$ as a percentage. However, we cannot divide the influence by $\abs{o_i - o_j}$ since using the actual data values will incur additional privacy loss, hence, for SUM and COUNT we divide the true influence by $\abs{\hat{o}_i - \hat{o}_j}$ as an approximation since the normalizer in the second term is bounded in $[0, 1]$.
However, when $agg = AVG$, the normalizer $\min(\abs{g_i(\neg p (D))}, \abs{g_j(\neg p (D)})$ (second term) is not bounded in $[0, 1]$, so
we further divide the influence by another constant, the minimum of the noisy counts/sizes of the groups, i.e., $\abs{\min(\hat{o}_i^C, \hat{o}_j^C)}$ (approximating the upper bound $\min(\abs{g_i(D)}, \abs{g_j(D)})$ of the normalizer to avoid additional privacy loss).
As a summary, we define the relative influence $\relimp(p; (\alpha_i, \alpha_j, >), D)$, or simply $\relimp(p)$, as follows, which is only used for display purposes. 
\begin{footnotesize}
    \begin{equation}
        \relimp(p) = 
        \imp(p) /
        \begin{cases}
            \abs{\hat{o}_i - \hat{o}_j} & \textrm{for } agg \in \{COUNT, SUM\} \\
            \abs{\hat{o}_i - \hat{o}_j} \times \abs{\min(\hat{o}_i^C, \hat{o}_j^C)} & \textrm{for } agg = AVG
        \end{cases}
        \label{eqn:rel_influ}
    \end{equation}
    \end{footnotesize}

\paratitle{Explanation Table}
We define the explanation table as follows.

\begin{definition}[\exptab\ containing top-$k$ explanations]\label{def:exp-table}
Given a database $D$, a group-by aggregate query $q$ as shown in \Cref{fig:query}, a user question $(\alpha_i, \alpha_j, >)$, 
a predicate space $\mathcal{P}$, a confidence level $\gamma$, and an integer $k$, a table of top-$k$ explanations is a list of $k$ 5-element tuples $(p_{u}, {\I^L_{\relinflu}}_u, {\I^U_{\relinflu}}_u, {\I^L_{\rnk}}_u, {\I^U_{\rnk}}_u)$ for $u = 1, 2, \ldots, k$ such that $p_{u}$ is an explanation predicate, 
$({\I^L_{\relinflu}}_u,$ ${\I^U_{\relinflu}}_u)$ is a confidence interval of relative influence $\relimp(p_{u})$ with confidence level $\gamma$,
and $ ({\I^L_{\rnk}}_u,$ ${\I^U_{\rnk}}_u)$ is a confidence interval of $\rank(p_{u})$ with confidence level $\gamma$ 
\end{definition}

\paratitle{Sorting the explanations in the explanation table} Since this table contains the bounds of the influences and ranks (see the last four columns in \Cref{fig:phase3} for an example), it is natural to present the table as a sorted list. 
Since the numbers in the table are generated by random processes, each column may imply a different sorting. 
In this paper, we sort the selected top-k explanations by the upper bound of the relative influence CI (the third column in \Cref{fig:phase3}) in a descending order; if there is a tie, we break it using the upper bound of the rank confidence interval (the fifth column in \Cref{fig:phase3}). Finding a principled way for sorting the explanation predicates is an intriguing subject of future work. 



\paratitle{Overall DP guarantee} We summarize the privacy guarantee of our \oursys\ framework as follows: (i) the private noisy query answers returned by Gaussian mechanism in Phase-1 satisfy $\rhoQuery$-zCDP together 
(see \Cref{sec:preliminary})
; (ii) Phase-2 only returns the confidence intervals of the noisy answers in Phase-1 as defined by the Gaussian mechanism and does not have any additional privacy loss 
(discussed in \Cref{sec:private_question_ci})
; (iii) Phase-3 returns $k$ explanation predicates and their upper and lower bounds on relative influence and ranks given a required confidence interval,
and uses three privacy parameters $\rhoTopk, \rhoInflu, \rhoRank$ (discussed in Section \ref{sec:private_topk}, \ref{sec:private_influence_ci} and \ref{sec:private_rank_ci}). 
The following theorem summarizes the total privacy guarantee.

\begin{theorem}\label{thm:main}
Given a group-by query $q$ and a user question comparing two aggregate values in the answers of $q$, 
the \oursys\ framework guarantees $(\rhoQuery + \rhoTopk + \rhoInflu + \rhoRank)$-zCDP.
\end{theorem}

%% file: evaluation.tex
\section{Experiments}\label{sec:experiments}


In this section, we evaluate the quality and efficiency of the explanations generated by \oursys\ 
with the following questions:


\begin{enumerate}[leftmargin=0.15in]
\itemsep0em
    \item How accurate are the confidence intervals (CI)  generated in Phase-2 by \oursys\ in 
    validating the user question?
    \item How accurate are the noisy top-k explanations compared to true top-k, along with the CI of influence and rank in Phase-3?
    \item How efficient is \oursys\ in computing the explanation table?
\end{enumerate}

\rev{To our knowledge, there are no existing benchmarks for explanations for query answers (even without privacy consideration) in the database research literature. Therefore, we evaluate the quality of our explanations with case studies along with accuracy and performance analysis with varying parameters. Developing a benchmark for usefulness of explanations to human analysts will require subjective evaluation and is a challenging direction for future work.}

We have implemented \oursys\ in Python 3.7.4 using the Pandas \cite{reback2020pandas}, NumPy \cite{harris2020array}, and SciPy \cite{2020SciPy-NMeth} libraries. All experiments were run on Intel(R) Core(TM) i7-7700 CPU @ 3.60GHz with 32 GB of RAM. The source code can be found in \cite{fullversion}.


\subsection{Experiment Setup}

We first
detail the data, queries, questions, and parameters. 

\paratitle{Datasets} 
We consider two datasets in our experiments. 
\begin{itemize}[leftmargin=0.1in]
    \item {\bf \dataset{IPUMS-CPS} (real data)}: A dataset of Current Population Survey from the U.S. Census Bureau \cite{flood2021ipums}. 
    We focus on the survey data from year 2011 to 2019. The dataset contains 8 categorical attributes where domain sizes vary from 3 to 36 and one numerical attribute. 
    The attribute \texttt{AGE} is discretized as 10 years per range, e.g., [0,10] is considered a single value. To set the domain of numerical attributes, we only include tuples with attribute \texttt{INCTOT} (the total income) smaller than 200k as a domain bound. 
    The total size of the dataset is 1,146,552. 
    
    \item {\bf \dataset{Greman-Credit} (synthetic data)}: A corrected collection of credit data \cite{gromping2019south}. It includes 20 attributes where the domain sizes vary from 2 to 11 and a numerical attribute. 
    Attributes \texttt{duration}, \texttt{credit-amount}, and \texttt{age} are discretized. The domain of attribute \texttt{good-credit} is zero or one. We synthesize the dataset to 1 million rows by combining a Bayesian network learner \cite{ankan2015pgmpy} and XGBoost \cite{brownlee2016xgboost} following the strategy of QUAIL \cite{rosenblatt2020differentially}.
\end{itemize}

\paratitle{Queries and Questions} 
The queries and questions used on the experiments are shown in \Cref{tab:exp-questions}. 

\begin{table}[h]
    \footnotesize	
    \setlength\tabcolsep{2pt}
    \def\datasetwidth{0.75cm}
    \def\querywidth{2.3cm}
    \centering
    \caption{Queries and questions for the experiments; 
   {\tt Valid} indicates if it is a valid question on the hidden true data.}
    \label{tab:exp-questions}
    \begin{tabular}{@{} >{\scriptsize}p{\datasetwidth} >{\scriptsize}p{\querywidth} l l @{}}
        \toprule
        Data & Query & Question  & Valid\\
        \midrule
        \multirow{5}{\datasetwidth}{\dataset{IPUMS-} \dataset{CPS}} & $q_1$: AVG({\tt INCTOT}) by {\tt SEX} & I1: Why {\tt Male} > {\tt Female} ? & Yes \\
        \cmidrule{2-4}
        & \multirow{2}{\querywidth}{$q_2$: {\tt INCTOT} by {\tt RELATE}} & I2: Why {\tt Grandchild} > {\tt Foster children} ?  & Yes\\
        & & I3: Why {\tt Head/householder} > {\tt Spouse} ? & No\\
        \cmidrule{2-4}
        & \multirow{2}{\querywidth}{$q_3$: {\tt INCTOT} by {\tt EDUC}} & I4: Why {\tt Bachelor} > {\tt High school} ? & Yes \\
        & &  I5: Why {\tt Grade 9} > {\tt None or preschool} ? & No \\
        
        \midrule
        \multirow{5}{\datasetwidth}{\dataset{German-} \dataset{Credit}} & $q_4$: AVG({\tt good-credit}) by {\tt status} & G1: Why {\tt no balance} > {\tt no chk account} ?  & Yes \\
        \cmidrule{2-4}
        & \multirow{2}{\querywidth}{$q_5$: AVG({\tt good-credit}) by {\tt purpose}} & G2: Why {\tt car (new)} > {\tt car (used)}?  & Yes \\
        & & G3: Why {\tt business} > {\tt vacation} ? & No  \\
        \cmidrule{2-4}
        & \multirow{2}{\querywidth}{$q_6$: AVG({\tt good-credit}) by {\tt residence}} & G4: Why {\tt "< 1 yr"} > {\tt ">= 7 yrs"}  ? & Yes \\
        & & G5: Why {\tt "[1, 4) yrs"} > {\tt "[4, 7) yrs"}  ? & No \\
        \bottomrule
    \end{tabular}
\end{table}

\paratitle{Default setting of \oursys}
Unless mentioned otherwise, the following default parameters are used: 
$\rhoQuery = 0.1$, $\rhoTopk = 0.5$, $\rhoInflu = 0.5$, $\rhoRank = 1.0$, $\gamma = 0.95$, $k = 5$, \rev{$\eta = 0.1$}
and the number of conjuncts in explanation predicates $l = 1$ (\Cref{def:explanation_predicate}). \rev{These settings are also used in the motivating example.} \rev{For $\eta = 0.1$, it means that for each selected explanation predicate, we allocate 90\% of the privacy budget for finding the rank upper bound and 10\% for the lower bound. The reason for this being our observation that the scores of explanation predicates have a long and flat tail, which intuitively means that a tight rank upper bound can be used to infer a precise score and, thus, costs more privacy.} \rev{For the total privacy budget, while it is 2.1 by default, we provide experiments to show that changing the partial budget of each component to a much smaller number can still lead to a high utility for most of the questions except I2 and I5 in \Cref{tab:exp-questions} (Figures \ref{fig:exp-noise}, \ref{fig:exp-topk-budget}, \ref{fig:exp-influci-budget}, \ref{fig:exp-rankci-budget}).}

\subsection{Case Studies} 
\label{sec:case_studies}

\paratitle{Case-1, IPUMS-CPS} We present a case study with the dataset \dataset{IPUMS-CPS} and default parameters. 
In \textbf{Phase-1}, the user submits a query 
$q_1$ from \Cref{tab:exp-questions}, and gets a noisy result: ({\tt "Female"}, 31135.25) and ({\tt "Male"}, 45778.46). The hidden true values are ({\tt "Female"}, 31135.78) and ({\tt "Male"}, 45778.39).
Next, in \textbf{Phase-2}, since there is a gap of 14643.21 between the noisy avg-income from two groups, the user asks 
a question I1 from \Cref{tab:exp-questions}. The framework then quantifies the noise in the question by reporting a confidence interval of the difference between two groups as (14636.63, 14649.79). 
Since the interval does not include zero, the framework suggests that this is a valid question, which is correct.
Finally, in \textbf{Phase-3}, the framework presents top-5 explanations to the user as \Cref{fig:exp-phase3-ipumscps} shows. The last two columns are the true relative influences and true ranks. 
We correctly find the top-5 explanation predicates, and the first and fourth explanations together suggests that a married man tends to earn more than a married woman, which is supported by the wage disparities in the labor market \cite{marriedmanincome}. The second and third explanations also match the wage disparities within the educated group and white people.
The total runtime for preparing the explanations in Phase-2 and Phase-3 is 67 seconds. 

\begin{figure}[t]
    \centering
\begin{tcolorbox}[colback=white,left=-1pt,right=1.5pt,top=0pt,bottom=0pt]
    \footnotesize
    \setlength\tabcolsep{1.5pt}
    \begin{tabular}{ >{\raggedright}p{3cm} c c c c | c c}
        \hline
        \multirow{2}{*}{explanation predicate} &
        \multicolumn{2}{c}{Rel Influ 95\%-CI} &
        \multicolumn{2}{c|}{Rank 95\%-CI} &
        {\EFFORT{25} Rel Influ} &
        {\EFFORT{25} Rank}
        \\
        \cline{2-3}
        \cline{4-5}
        & L & U & L & U & \EFFORT{25} (hidden) & \EFFORT{25} (hidden)\\
        \hline
\tt \scriptsize RELATE = "Head/householder" & 12.18\% & 12.52\% & 1 & 1 & \EFFORT{25} 12.41\% & \EFFORT{25} 1 \\
\tt \scriptsize  EDUC = "Bachelor's degree" & 7.10\% & 7.45\% & 2 & 3   & \EFFORT{25} 7.32\%  & \EFFORT{25} 2 \\
\tt \scriptsize              RACE = "White" & 6.41\% & 6.75\% & 2 & 5   & \EFFORT{25} 6.54\%  & \EFFORT{25} 3 \\
\tt \scriptsize           RELATE = "Spouse" & 5.70\% & 6.04\% & 2 & 5   & \EFFORT{25} 6.01\%  & \EFFORT{25} 4 \\
\tt \scriptsize            CLASSWKR = "NIU" & 3.83\% & 4.18\% & 2 & 6   & \EFFORT{25} 4.22\%  & \EFFORT{25} 5 \\
        \hline
    \end{tabular}
\end{tcolorbox}    
    \caption{Phase-3 of \oursys~ for the case IPUMS-CPS.} 
    \label{fig:exp-phase3-ipumscps}
\end{figure}

\paratitle{Case-2, German-Credit}
We now present a case study over the  \dataset{German-Credit} dataset
with default parameters. 
In \textbf{Phase-1}, the user submits a query 
$q_4$ from \Cref{tab:exp-questions}, and gets a noisy result: ({\tt "no checking account"}, 0.526571) and ({\tt "no balance"}, 0.574447). The true hidden result is ({\tt "no checking account"}, 0.526574) and ({\tt "no balance"}, 0.574466). Next, in \textbf{Phase-2}, since there is a gap of 0.047876 between the noisy avg-credit from two groups, the user asks 
a question G1 from \Cref{tab:exp-questions}. The framework then quantifies the noise in the question by reporting a confidence interval of the difference between two groups as (0.047786, 0.047967). 
Since the interval does not include zero, the framework suggests that this is a valid question, which is correct.
Finally, in \textbf{Phase-3}, the framework presents top-5 explanations to the user as \Cref{fig:exp-phase3-german} shows. The last two columns are the true relative influences and true ranks. 
We correctly find the top-5 explanations, and the first explanation suggests that for a person who  already has a credit in the bank, the bank tends to mark its credit as good if she has a checking account even with zero balance with higher probability than the case of no account, which is consistent to the intuition that a person having a credit account but no checking account is risky to the bank.
The total runtime for preparing the explanations in Phase-2 and Phase-3 is 40 seconds. 

\begin{figure}[t]
    \centering
\begin{tcolorbox}[colback=white,left=-2pt,right=1.5pt,top=0pt,bottom=0pt]
    \footnotesize
    \setlength\tabcolsep{1pt}
    \begin{tabular}{ >{\raggedright}p{3.25cm} c c c c | c c}
        \hline
        \multirow{2}{*}{explanation predicate} &
        \multicolumn{2}{c}{Rel Influ 95\%-CI} &
        \multicolumn{2}{c|}{Rank 95\%-CI} &
        {\EFFORT{25} Rel Influ} &
        {\EFFORT{25} Rank}
        \\
        \cline{2-3}
        \cline{4-5}
        & L & U & L & U & \EFFORT{25} (hidden) & \EFFORT{25} (hidden)\\
        \hline
        
\tt\tiny                                         existing-credits = "1" & 77.90\% & 78.99\% & 1 & 1 &\EFFORT{25} 78.16\% &\EFFORT{25} 1 \\
\tt\tiny                            job = "skilled employee / official" & 71.21\% & 72.29\% & 1 & 2 &\EFFORT{25} 71.83\% &\EFFORT{25} 2 \\
\tt\tiny                           sex-marst = "male : married/widowed" & 54.34\% & 55.42\% & 2 & 4 &\EFFORT{25} 55.10\% &\EFFORT{25} 3 \\
\tt\tiny                                  credit-amount = "(500, 2500]" & 50.01\% & 51.10\% & 2 & 5 &\EFFORT{25} 50.27\% &\EFFORT{25} 4 \\
\tt\tiny credit-history = "no credits taken/all credits paid back duly" & 49.07\% & 50.16\% & 4 & 5 &\EFFORT{25} 49.14\% &\EFFORT{25} 5 \\
        \hline
    \end{tabular}
\end{tcolorbox}    
    \caption{Phase-3 of \oursys\ for the case German-Credit.}
    \label{fig:exp-phase3-german}
    \vspace{-3mm}
\end{figure}

\subsection{Accuracy and Performance Analysis}
We detail our experimental analysis for the different questions and configurations of \oursys. All results are averaged over 10 runs.

\begin{figure}
    \centering
    \includegraphics[width=\linewidth]{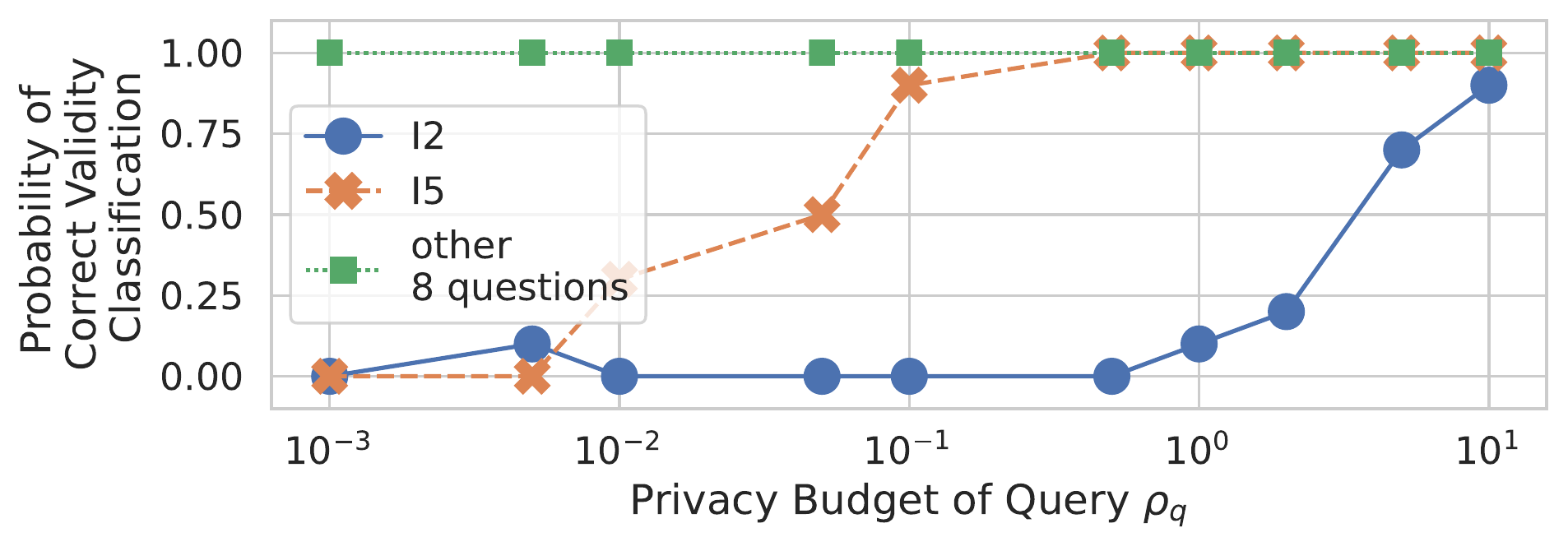}
    \caption{The probability of correctly validating user questions. All questions except I2 and I5 (\Cref{fig:exp-noise}) are at 100\%.}
    \label{fig:exp-noise}
\end{figure}

\paratitle{Correctness of noise interval} In Phase-2 of \oursys, user has the option to either stop or proceed based on the confidence interval of the question. We evaluate a simple strategy for judging the validity of the question: 
if the interval contains non-positive numbers, the question is invalid, otherwise valid.
From \Cref{fig:exp-noise}, we find that 8 out of 10 questions (plotted together for clarity) from \Cref{tab:exp-questions} 
are classified correctly with accuracy $100\%$ given a wide range of privacy budget of query $\rhoQuery$. However, there are two questions, I2 and I5, only show high accuracy given a large privacy budget of $\rhoQuery = 10$. One reason is that the minimum group size involved in these questions is small compared to other questions, and, therefore, the partial confidence intervals in the denominators of the $AVG$ query are low, which makes the final confidence intervals wider 
including negative numbers when it should not. For I2, the minimum group size is at least 600 times smaller than the other questions of \dataset{IPUMS-CPS} while this number for I5 is 60. 


\begin{figure}[h]
    \centering
    \begin{subfigure}[t]{0.49\linewidth}
        \includegraphics[width=\linewidth]{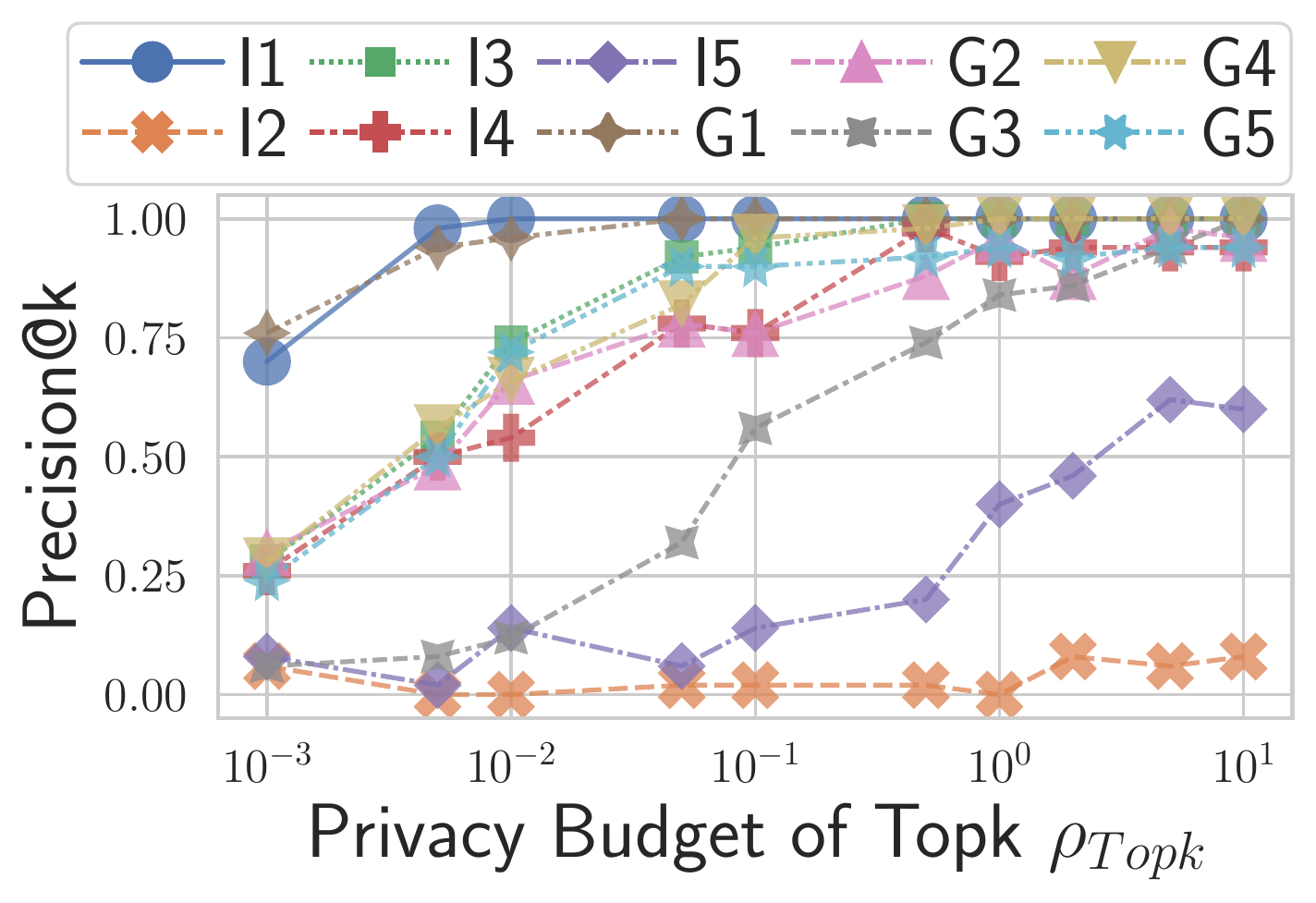}
        \vspace{-0.65cm}
        \caption{Trend by $\rhoTopk$. }
        \label{fig:exp-topk-budget}
    \end{subfigure}
    \hfill
    \begin{subfigure}[t]{0.49\linewidth}
        \includegraphics[width=\linewidth]{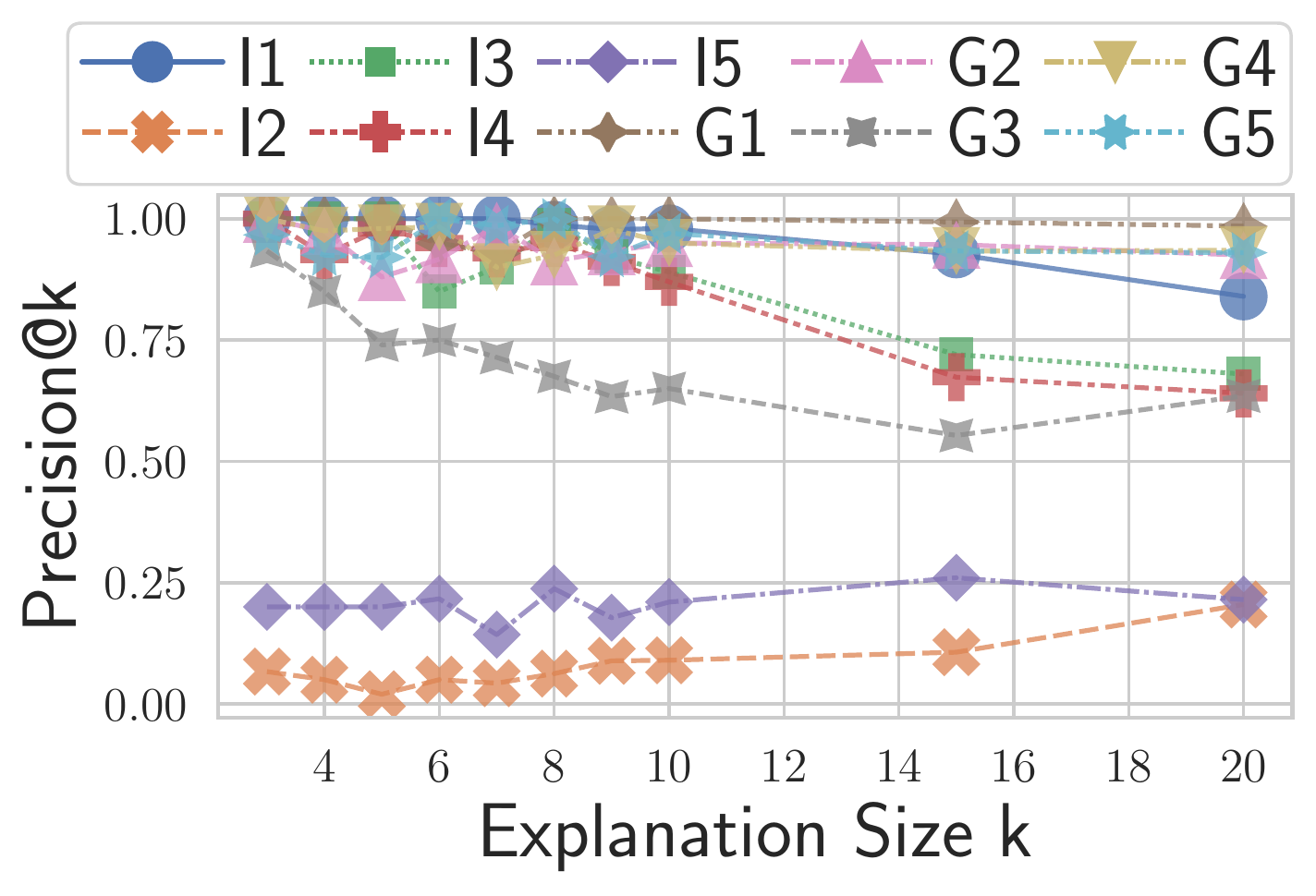}
        \vspace{-0.65cm}
        \caption{Trend by $k$.}
        \label{fig:exp-topk-k}
    \end{subfigure}
    \caption{Precision@k of top-k selection by \oursys.}
    \label{fig:exp-topk}
\end{figure}

\paratitle{Accuracy of top-k explanation predicates} 
In Phase-3 of \oursys, we first select top-k explanation predicates. We measure the accuracy of the selection by Precision@k \cite{herlocker2004evaluating}, the fraction of the selected top-k explanation predicates that are actually ranked within top-k.
Another experiment on the full ranking is included in 
\iffullversion
\Cref{sec:full_ranking}
\fi
From \Cref{fig:exp-topk-budget}, we find that the privacy budget of top-k selection $\rhoTopk$ has a positive effect to Precision@k at k = 5 for various questions. When $\rhoTopk = 1.0$, all the questions except I2 and I5 have Precision@k $\geq 0.8$. The selection accuracy of question I2 and I5 are generally lower 
because of small group sizes, and, therefore, the influences of explanation predicates are small and the rankings are perturbed by the noise more significantly.

From \Cref{fig:exp-topk-k}, we find that the trend of Precision@k by k is different across questions and there is no clear trend that Precision@k increases as k increases. 
For example, for G3, it first decreases from k=3 to k=5, but increases from k=5 to k=6.
When k = 3, most questions have high Precision@k; this is because the highest three influences are much higher than the others, which makes the probability high to include the true top three. When k is large, for explanation predicates that have similar scores, they have equal probability to be included in top-k and therefore the top-k selected by the algorithm are different from the true top-k selections. The relationship between Precision@k and k depends on the distribution of all the explanation predicate influences.
We also study the relationship between Precision@k at k=5 with the conjunction size $l$. For two questions I1 and G1, their Precision@k stays at 1.0 when $l$ varies in $1, 2$, and $3$. Although the size of explanation predicates grows exponentially with the conjunction size $l$, \oursys~ can select the top-k explanation predicates with high accuracy even from a large space. 

\begin{figure}[h]
    \centering
    \begin{subfigure}[b]{0.49\linewidth}
        \includegraphics[width=\linewidth]{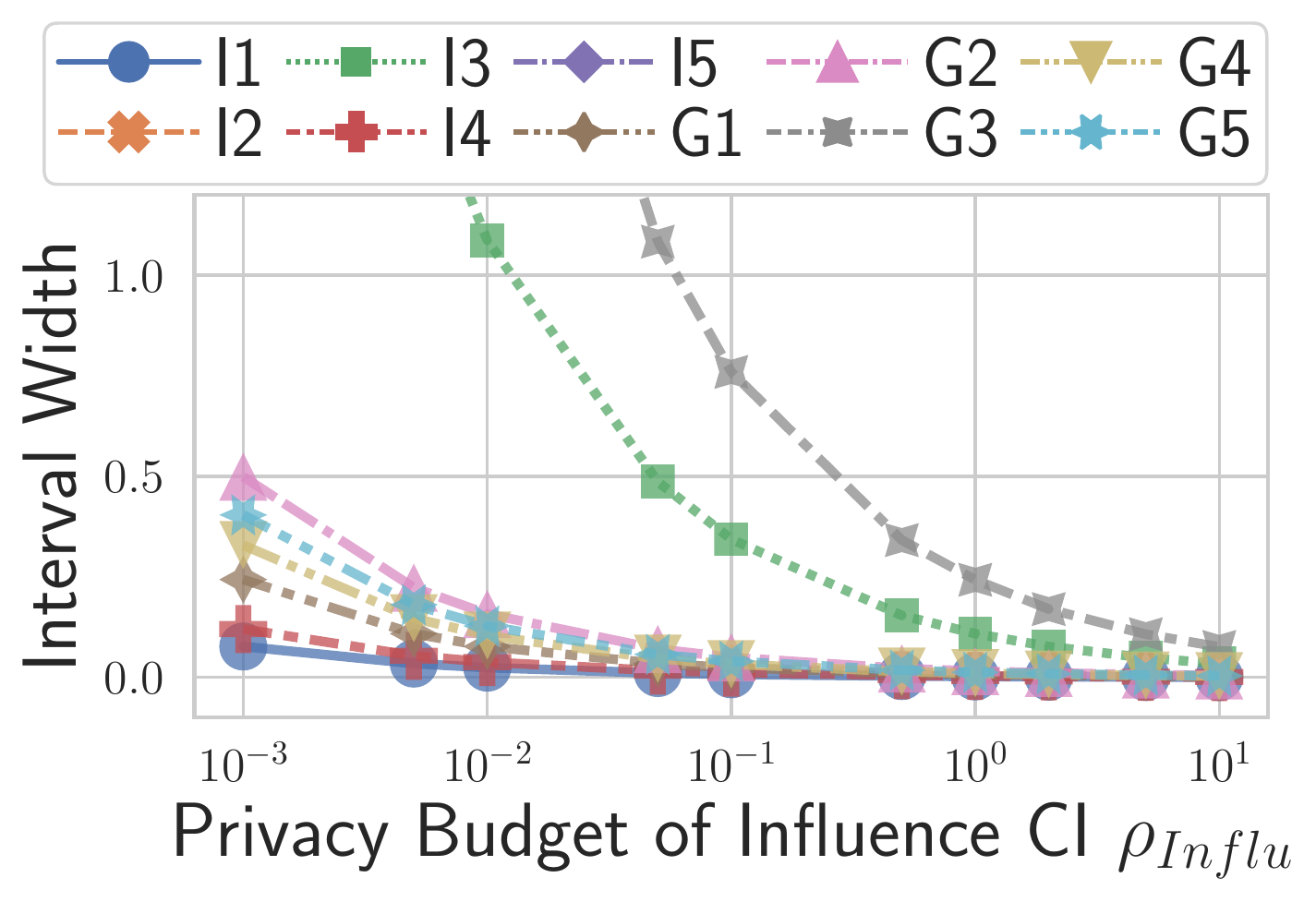}
        \vspace{-0.65cm}
        \caption{Relative Influence}
        \label{fig:exp-influci-budget}
    \end{subfigure}
    \hfill
    \begin{subfigure}[b]{0.49\linewidth}
        \includegraphics[width=\linewidth]{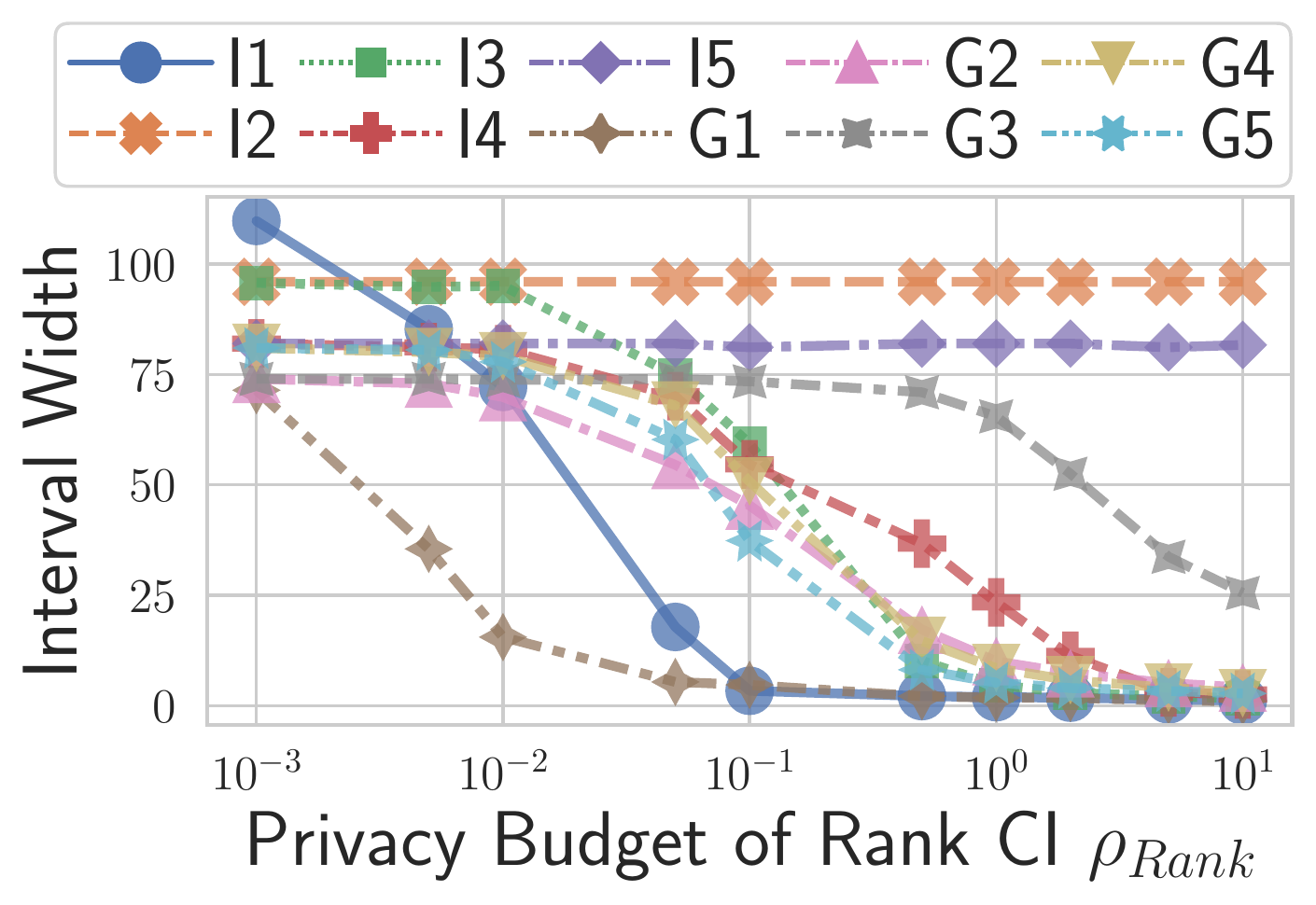}
        \vspace{-0.65cm}
        \caption{Rank}
        \label{fig:exp-rankci-budget}
    \end{subfigure}
    \caption{The width of confidence intervals by \oursys. \rev{For I2 and I5, their confidence interval width of relative influence are all beyond 2 across various privacy budgets.}}
    \label{fig:exp-influ-rank-ci}
\end{figure}

\paratitle{Precision of relative influence and rank confidence Interval (CI)}
In Phase-3, the last step is to describe the selected top-k explanation predicates by a CI of relative influence and rank for each. To measure the precision of the description, we adopt the measure of {\bf interval width} \cite{ferrando2021parametric}. \Cref{fig:exp-influ-rank-ci} illustrates the average width of $k$ CIs of relative influence and rank. From Figure \ref{fig:exp-influci-budget} and \ref{fig:exp-rankci-budget}, we find that the increase of privacy budget 
$\rhoInflu$ and 
$\rhoRank$ shrinks the interval width of relative influence CI and rank CI separately. In particular, when $\rhoInflu \geq 0.5$, 6 out of 10 questions have the interval width of relative influence CI $ \leq 0.025$ 
; when $\rhoRank \geq  1.0$, 2 questions have the interval width of rank CI $\leq 2$ and 6 questions have this number $\leq 10$. 
%
We also measure the {\bf effect of confidence level $\gamma$} to the CI by changing $\gamma$ from 0.1 to 0.9 by step size 0.1 and from 0.95 and 0.99. 
Figures can be found in 
\iffullversion
\Cref{sec:confidence_level}.
\fi
The results show that it has a non-significant effect to the interval width, 
as it changes $< 0.03$ for the influence CI of 6 questions,  and changes $< 5$ for the rank CI of 8 questions.

\begin{figure}[h]
    \centering
    \begin{subfigure}[b]{0.45\linewidth}
        \includegraphics[width=\linewidth]{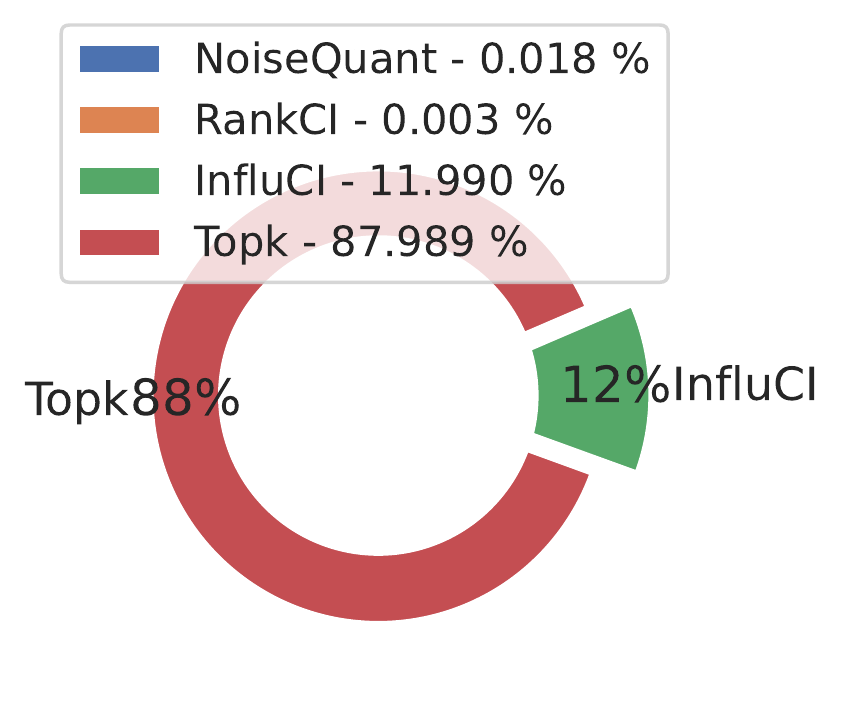}
        \vspace{-0.65cm}
        \caption{Breakdown of time}
        \label{fig:exp-runtime-breakdown}
    \end{subfigure}
    \begin{subfigure}[b]{0.54\linewidth}
        \includegraphics[width=\linewidth]{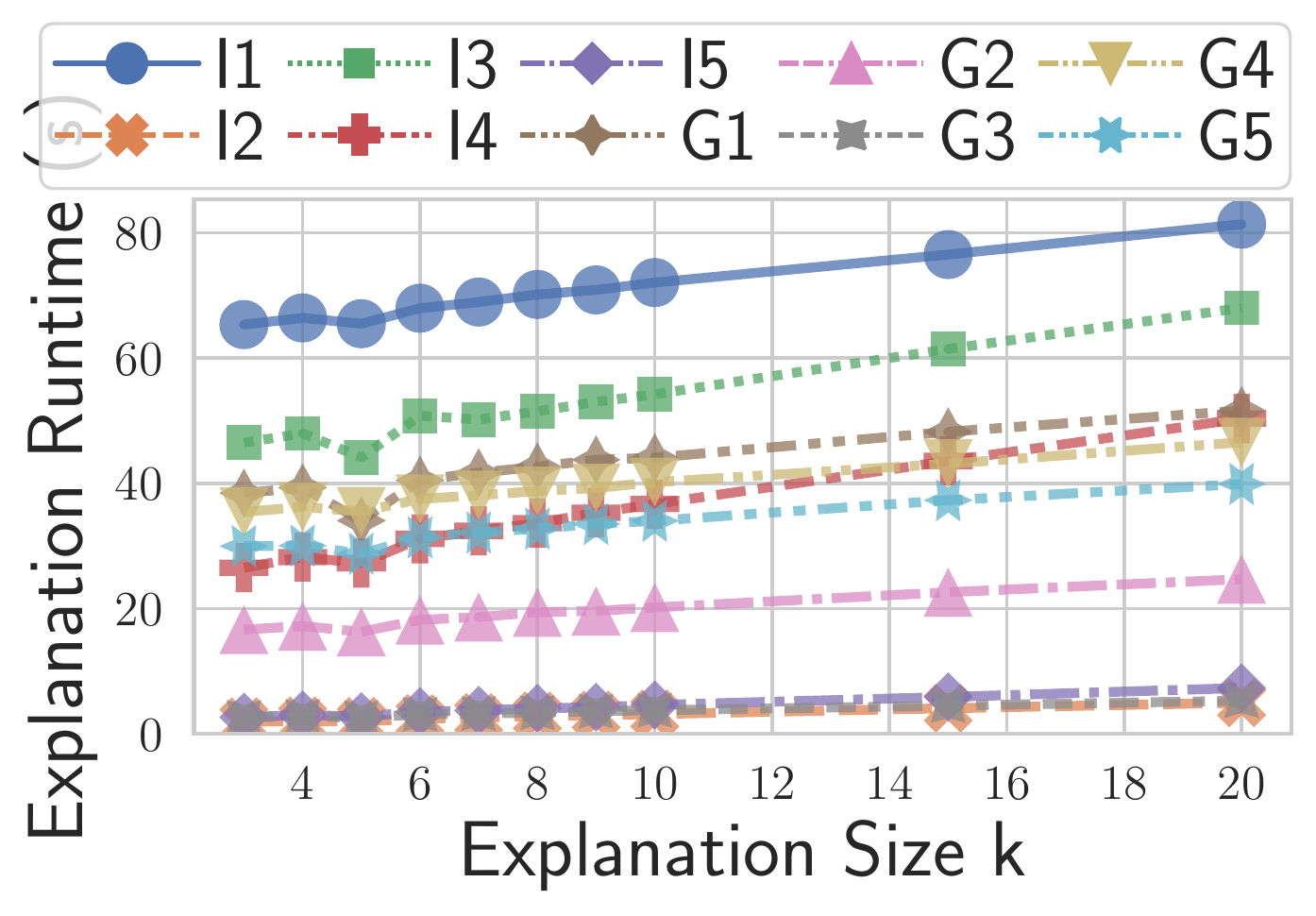}
        \vspace{-0.65cm}
        \caption{Trend by $k$}
        \label{fig:exp-runtime-k}
    \end{subfigure}
    \caption{Runtime analysis of \oursys.}
    \label{fig:exp-runtime}
\end{figure}

\paratitle{Runtime analysis}
Finally, we analyze the runtime of \oursys\ for preparing the explanations from Phase-2 and Phase-3. From \Cref{fig:exp-runtime-breakdown}, a runtime breakdown in average for all the questions from \Cref{tab:exp-questions} with total runtime 32 seconds in average, shows that $88\%$ of the time is used for the top-k explanation predicate selection procedure, especially on computing the influences for all the explanation predicates. The next highest time usage is for computing the confidence interval of influence, which needs to evaluate each sub queries. For the step noise quantification and confidence interval of rank, the time usage is not significant since the first only needs to find the image of two intervals and the second is a binary search.
From \Cref{fig:exp-runtime-k}, we find the runtime is linearly proportional to the size of explanations $k$, and the difference between questions is due to the difference of group sizes. We also 
find the runtime grows exponentially with the number of conjuncts $l$ as the number of explanation predicates grows exponentially: for $l = 1,2,3$, the runtime about question I1 is 67, 3078 and 79634, and for question G1 it is 40, 1587 and 39922  seconds.

The \textbf{summary of our findings} in the experiments is listed below:
\begin{enumerate}[leftmargin=0.15in]
\itemsep0em
    \item \oursys~ can correctly suggests the validity of the question with 100\% accuracy for 8 out of 10 questions at $\rhoQuery = 0.1$.
    \item \oursys~ can select at least 80\% of the true top-5 explanation predicates correctly for 8 out of 10 questions at $\rhoTopk = 0.5$, and the associated confidence intervals of relative influence have width $\leq 0.015$ for 6 questions at $\rhoInflu = 0.5$ and the confidence intervals of rank have width $\leq 10$ for 6 questions at $\rhoRank = 1.0$. 
    \item The runtime of \oursys~ for preparing the explanations is in average 32 seconds for the 10 examined questions.
\end{enumerate}

%% file: related_work.tex
\section{Related Work}\label{sec:related}
We next survey related work in the fields of DP and explanations for query results. 
{\em To the best of our knowledge, \oursys\ is the first work that explains aggregate query results while satisfying DP.}

\paratitle{Explanations for query results}
The database community has proposed several approaches to explaining aggregate and non-aggregate queries in multiple previous works. 
Proposed approaches include provenance \cite{DeutchFG20,DBLP:journals/pvldb/HuangCDN08,
DBLP:journals/pvldb/HerschelH10,LS17,DBLP:conf/sigmod/ChapmanJ09,DBLP:conf/sigmod/TranC10,DeutchFGH20, LeeLG19}, intervention  \cite{wu2013scorpion,RoyS14,RoyOS15},  entropy \cite{Gebaly+2014-expltable}, responsibility \cite{MeliouGMS11,MeliouGNS11}, Shapley values \cite{LivshitsBKS20,ReshefKL20}, counterbalance  \cite{miao2019going} and augmented provenance \cite{li2021putting}, and several of these approaches have used predicates on tuple values as explanations like \oursys, e.g., \cite{wu2013scorpion, RoyS14, Gebaly+2014-expltable, li2021putting}.
We note that any approaches that consider individual tuples or explicit tuple sets in any form as explanations (e.g., \cite{MeliouGMS11,LeeLG19,DeutchFG20,LivshitsBKS20}) cannot be applied in the DP setting since they would violate privacy. 
Among the other summarization or predicate-based approaches, Scorpion \cite{wu2013scorpion} explains outliers in query results with the intervention of most influential predicates. Our influence function (Section \ref{sec:influence}) is inspired by the influence function of Scorpion, but has been modified to deliver accurate results while satisfying DP. Another intervention-based work \cite{RoyS14} that also uses explanation predicates, models inter-dependence among tuples from multiple relations with causal paths. \oursys\ does not support joins in the queries, which is a challenging future work (see Section~\ref{sec:futurework}).


\paratitle{Differential privacy}
Private SQL query answering systems \cite{kotsogiannis2019privatesql, wilson2019differentially, kotsogiannis2019architecting, johnson2018towards, DBLP:journals/pvldb/McKennaMHM18, tao2020computing, dong2022r2t} 
consider a workload of aggregation queries with or without joins on a single or multi-relational database, but none supports explanation under differential privacy. 
The selection of private top-k candidates is well-studied by the community \cite{qiao2021oneshot, DBLP:conf/nips/DurfeeR19, bafna2017price, mcsherry2007mechanism, chaudhuri2014large, mckenna2020permute, ding2021permute, bonomi2013mining, DBLP:journals/pvldb/LiQSC12, carvalho2020differentially, lee2014top, thakurta2013differentially, bhaskar2010discovering}. We adopt One-shot Top-k mechanism \cite{qiao2021oneshot} since it is easy to understand. 
Private confidence interval is a new trend of estimating the uncertainty under differential privacy \cite{ferrando2020general, covington2021unbiased, brawner2018bootstrap}, however the current bootstrap based methods measure the uncertainty from both the sampling process and the noise injection, while we only focus on the second part which is likely to give tighter intervals.
The most relevant work to the private rank estimation in our framework is private quantile \cite{DBLP:conf/icml/GillenwaterJK21, DBLP:journals/corr/abs-2110-05429, DBLP:journals/corr/abs-2201-03380, cormode2019answering, dwork2009differential, lei2011differentially, smith2011privacy}, which is to find the value given a position such as median, but the problem of rank estimation in our setting is to find the position given a value. 


\paratitle{Privacy and provenance}
As mentioned earlier, data provenance is often used for explaining query results, mainly for non-aggregate queries. 
Within the context of provenance privacy \cite{DavidsonKMPR11,DavidsonKRSTC11,BertinoGKNPSSTX14,SanchezBS18,Ruan0DLOZ19,AndersonC12,TanKH13,Cheney11}, 
one line of work \cite{DavidsonKMPR11,DavidsonKRSTC11,DavidsonKTRCMS11} studied the preservation of workflow privacy (privacy of data transferred in a workflow with multiple modules or functions), with a privacy criterion inspired by $l$-diversity \cite{MachanavajjhalaKGV07}.
A recent work \cite{DeutchG19} explored what can be inferred about the {\em query} from provenance-based explanations and found that the query can be reversed-engineered from the provenance in various semirings \cite{GKT-pods07}. To account for this, a follow-up paper \cite{DeutchFGM21} proposed an approach for provenance obfuscation that is based on abstraction. This work uses $k$-anonymity \cite{KAnonymity2002} to measure how many `good' queries can generate concrete provenance that can be mapped to the abstracted provenance, thus quantifying the privacy of the underlying query. 
Devising techniques for releasing provenance 
of non-aggregate and aggregate queries while satisfying DP is an interesting research direction.


%% file: future.tex
\section{Future Work} 
\label{sec:futurework}
\cut{
In this paper we proposed a 
novel three-phase framework 
to explain comparisons of two aggregates output by  a group-by aggregate query (COUNT, SUM, or AVG) while satisfying DP. 
We defined an influence function with low sensitivity to rank top explanations, and proposed algorithms for  computing the confidence intervals of noisy query answers, and influences  and ranks of explanation predicates all with DP. Our experiments show the efficiency and accuracy of \oursys\ to find interesting explanations for query answers with the guarantee of DP, which would facilitate providing explanations for queries on sensitive data.  
}

There are several interesting future directions. First,  the current \oursys\ framework does not support queries with joins. Unlike standard explanation frameworks like \cite{wu2013scorpion} where the join results can be materialized before running the explanation mechanism, in the DP settings with multiple relations, a careful sensitivity analysis of adding/removing tuples from different tables is needed \cite{tao2020computing}. Extending the framework to more general queries and questions is an important future work. \cut{Second, a tighter analysis to the sensitivity of the influence function is the key to the successful application of differential privacy.} Second, the complexity of the top-k selection algorithm is high since it needs to iterate over all the explanation predicates, leaving room for future improvements.  
Additionally, other interesting notions of explanations for query answers  (e.g., \cite{LivshitsBKS20, miao2019going, li2021putting}) can be explored in the DP setting.
\rev{Finally, evaluating our approach with a comprehensive user study and examining different metrics of understandability of the explanations generated by \oursys\ is also an important direction for the future investigation. }







%% file: appendix.tex
\appendix

\input{appendix-proof}

\clearpage
\input{appendix-extensions}

\clearpage
\input{appendix-algorithm}

\clearpage
\input{appendix-experiment}

%% file: appendix-proof.tex
\section{Theorems and Proofs}









\subsection{Influence Function}
\label{proof:influence}






\influSens*

\begin{proof}
\textbf{(1) COUNT.}
    Recall the definition of influence function:
    \begin{align*}
    \imp(p; Q, D) = &
        \Big( 
        \left(q(g_i(D)) - q(g_j(D))\right)
        \\ & -
        \left(q(g_i(\neg p (D))) - q(g_j(\neg p (D)))\right)
        \Big)
        \\ & \times
        \frac{
            \min\limits_{t \in \{i, j\}} \abs{g_t(\neg p (D))}
        }{
            \max\limits_{t \in \{i, j\}} \abs{g_t(D)} + 1
        }
    \end{align*}
    We interpret and consider the following equations or notations:
    \begin{align*}
        q(D)   &= \abs{D} \\
        \phi_i &= (\phi \land \Agb = \alpha_i) \\
        g_i(D) &= \phi_i(D) \\
        g_i(p(D)) &= (\phi_i \land p)(D) \\
        g_i(\neg p(D)) &= (\phi_i \land \neg p)(D) \\
        f(D)   &= \min\nolimits_{t \in \{i, j\}} \abs{g_t(\neg p (D))} \\
        g(D)   &= \max\nolimits_{t \in \{i, j\}} \abs{g_t(D)} + 1 \\
        h_i(D) &= q((\phi_i \land p)(D)) f(D) / g(D) 
    \end{align*}
    
    Since $q$ is a counting query 
    , we have $q(g_i(D)) - q(g_i(\neg p (D))) = q(g_i(p(D)))$, and by replacing $g_i(p(D))$ with $(\phi_i \land p)(D)$ we have $q(g_i(D)) - q(g_i(\neg p (D))) = q((\phi_i \land p)(D))$.
    By further replacing the last numerator and denominator in the influence function with $f(D)$ and $g(D)$, we have $\imp(p; Q, D) = h_i(D) - h_j(D)$. 
    
    We prove the sensitivity bound by the following inequality chains.
    
    \begin{align}
        \Delta_{\imp} = & ~ max_{D \neighbor D'} \abs{\imp(p; Q_{CNT}, D) - \imp(p; Q_{CNT}, D'}
        \\
        \intertext{
        We first replace $\imp$ according to $\imp(p; Q, D) = h_i(D) - h_j(D)$, and then apply  \Cref{lem:sum_of_sensitivity} (see \Cref{sec:proof_supple}) to bound the sensitivity by the sum of sensitivities of $h_i$ and $h_j$.}
        \leq & \sum_{t \in \{i, j\}} \max_{\abs{D'} = \abs{D}+1} \abs{h_t(D') - h_t(D)} 
        \\
        \intertext{The second inequality is by \Cref{lem:basic_influ_sensitivity} (see \Cref{sec:proof_supple}), since $f$ is a non-negative query with sensitivity 1 and $g$ is a monotonic positive and positive query with sensitivity 1.}
        \leq & \sum_{t \in \{i, j\}} \frac{ 2 \abs{(\phi_t \land p)(D)} + f(D) + 1}{g(D)} \Delta_q
        \label{eqn:influ_sens_proof_eq1}
        \\
        \intertext{The next equality is by replacing the variables. Since $q$ is a counting query, it has sensitivity $\Delta_q = 1$.}
        = & \sum_{t \in \{i, j\}} \frac{ 2 \abs{(\phi_t \land p)(D)} + \min\limits_{s \in \{i, j\}} \abs{(\phi_s \land \neg p )(D)} + 1}
        {\max\nolimits_{s \in \{i, j\}} \abs{g_s(D)} + 1 } 
        \\ 
        \intertext{The third inequality is by the property of $min$ and $max$, since $\min\limits_{s \in \{i, j\}} \abs{(\phi_s \land \neg p )(D)} \leq \abs{(\phi_t \land \neg p )(D)}$ and $\max\nolimits_{s \in \{i, j\}} \abs{g_s(D)} \geq \abs{g_t(D)}$.}
        \leq & \sum_{t \in \{i, j\}} \frac{  \abs{(\phi_t \land p)(D)} + \abs{(\phi_t \land p)(D)} +  \abs{(\phi_t \land \neg p)(D)} + 1}
        {\abs{g_t(D)} + 1 } 
        \\ 
        \intertext{The next equality is due to that $\phi_t = (\phi_t \land p) \lor (\phi_t \land \neg p)$.}
        = & \sum_{t \in \{i, j\}}  \frac{\abs{(\phi_t \land p)(D)} + (\abs{\phi_t(D)} + 1) }{\abs{g_t(D)} + 1}
        \\ 
        \intertext{The fourth inequality is due to that $\abs{(\phi_t \land p)(D)} \leq \abs{\phi_t(D)} = \abs{g_t(D)} \leq \abs{g_t(D)} + 1$.}
        \leq & \sum_{t \in \{i, j\}}  \frac{(\abs{g_t(D)} + 1) + (\abs{g_t(D)} + 1)}{\abs{g_t(D)} + 1}
        \\
        \leq  & 4
    \end{align}

\textbf{(2) SUM.}
    Similar to the proof of the sensitivity of $CNT$ influence, but with $\Delta_q = \dommax{\Aagg}$, which should be replaced at \Cref{eqn:influ_sens_proof_eq1}.

\textbf{(3) AVG.}
    \begin{align*}
        & \imp(p; Q_{AVG}, D) \\
        = & \left(
            (
            \frac{SUM(\phi_i(D), \Aagg)}{\abs{\phi_i(D)}} 
            - 
            \frac{SUM(\phi_j(D), \Aagg)}{\abs{\phi_j(D)}}
            )
            - \right. \\ & \left. 
            (
            \frac{SUM((\phi_i \land \neg p)(D), \Aagg)}{\abs{(\phi_i \land \neg p)(D)}} 
            - 
            \frac{SUM((\phi_j \land \neg p)(D), \Aagg)}{\abs{(\phi_j \land \neg p)(D)}}
            )
            \right)
            \\ & 
            \min\limits_{t \in \{i, j\}} \abs{(\phi_t \land \neg p )(D)}
    \end{align*}
    Now we consider decompose this query into four parts (for example, $\frac{SUM(\phi_i(D), \Aagg)}{\abs{\phi_i(D)}} \min\limits_{t \in \{i, j\}} \abs{(\phi_t \land \neg p )(D)}$ as one part) , and analyze the sensitivity for each part and finally sum up. Consider query $q$ as summing up $\Aagg$ with sensitivity $\Delta_q = \dommax{\Aagg}$. By \Cref{lem:basic_influ_sensitivity}, we can show that the sensitivity of each part is 4 $\Delta_q$. Together, the total sensitivity is bounded by 16 $\Delta_q$.
\end{proof}

\subsection{Private Explanations}
\label{sec:private_explanation}

\PrivateTopk*

\begin{proof}  
\textbf{(1) Differential Privacy.} 
it is equivalent to iteratively applying $k$ exponential mechanisms \cite{dwork2014algorithmic} that satisfies $\epsilon^2/8$-zCDP \cite{DBLP:conf/nips/DurfeeR19, dong2020optimal, cesar2021bounding, DPorg-exponential-mechanism-bounded-range} for each, where $\epsilon = \sqrt{8\rhoTopk/k}$ \cite{DBLP:conf/nips/DurfeeR19, DPorg-one-shot-top-k}, therefore in total it satisfies $(k\epsilon^2/8)$-zCDP which is also $\rhoTopk$-zCDP.

\textbf{(2) Utility Bound.}  It is extended from the utility theorem of EM in Thm 3.11 of \cite{dwork2014algorithmic}, which states that
\begin{small}
    $$Pr\left[\imp(\mathcal{M}^{(1)}) \leq OPT^{(1)} - \frac{2\Delta_{\imp}}{\epsilon}(\ln(|\mathcal{P}|) + t)\right] \leq e^{-t}$$
    \end{small}
where $\epsilon = \sqrt{8\rhoTopk/k}$.
To extend from $i=1$ to $\forall i \in \{1, 2, \ldots, k\}$, we follow the original proof:
\begin{align*}
    Pr[\imp(\mathcal{M}^{(i)}) \leq c] \leq 
    \frac{|\mathcal{P}|\exp(\epsilon c / (2\Delta_{\imp}))}
        {\exp(\epsilon OPT^{(i)} / (2\Delta_{\imp}))}
\end{align*}
by giving a upper bound and lower bound of the numerator and denominator. Replacing $c$ with the appropriate value will give this theorem.
\end{proof}





\rankInfluSens*

\begin{proof}
Drop $\mathcal{P}$ and $\imp$ from $\rank^{-1}(t; D, \mathcal{P}, \imp)$ for simplicity. 
Next we show that for any two neighboring datasets $D' \sim D$, we have $\abs{\imp(\rank^{-1}(t; D'); D') - \imp(\rank^{-1}(t; D); D)} \leq \Delta_{\imp}$
, which is equivalent to showing $-\Delta_{\imp} \leq  \imp(\rank^{-1}(t; D'); D') - \imp(\rank^{-1}(t; D); D) \leq \Delta_{\imp}$.

\paratitle{Case 1, lower bound} This is to show that for any $D' \neighbor D$, we have $ \imp(\rank^{-1}(t; D'); D') - \imp(\rank^{-1}(t; D); D) \geq - \Delta_{\imp}$.

By the definition of global sensitivity, for any explanation predicate $p$, we have 
$\abs{\imp(p; D') - \imp(p; D)} \leq \Delta_{\imp}$, and therefore
$\imp(p; D') \geq \imp(p; D) - \Delta_{\imp}$. By replacing $p$ with $\rank^{-1}(j; D)$ for some $j$, we have $\imp(\rank^{-1}(j; D); D') \geq \imp(\rank^{-1}(j; D); D) - \Delta_{\imp}$.
For any $j \leq t$, by the property of ranking , we have
$\imp(\rank^{-1}(j; D); D)  \geq \imp(\rank^{-1}(t; D); D)$. Together, for any $j \leq t$, we have 
$\imp(\rank^{-1}(j; D); D') \geq \imp(\rank^{-1}(j; D); D) - \Delta_{\imp} \geq \imp(\rank^{-1}(t; D); D) - \Delta_{\imp}$.
This means there are at least $t$ elements in $D'$ such that their scores are above $\imp(\rank^{-1}(t; D); D) - \Delta_{\imp}$, which implies for the $t$-th largest score in $D'$ we have $\imp(\rank^{-1}(t; D'); D') \geq \imp(\rank^{-1}(t; D); D)  - \Delta_{\imp}$.

\paratitle{Case 2, upper bound} This is to show that for any $D' \neighbor D$, we have $\imp(\rank^{-1}(t; D'); D') - \imp(\rank^{-1}(t; D); D)  \leq \Delta_{\imp}$. 

By the definition of global sensitivity, for any explanation predicate $p$, we have 
$\abs{\imp(p; D') - \imp(p; D)} \leq \Delta_{\imp}$, and therefore
$\imp(p; D') \leq \imp(p; D) + \Delta_{\imp}$. By replacing $p$ with $\rank^{-1}(j; D)$ for some $j$, we have $\imp(\rank^{-1}(j; D); D') \leq \imp(\rank^{-1}(j; D); D) + \Delta_{\imp}$.
For any $j \geq t$, by the property of ranking , we have
$\imp(\rank^{-1}(j; D); D)  \leq \imp(\rank^{-1}(t; D); D)$. Together, for any $j \geq t$, we have 
$\imp(\rank^{-1}(j; D); D') \leq \imp(\rank^{-1}(j; D); D) + \Delta_{\imp} \leq \imp(\rank^{-1}(t; D); D) + \Delta_{\imp}$.
This means there are at most $t-1$ elements in $D'$ such that their scores can be above $\imp(\rank^{-1}(t; D); D) + \Delta_{\imp}$, which implies for the $t$-th largest score in $D'$ we have $\imp(\rank^{-1}(t; D'); D') \leq \imp(\rank^{-1}(t; D); D) + \Delta_{\imp}$.
\end{proof}

\begin{lemma}
\label{lem:indicator_sens}
Given a set of predicates $\mathcal{P}$, an influence function $\imp$ with global sensitivity $\Delta_{\imp}$ and a number $t$, then the function $s(D) = \imp(p; D) - \imp(\rank^{-1}(t; D, \mathcal{P}, \imp); D)$ has global sensitivity $2\Delta_{\imp}$.
\end{lemma}

\begin{proof}
The sensitivity of $\imp$ is $\Delta_{\imp}$ by definition and the sensitivity of $\imp(\rank^{-1}(t; D, \mathcal{P}, \imp)$ is $\Delta_{\imp}$ by \Cref{thm:indicator_sens}. By \Cref{lem:sum_of_sensitivity}, together it has sensitivity $2\Delta_{\imp}$. 
\end{proof}

\rankCI*

\begin{proof}
Please find \Cref{sec:appendix_private_rank_ci} for the full description of the noisy binary search mechanism as \Cref{alg:find_rank_ci}.

\paratitle{(1) Differential Privacy}
Now we discuss why \Cref{alg:find_rank_ci} satisfies $\rhoRank$-zCDP.

The main structure of \Cref{alg:find_rank_ci} is a for-loop of $k$ explanation predicates from line \ref{lin:cirank_for_start} to \ref{lin:cirank_for_end}, and within each for-loop, we first prepare the parameters at line \ref{lin:cirank_rho} and \ref{lin:cirank_beta}, make two calls to the sub-routine \textsc{RankBound} and construct the confidence interval by the sub-routine outputs. We first show below that each call to the sub-routine \textsc{RankBound} with parameters $(p, \rho, \beta, dir)$ satisfies $\rho$-zCDP. Given this is true, we then show that our two calls \Call{RankBound}{$p_{u}, 0.1\rho, \beta, -1$} and \Call{RankBound}{$p_{u}, 0.9\rho, \beta, +1$} at \Cref{lin:cirank_ci} satisfies $0.1\rho$-zCDP and $0.9\rho$-zCDP, which together satisfies $\rho$-zCDP by the composition rule (\Cref{lem:composition}). By line \ref{lin:cirank_rho}, we set $\rho = \rhoRank / k$, therefore each loop satisfies $(\rhoRank / k)$-zCDP, and after in total $k$ loops, it satisfies $\rhoRank$-zCDP by the composition rule (\Cref{lem:composition}).

Next we show that \Call{RankBound}{$p, \rho, \beta, dir$} satisfies $\rho$-zCDP, from line \ref{lin:cirank_rankbound} to \ref{lin:cirank_rankbound_end}. 
We first prepare some parameters at the start of the sub-routine, which does not touch the data, and then enters a while loop with at most $N = \lceil\log_2\abs{\mathcal{P}}\rceil$ loops. Denote $s = \imp(p) - \imp(\rank^{-1}(t))$. Within each loop, we add a Gaussian noise to a secret $s$ at \cref{lin:cirank_noisy_indicator}. The value of $s$ touches the sensitive data, but by adding a Gaussian noise to $s$, the release of $\hat{s}$ satisfies zCDP. By \Cref{thm:gaussian_mechanism}, with noise scale $\sigma$, it satisfies $(\Delta_q^2)/2\sigma^2$-zCDP where $\Delta_q$ is the sensitivity of the function that we want to release. Since we set $\sigma = (2 \Delta_{\imp}) / \sqrt{2(\rho/N)}$ at line \ref{lin:cirank_noise_scale} and the sensitivity of $s$ is $2\Delta_{\imp}$ by \Cref{lem:indicator_sens}, it satisfies $(\rho / N)$-zCDP. Since we have at most $N$ noisy releases of $S$ using the Gaussian mechanism, by composition rule (\Cref{lem:composition}), the entire while loop satisfies $\rho$-zCDP, and so is the sub-routine.

    
\paratitle{(2) Confidence Interval}
Now we discuss that the confidence interval outputted from the sub-routine \Call{RankBound}{$p, \rho, \beta, dir$}, from line \ref{lin:cirank_rankbound} to \ref{lin:cirank_rankbound_end}, is a $\gamma$-level confidence interval. 

The sub-routine \textsc{RankBound} with direction as upper is mirror to the sub-algorithm \textsc{RankBound} with direction as lower. We first show that \textsc{RankBound} returns a bound in either upper or lower case such that it is a true bound with probability $\beta = \frac{\gamma+1}{2}$, therefore the target rank is within two bounds with probability $\gamma$. We give the proof for the case when direction is upper for the sub-algorithm \textsc{RankBound}, and skip the proof for the case when direction is lower due to the similarity.

    The sub-routine \textsc{RankBound} is a random binary search algorithm with in total $N$ loops. 
    To ensure that the final $t_{high}$ is a rank bound, one sufficient condition is that $t_{high}$ is always an upper bound of rank during all the loops. 
    Recall that in the noisy binary search, in each loop we first find $t$ as the middle of $t_{high}$ and $t_{low}$, check $s =  \imp(p) - \imp(\rank^{-1}(t)) \leq 0$, add noise a Gaussian noise to $s$ to get $\hat{s}$ and compare $\hat{s}$ with margin, which is $\xi$ in this case. 
    If $\hat{s} \geq \xi$, notice that at line \ref{lin:cirank_move_high}, we change $t_{high}$ to $t$.
    If in this case, $s \leq 0$, which means $t$ is not an upper bound of rank, we never have chance to make $t_{high}$ to be a valid upper bound of rank since it will only decrease in the further loops. Therefore, 
    We say a loop is a failure if during that loop, $s \leq 0$ but $\hat{s} > \xi$. To have a valid rank upper bound, it is necessary to have no loop failure during the entire noisy binary search. We next show that the probability of no such a failure occur is at least $\beta$. 
    
    See the chain of inequalities below. 
    
    \begin{align}
        ~& Pr[\I^U_u ~\text{is an upper bound of $\rank(p_{u}; D, \mathcal{P}, I)$}]
        \\
        \intertext{ The first inequality is due to the bound of the number of while loops. To be a rank bound, it cannot fail at each loop, therefore it has to success for all the $N$ loops. These are independent events, so we can use a product for all the events happen together.}
        \geq & (1-Pr[\text{loop failure}])^{N} \\
        \intertext{The second inequality is due to the bound of $Pr[\text{loop failure}]$. Since any case such that $S \leq 0$ but $\hat{s} > \xi$ is considered as a loop failure, $\hat{s}$ is achieved by adding a Gaussian noise to $s$ and $\xi$ is a constant, the probability of a loop failure only depends on the value of $s$. Since here we have a condition about $s \leq 0$, $\sup_{s \leq 0} Pr[\hat{s} > \xi]$ is an upper bound of $Pr[\text{loop failure}]$.}
        \geq& (1 - \sup_{s \leq 0} Pr[\hat{S} > \xi])^{N} \\
        \intertext{The next equality is because $\sup_{s \leq 0} Pr[\hat{s} > \xi] = Pr[N(0, \sigma^2) > \xi]$. Recall that $\hat{s} = s + N(0, \sigma^2)$ in \cref{lin:cirank_noisy_indicator}, therefore $\sup_{s \leq 0} Pr[\hat{s} > \xi] = \sup_{s \leq 0} Pr[s + N(0, \sigma^2) > \xi] = \sup_{s \leq 0} Pr[N(0, \sigma^2) > \xi - s]$. Since $Pr[N(0, \sigma^2) > \xi - s]$ increases as $s$ increases, it achieves maximum at $s = 0$ for $s \leq 0$. Therefore, $\sup_{s \leq 0} Pr[\hat{s} > \xi] = Pr[N(0, \sigma^2) > \xi]$.}
        =& Pr[\mathcal{N}(0, \sigma^2) \leq \xi]^{N} \\
        \intertext{The third bound is due to Chernoff bound of the Q-function (\Cref{lem:chernoff_q_bound}). Since $Pr[\mathcal{N}(0, \sigma^2) \leq \xi] = 1 - Pr[\mathcal{N}(0, 1) > \xi / \sigma]$, by Chernoff bound we have $Pr[\mathcal{N}(0, 1) > \xi / \sigma] \leq \exp(-(\xi / \sigma)^2/2)$ and therefore $Pr[\mathcal{N}(0, \sigma^2) \leq \xi] \geq 1 - \exp(-(\xi / \sigma)^2/2)$.}
        \geq& \left(1 - \exp(-(\xi/\sigma)^2/2)\right)^{N} \\
        \intertext{The fourth bound is due to $(1+x)^r \geq 1 + rx$ for $x \geq -1$ and $r \geq 1$.}
        \geq& 1 - N \exp(-(\xi/\sigma)^2/2) \\
        \intertext{The final equality is by plugging $\xi = \sigma\sqrt{2\ln(N / (1-\beta))}$.}
        =& \beta
    \end{align}
    
    \sloppy
    Similarly, we have $Pr[\I^L_u ~\text{is a lower bound of $\rank(p_{u}; D, \mathcal{P}, I)$}] \geq \beta$. Together, the probability of $\I_u$ is a $\gamma$ level confidence interval of $\rank(p_{u}; D, \mathcal{P}, I)$
    equals to both events $\I^U_u$ is an upper bound of $\rank(p_{u}; D, \mathcal{P}, I)$ and $\I^L_u$ is a lower bound of $\rank(p_{u}; D, \mathcal{P}, I)$ happen together, which is greater than or equal to the probability sum of each single event minus one (\Cref{lem:event_intersection}, which is $\beta + \beta - 1 = 2 \beta - 1$. By plugging $\beta = (\gamma + 1)/2$ from \cref{lin:cirank_beta}, we have $2 \beta - 1 = \gamma$, which is the confidence interval level for the final confidence interval.
\end{proof}

\begin{theorem}
Given a database $D$, a predicate space $\mathcal{P}$, an influence function $\imp$ with sensitivity $\Delta_{\imp}$, explanation predicates $p_{1}, p_{2}, \ldots, p_{k}$, a confidence level $\gamma$, and a privacy parameter $\rhoRank$,  noisy binary search mechanism returns confidence intervals $\I_1, \I_2, \ldots, \I_k$ such that for $\forall u \in \{1, 2, \ldots, k\}$ and $\forall x \geq 0$, the confidence interval $\I_u = (\I^L_u, \I^U_u)$ satisfies $$Pr[A \leq \imp(p_u) \leq B] \geq 1 - 2e^{-x}$$ where $A = \imp(\rank^{-1}(\I_{u}^L)) - (\abs{\xi_{-1}} + \sigma_{-1}\sqrt{2(x+\ln N)})$ and $B = \imp(\rank^{-1}(\I_{u}^U)) + (\abs{\xi_{+1}} + \sigma_{+1}\sqrt{2(x+\ln N)})$.
\end{theorem}

\begin{proof}

We first show that each utility bound has probability $\geq 1-e^{-x}$, then use the union probability rule to show together it is bounded by $\geq 1-2e^{-x}$.

    Consider the upper utility bound.
    One sufficient condition for the upper utility bound to be true is that $\imp(\rank^{-1}(\I_{rank}^U; D, \mathcal{P}, \imp); D) \geq \imp(p, D) - (\xi + \sigma\sqrt{2(x+\ln N)})$ is always true. Similar to the proof of confidence rank bound, here we say a loop is a failure if $S > \xi + \sigma\sqrt{2(x+\ln N)}$ but $\hat{S} \leq \xi$.
    The proof of the inequality chain below is similar to the proof of confidence interval, except for that the third inequality is due to that $(1-a)^x \geq 1-ax$ for $a \in (0, 1)$ and $x \geq 1$.
    \begin{align*}
        ~& Pr[\imp(\rank^{-1}(\I_{rank}^U; D, \mathcal{P}, \imp)) - \sigma(\xi + \sqrt{2(x+\ln N)})] \\
        \geq & (1-Pr[\text{loop failure}])^{N} \\
        \geq& (1 - \sup_{S > \xi + \sigma\sqrt{2(x+\ln N)}} Pr[\hat{S} \leq \xi])^{N} \\
        =& (1 - Pr[N(0, \sigma^2) \leq -\sigma\sqrt{2(x+\ln N)}])^{N} \\
        \geq& 1 - N Pr[N(0, \sigma^2) \leq -\sigma\sqrt{2(x+\ln N)}] \\
        \geq& 1 - N \exp(-(\sqrt{2(x+\ln N)})^2/2) \\
        =& 1 - \exp(-x)
    \end{align*}
    
\end{proof}

\subsection{Supplementary}
\label{sec:proof_supple}

\begin{lemma}[Chernoff bound of Q function]
\label{lem:chernoff_q_bound}
Given a $Q$ function: $Q(x) = Pr[X > x]$, where $X \sim N(0, 1)$ is a standard Gaussian distribution, if $x \geq 0$, we have
\begin{align*}
    Q(x) \leq \exp(-x^2 / 2)
\end{align*}
\end{lemma}

\begin{proof}
By Chernoff bound, we have $Pr[X > x] \leq E[e^{t X}] / e^{t x}$ for any $t \geq 0$. By the property of Gaussian distribution, we have $E[e^{t X}] = e^{t^2/2}$. Together, we have $Pr[X > x] \leq e^{t^2/2 - tx}$. Since $x \geq 0$, we can choose $t = x$, and have $Pr[X > x] \leq e^{-t^2/2}$.
\end{proof}

\begin{lemma}[Composition \cite{bun2016concentrated}]
\label{lem:composition}
Let $\mechanism: \mathcal{X}^n \rightarrow \mathcal{Y}$ and  $\mechanism': \mathcal{X}^n \rightarrow \mathcal{Z}$ be randomized algorithms. Suppose $\mechanism$ satisfies $\rho$-zCDP and $\mechanism'$ satisfies $\rho'$-zCDP. Define $\mechanism'': \mathcal{X}^n \rightarrow \mathcal{Y} \times \mathcal{Z}$ by $\mechanism''(x) = (\mechanism(x), \mechanism'(x))$. Then $\mechanism''$ satisfies $(\rho + \rho')$-zCDP.
\end{lemma}

\begin{lemma}[Postprocessing \cite{bun2016concentrated}]
\label{lem:postprocessing}
Let $\mechanism: \mathcal{X}^n \rightarrow \mathcal{Y}$ and  $f: \mathcal{Y} \rightarrow \mathcal{Z}$ be randomized algorithms.
Suppose $\mechanism$ satisfies $\rho$-zCDP. Define $\mechanism': \mathcal{X}^n \rightarrow \mathcal{Z}$ by $\mechanism'(x) = f(\mechanism(x))$. Then $\mechanism'$ satisfies $\rho$-zCDP.
\end{lemma}

\begin{lemma}
\label{lem:sum_of_sensitivity}
Given two functions $f$ ang $g$ with sensitivities $\Delta_f$ and $\Delta_g$, the sum of two functions have sensitivity $\Delta_f + \Delta_g$
\end{lemma}
\begin{proof}
BY definition, we have $\max_{D \neighbor D'} \abs{f(D) - f(D')} \leq \Delta_f$ and $\max_{D \neighbor D'} \abs{g(D) - g(D')} \leq \Delta_g$. Therefore, $\max_{D \neighbor D'} \abs{(f(D) + g(D)) - (f(D') + g(D'))} = \max_{D \neighbor D'} \abs{(f(D) - f(D') + (g(D) - g(D'))} \leq \max_{D \neighbor D'} \abs{(f(D) - f(D')} + \max_{D \neighbor D'}\abs{(g(D) - g(D'))} = \Delta_f + \Delta_g$. The inequality is due to the property of absolute.
\end{proof}

\begin{lemma}[Gaussian Confidence Interval \cite{wasserman2004all}]
\label{lem:gaussian_ci}
Given a Gaussian random variable $Z \sim N(\mu, \sigma^2)$ with unknown location parameter $\mu$ and known scale parameter $\sigma$. Let $\I^L = Z - \sigma\sqrt{2}\erf^{-1}(\gamma)$ and $\I^U = Z + \sigma\sqrt{2}\erf^{-1}(\gamma)$, then $\I=(\I^L, \I^U)$ is a $\gamma$ level confidence interval of $\mu$.
\end{lemma}

\begin{proof}
By Theorem 6.16 from the text book \cite{wasserman2004all}. 
\end{proof}


\begin{lemma}
\label{lem:event_intersection}
Given events $A_1, A_2, \ldots, A_\ell$, the following inequality holds: 
$$Pr[\bigwedge_{i=1}^{\ell} A_i] \geq \sum_{i=1}^{\ell} Pr[A_i] - (\ell-1)$$
\end{lemma}
\begin{proof}
First we show that given events $A$ and $B$, we have $Pr[A \land B] \geq Pr[A] + Pr[B] - 1$ since $1 \geq Pr[A \lor B] = Pr[A] + Pr[B] - Pr[A \land B]$. Next we show that 
$$Pr[\bigwedge_{i=1}^{\ell} A_i] \geq Pr[\bigwedge_{i=1}^{\ell-1} A_i] + Pr[A_\ell] - 1$$ using the previous rule. This gives a recursive expression and can be reduced to the final formula in the lemma.
\end{proof}


\begin{restatable}{lemma}{basicInfluSensitivity}
\label{lem:basic_influ_sensitivity}
Given a COUNT or SUM query $q$ with sensitivity $\Delta_q$, a predicate $\phi$, a non-negative query $f: \mathcal{D} \rightarrow \mathbb{N}_0$ with sensitivity 1 and another monotonic
\footnote{A query $q$ is monotonic if for any two databases $D'$ and $D$ such that $\abs{D'} \geq \abs{D}$, we have $q(D') \geq q(D)$.}
and positive query $g: \mathcal{D} \rightarrow \mathbb{N}^+$ with sensitivity 1. Denote $h(D)$ as
\begin{align*}
    h(D) = q(\phi(D)) \frac{f(D)}{g(D)}
\end{align*}
For any two neighboring datasets $D$ and $D'$ such that $\abs{D'} = \abs{D} + 1$, we have
\begin{align*}
    \abs{h(D') - h(D)} \leq \frac{2 \abs{\phi(D)} + f(D) + 1}{g(D)} \Delta_q
\end{align*}
\end{restatable}

\begin{proof}
Denote $x = q(\phi(D))$, $x' = q(\phi(D'))$, $n = \abs{\phi(D)}$. Since $x$ is the aggregation over tuples from $\phi(D)$ and $x$ has sensitivity $\Delta_q$, we have $\abs{x} \leq n \Delta_q$ 
. Denote $\delta_x = x' - x$. Since $x$ has sensitivity $\Delta_q$, we have $\abs{\delta_x} \leq \Delta_q$. Since $g(D)$ is monotonic and has sensitivity 1, we have $g(D) \leq g(D') \leq g(D) + 1$. Since $f$ has sensitivity 1, we have $\abs{f(D) - f(D')} \leq 1$. 
\begin{align*}
    & \abs{h(D') - h(D)} \\
    =& \abs{
        x'
        \frac{
            f(D')
        }{
            g(D')
        }
        -
        x
        \frac{
            f(D)
        }{
            g(D)
        }
    } \\
    =& \abs{
        (x+\delta_x)
        \frac{
            f(D')
        }{
            g(D')
        }
        -
        x
        \frac{
            f(D)
        }{
            g(D)
        }
    } \\
    =& \abs{
        x \left(
        \frac{
            f(D')
        }{
            g(D')
        }
        -
        \frac{
            f(D)
        }{
            g(D)
        }
        \right)
        + 
        \delta_x 
        \frac{
            f(D')
        }{
            g(D')
        }
    }
\end{align*}
Now we divide into two cases depending on the sign of the factor of x in the formula above.

\textbf{Case 1, the factor of x is non-negative.}
\begin{align*}
    & \abs{h(D') - h(D)} \\
    \leq &
        n \Delta_q \left(
        \frac{
            f(D')
        }{
            g(D')
        }
        -
        \frac{
            f(D)
        }{
            g(D)
        }
        \right)
        + 
        \Delta_q 
        \frac{
            f(D')
        }{
            g(D')
        }
    \\
    =& \left[
        (n + 1) 
        \frac{
            f(D')
        }{
            g(D')
        }
        -
        n
        \frac{
            f(D)
        }{
            g(D)
        }
    \right] \Delta_q
    \\
    \leq & \left[
        (n + 1) 
        \frac{
            f(D) + 1
        }{
            g(D)
        }
        -
        n
        \frac{
            f(D)
        }{
            g(D)
        }
    \right] \Delta_q 
    \\
    \leq & \frac{
        f(D) + n + 1
    }{
        g(D)
    } \Delta_q
\end{align*}
\textbf{Case 2, the factor of x is non-positive.}
\begin{align*}
    & \abs{h(D') - h(D)} \\
    \leq & 
        n \Delta_q \left(
        \frac{
            f(D)
        }{
            g(D)
        }
        -
        \frac{
            f(D')
        }{
            g(D')
        }
        \right)
        + 
        \Delta_q 
        \frac{
            f(D')
        }{
            g(D')
        }
    \\
    \leq & 
        \left[ n \left(
        \frac{
            f(D')+2
        }{
            g(D')
        }
        -
        \frac{
            f(D')
        }{
            g(D')
        }
        \right) \right. 
        + \left.
        \frac{
            f(D')
        }{
            g(D')
        }
        \right] \Delta_q
    \\ 
    \leq & 
    \frac{
        2 n  + f(D')
    }{
        g(D')
    } \Delta_q
    \\
    \leq & 
    \frac{
        2 n  + f(D) + 1
    }{
        g(D)
    } \Delta_q
\end{align*}
In conclusion, $\abs{h(D') - h(D)} \leq \frac{2n + f(D) + 1}{g(D)}\Delta_q$.
\end{proof}

%% file: appendix-extensions.tex
\section{Extra Algorithm Descriptions}
\label{sec:extensions}

\subsection{Confidence Interval of Question}
\label{sec:ext_ci_question}

In this section, we elaborate the algorithm of \Cref{sec:private_question_ci} in the form of pseudo codes.

\paratitle{Confidence interval for COUNT and SUM}
In Algorithm \ref{alg:ci_of_question}, at  \cref{lin:questionci-cnt-sigma}, 
we set the noise scale $\sigma$ according to aggregation as $COUNT$ ($SUM$), and at line \ref{lin:questionci_gau_l} and \ref{lin:questionci_gau_u}, 
we set the confidence interval from the standard properties of Gaussian distribution by a margin as $\sqrt{2}(\sqrt{2}\sigma)\erf^{-1}(\gamma)$ for both bounds
\footnote{$\erf^{-1}$ is the inverse function of the error function $\erf$.}
\cite{wasserman2004all}
. 

\paratitle{Confidence interval for AVG} 
In \Cref{alg:ci_of_question}, at line \ref{lin:questionci_beta}, we set the sub confidence level $\beta = 1-(1-\gamma)/4 $ for each individual confidence interval, so that the final confidence level for $o_i - o_j$ is $\gamma$. 
At line \ref{lin:questionci_sigmaS} and \ref{lin:questionci_sigmaC}, 
we set the noise level $\sigma$ for $SUM$ and $COUNT$.
From line \ref{lin:questionci_avg_for_start} to \ref{lin:questionci_avg_for_end}, we extract all the intermediate numerators and denominators, and construct individual confidence intervals.
At line \ref{lin:questionci_image_l} and \ref{lin:questionci_image_u}, we compute the infimum and supremum of the image of the cross product of individual confidence intervals, which is also the confidence interval at level $\gamma$.


\begin{algorithm}
\caption{Compute Confidence Interval of User Question }
\begin{algorithmic}[1]
\Require A user question $Q = (\alpha_i, >, \alpha_j)$ with respect to the query \texttt{SELECT $\Agb$, agg($\Aagg$) FROM R WHERE $\phi$ GROUP BY $\Agb$}, the noisy results $\hat{o}_i$ and $\hat{o}_j$, the privacy budget $\rhoQuery$ for the private query answering, and the confidence level $\gamma$.
\Ensure A $\gamma$-level confidence interval of $o_i - o_j$.
\If{agg $= COUNT$ or agg $= SUM$}
    \If{agg = $COUNT$}
        \label{lin:questionci-cnt-sigma}
        \State $\sigma \gets 1 / \sqrt{2\rhoQuery}$
    \ElsIf{agg = $SUM$}
        \State $\sigma \gets \dommax{\Aagg} / \sqrt{2\rhoQuery}$
    \EndIf
    \State $\I^L \gets \hat{o}_i - \hat{o}_j - 2\sigma\erf^{-1}(\gamma)$
    \label{lin:questionci_gau_l}
    \State $\I^U \gets \hat{o}_i - \hat{o}_j + 2\sigma\erf^{-1}(\gamma)$
    \label{lin:questionci_gau_u}
\ElsIf{agg $= AVG$}
    \State $\beta \gets 1 - (1-\gamma)/4$
    \label{lin:questionci_beta}
    \State $\sigma_{S} \gets \dommax{\Aagg}/ \sqrt{2\rhoQuery/2}$ 
    \label{lin:questionci_sigmaS}
    \State $\sigma_{C} \gets 1 / \sqrt{2\rhoQuery/2}$
    \label{lin:questionci_sigmaC}
    \For{$t \in \{i, j\}$} \emph{/* Recall that $\hat{o}_t= \hat{o}_t^{S} / \hat{o}_t^{C} $*/}
        \label{lin:questionci_avg_for_start}
        \State $\hat{o}_t^{S} \gets$ numerator of $\hat{o}_t$.
        \State $\I_t^{S} \gets (\hat{o}_t^{S} - \sigma_{S}\sqrt{2}\erf^{-1}(\beta), \hat{o}_t^{S} + \sigma_{S}\sqrt{2}\erf^{-1}(\beta))$
        \State $\hat{o}_t^{C} \gets$ denominator of $\hat{o}_t$.
        \State $\I_t^{C} \gets (\hat{o}_t^{C} - \sigma_{C}\sqrt{2}\erf^{-1}(\beta), \hat{o}_t^{C} + \sigma_{C}\sqrt{2}\erf^{-1}(\beta))$
    \EndFor
    \label{lin:questionci_avg_for_end}
    \State $\I^L \gets \inf \{\I_i^{S} / \I_i^{C} - \I_j^{S} / \I_j^{C} \}$
    \label{lin:questionci_image_l}
    \State $\I^U \gets \sup \{\I_i^{S} / \I_i^{C} - \I_j^{S} / \I_j^{C} \}$
    \label{lin:questionci_image_u}
\EndIf
\State $\I \gets (\I^L, \I^U)$
\State \Return $\I$
\end{algorithmic}
\label{alg:ci_of_question}
\end{algorithm}

\subsection{Influence Function Monotonicity}
\label{sec:influence_monotonicity}

The influence function $\imp(p)$ from \Cref{sec:influence} is not monotone w.r.t. predicate implication even without the normalizing factor in the function. We demonstrate this property in the  example below. 

\begin{example}
    \sloppy
    Start with a database with three binary attributes: $A, B, C$ and two tuples: (0, 0, 0), (1, 0, 1). Consider an $agg = COUNT$ query with group by on $A$, so we have $agg(g_0(D)) = 1$ and $agg(g_1(D)) = 1$ for two groups $A = 0$ and $A = 1$. 
    Consider three explanation predicates for the user question $(\alpha_0, \alpha_1, >)$ (note that the noisy values can be different from the true values):
    $p_1: B = 0$, $p_2: B = 0 \land C = 0$ and $p_3: B = 0 \land C = 1$, which satisfy 
    $p_2 \Rightarrow p_1$ and $p_3 \Rightarrow p_1$. 
    However, while $\imp(p_1) = 0$, we have $\imp(p_2) = 1$ and $\imp(p_3) = -1$, i.e., $\imp(p_3) < \imp(p_1) < \imp(p_2)$. 
    
\end{example}

\subsection{Private Top-k Explanation Predicates}
\label{sec:ext_topk}

In this section, we restate the One-shot Top-k mechanism based on the exponential mechanism \cite{dwork2014algorithmic} from \Cref{sec:private_topk} with pseudo codes.

Given a score function $\score: \mathcal{P} \rightarrow \mathcal{R}$ that maps an explanation predicate $p$ to a number, the exponential mechanism (EM) \cite{dwork2014algorithmic} randomly samples $p$ from $\mathcal{P}$ with probability proportional to $exp(\epsilon \score(p) / (2 \Delta_{\score}))$ with some privacy parameter $\epsilon$ and satisfies $(\epsilon^2/8)$-zCDP \cite{DBLP:conf/nips/DurfeeR19, dong2020optimal, cesar2021bounding, DPorg-exponential-mechanism-bounded-range}.
The higher the score is, the more possible that an explanation predicate is selected. 
In \oursys, we use the influence function as the score function.

We denote the exponential mechanism as $\mathcal{M}_E$. 
To find `top-$k$' explanation predicate satisfying DP,
we can first apply $\mathcal{M}_E$ to find one explanation predicate, remove it from the entire explanation predicate space, and then apply $\mathcal{M}_E$ again until $k$ explanation predicates are found. 
It was shown by previous work \cite{DBLP:conf/nips/DurfeeR19, DPorg-one-shot-top-k} that this process is identical to adding i.i.d. Gumbel noise\footnote{For a Gumbel noise $Z \sim Gumbel(\sigma)$, its CDF is $Pr[Z \leq z] = \exp(-\exp(-z/\sigma))$.}
to each score and releasing the top-$k$ predicates by the noisy scores 
(i.e., there is no need to remove predicates after sampling)
. 
We, therefore, use this result to devise a similar solution that is presented in Algorithm \ref{alg:noisy_topk}.
In line 1, we set the noise scale.
In lines 2--4, we randomly sample Gumbel noise with scale $\sigma$ and add it to the influence of each explanation predicate from the space $\mathcal{P}$. In line 5, we sort the noisy scores in the descending order, and in line 6, we find the top-k explanation predicates by their noisy scores. This algorithm satisfies $\rhoTopk$-zCDP (as formally stated in 
\Cref{thm:private_top_k}), 
and can be applied to questions on SUM, COUNT, or AVG queries, with different score functions and sensitivity values for different aggregates. 

\begin{algorithm}
\caption{Noisy Top-k Predicates}
\begin{algorithmic}[1]
\Require 
An influence function $\imp$ with sensitivity $\Delta_{\imp}$, a set of explanation predicates $\mathcal{P}$, a privacy parameter $\rhoTopk$ and a size parameter $k$.
\Ensure Top-k explanation predicates.
\State $\sigma \gets 2\Delta_{\imp}\sqrt{k/(8\rhoTopk)}$
\For {$u \gets 1 \ldots \abs{\mathcal{P}}$}
    \State $s_u \gets \imp(p_u) + Gumbel(\sigma)$
\EndFor
\State Sort $s_1 \ldots s_{\abs{\mathcal{P}}}$ in the descending order.
\State Let $p_1, p_2, \ldots, p_k$ be the top-k elements in the list.
\State \Return $p_1, p_2, \ldots, p_k$
\end{algorithmic}
\label{alg:noisy_topk}
\end{algorithm}

\subsection{Private Confidence Interval of Influence}
\label{sec:ext_influence_ci}

In this section, we elaborate the algorithm of \Cref{sec:private_influence_ci} in the form of pseudo codes.

\Cref{alg:ci_of_influence} takes a privacy budget $\rhoInflu$ as input. 
In \Cref{lin:influ_split_rho} we divide the privacy budget $\rhoInflu$ into $k$ equal portions for each explanation predicate $p_{u}$ for $u \in \{1, \ldots, k\}$. In \Cref{lin:influ_set_scale}, we calibrate the noise scale according to the sensitivity of the influence function. In \Cref{lin:influ_add_noise}, we add a Gaussian noise to the influence $\imp(p_u)$ of explanation predicate $p_u$, and finally in Lines \ref{lin:influ_ci_l} and \ref{lin:influ_ci_u}, we derive the confidence interval based on the Gaussian property \cite{wasserman2004all}. 

\begin{algorithm}
\caption{Compute Confidence Interval of Influence}
\begin{algorithmic}[1]
\Require 
An influence function $\imp$ with respect to the question $(\alpha_i, >, \alpha_j)$,
$k$ explanation predicates $p_1, p_2, \ldots, p_k$, 
a private database $D$, a privacy budget $\rhoInflu$, and a confidence level $\gamma$.
\Ensure A list of $\gamma$-level confidence intervals of the influence $\imp(p_u) / (\hat{o}_i - \hat{o}_j)$ for $u \in \{1, 2, \ldots, k\}$.
\For{$u \in \{1, 2, \ldots, k\}$}
    \State $\rho \gets \rhoInflu / k$ \label{lin:influ_split_rho}
    \If{agg $= COUNT$}
        \label{lin:influ_set_scale}
        \State $\sigma \gets 4 / \sqrt{2\rho}$
    \ElsIf{agg $= SUM$}
        \State $\sigma \gets 4 \dommax{\Aagg} / \sqrt{2\rho}$
    \ElsIf{agg $= AVG$}
        \State $\sigma \gets 16 \dommax{\Aagg} / \sqrt{2\rho}$
    \EndIf
    \State 
    $\hat{\imp} \gets \imp(p_u) + N(0, \sigma^2)$
    \label{lin:influ_add_noise}
    \State $\I^L_u \gets \hat{\imp} - \sqrt{2}\sigma\erf^{-1}(\gamma)$
    \label{lin:influ_ci_l}
    \State $\I^U_u \gets \hat{\imp} + \sqrt{2}\sigma\erf^{-1}(\gamma)$
    \label{lin:influ_ci_u}
    \State $\I_u \gets (\I^L_u, \I^U_u)$
    \label{lin:influ_ci_merge}
\EndFor
\State \Return $\I_1, \I_2, \ldots, \I_k$
\end{algorithmic}
\label{alg:ci_of_influence}
\end{algorithm}

\input{appendix-rankci}

%% file: appendix-rankci.tex
\subsection{\bf \Cref{pro:private_rank_ci}: Private Confidence Interval of Rank}
\label{sec:appendix_private_rank_ci}


In this section, we elaborate the algorithm of \Cref{sec:private_influence_ci}, noisy binary search mechanism, in the form of pseudo codes as \Cref{alg:find_rank_ci} shows.

In \cref{lin:cirank_rankbound}, \textsc{RankBound} takes four parameters: an explanation predicate $p$, a privacy budget $\rho$, a sub confidence level $\beta$ and a direction $dir \in \{-1, +1\}$. It guarantees that it will find a lower ($dir = -1$) or upper ($dir = +1$) bound of rank with confidence $\beta$ for the explanation predicate $p$ using privacy budget $\rho$. 
In \cref{lin:cirank_depth}, we set the maximum depth $N$ of the binary search.
In \cref{lin:cirank_noise_scale}, we set the noise scale $\sigma_{-1}$ or $\sigma_{+1}$,  which depends on the sensitivity of $\imp(p) - \imp(\rank^{-1}(t))$ (in \cref{lin:cirank_noisy_indicator}), which is $2\Delta_{\imp}$; and the number of Gaussian mechanisms used in the binary search, which is $N$.  
In \cref{lin:cirank_bias}, we set the margin $\xi_{+1}$ or $\xi_{-1}$, which will be discussed in \cref{lin:cirank_compare_start}.
In \cref{lin:cirank_bs_init}, we initialize the binary search by setting two pointers, $t_{low}$ and $t_{high}$, as the first and last rank.
In lines \ref{lin:cirank_bs_start}--\ref{lin:cirank_bs_end} there is a while loop for the binary search. In \cref{lin:cirank_bs_middle}, we pick a rank that is at the middle of two pointers. In \cref{lin:cirank_noisy_indicator}, we add a Gaussian noise with scale $\sigma$ to the difference between the influence of the target explanation predicate $p$ and the influence of the explanation predicate that has rank $t$. From line \ref{lin:cirank_compare_start} to \ref{lin:cirank_compare_end} we update one of the pointer according to the relationship between the noisy difference and the margin $\xi_{dir}$. If we are trying to find a rank upper bound ($dir = +1$), we want the binary search to find the rank such that the difference (without noise) is above zero. Due to the noise injected, even if the noisy difference is above zero, the true difference could be negative. To secure the goal with high probability, we requires the noisy difference to be above a margin $\xi_{dir}$, as shown in \cref{lin:cirank_compare_start}. In this case, we narrow down the search space by moving $t_{high}$ to $\max\{t-1, 1\}$. The strategy is similar when we are looking for a rank lower bound ($dir = -1$).

Now, we describe the usage of the sub-routine \textsc{RankBound}.  We repeat the following for each explanation predicate.
In \cref{lin:cirank_rho}, we allocate an even portion from the total privacy budget $\rhoRank$,
and set the sub confidence level to $\beta = (\gamma+1)/2$ so the final confidence interval has confidence level $2 \beta - 1 = \gamma$ by the rule of union bound.
In lines \ref{lin:cirank_ci}, we divide the privacy budget $\rho$, and make two calls to the sub-routine \textsc{RankBound} to find a rank upper bound and a rank lower bound for the explanation predicate $p_{u}$, and finally merge them into a single confidence interval. We spend more budget for the rank upper bank since this is more important in the explanation.

\begin{algorithm}
\caption{Compute Confidence Interval of Rank}
\begin{algorithmic}[1]
\Require A dataset $D$, a predicate space $\mathcal{P}$, an influence function $\imp$ with sensitivity $\Delta_{\imp}$, explanation predicates $p_{1}, p_{2}, \ldots, p_{k}$, a confidence level $\gamma$, and a privacy parameter $\rhoRank$.
\Ensure A list of $\gamma$-level confidence intervals of the influence $\rank(p_{u}; D, \mathcal{P}, \imp)$ for $u \in \{1, 2, \ldots, k\}$.
\Function{RankBound}{$p, \rho, \beta, dir$}
\label{lin:cirank_rankbound}
\State $N \gets \lceil\log_2\abs{\mathcal{P}}\rceil$
\label{lin:cirank_depth}
\State $\sigma_{dir} \gets (2 \Delta_{\imp}) / \sqrt{2(\rho/N)}$
\label{lin:cirank_noise_scale}
\State $\xi_{dir} \gets \sigma_{dir}\sqrt{2\ln(N / (1-\beta))} \times dir$
\label{lin:cirank_bias} 


\State $t_{low}, t_{high} \gets 1, \abs{\mathcal{P}}$
\label{lin:cirank_bs_init}
\While{$t_{high} \geq t_{low}$}
\label{lin:cirank_bs_start}
    \State $t \gets \lfloor\frac{t_{high} + t_{low}}{2}\rfloor$
    \label{lin:cirank_bs_middle}
    \State $\hat{s} \gets \imp(p) - \imp(\rank^{-1}(t)) + \mathcal{N}(0, \sigma^2)$
    \label{lin:cirank_noisy_indicator}
    \If {$\hat{s} \geq \xi_{dir} $ }
    \label{lin:cirank_compare_start}
        $t_{high} \gets \max\{t-1, 1\}$
        \label{lin:cirank_move_high}
    \Else\
        $t_{low} \gets \min \{t+1, \abs{\mathcal{P}}\}$
        \label{lin:cirank_move_low}
    \EndIf
    \label{lin:cirank_compare_end}
\EndWhile
\label{lin:cirank_bs_end}
\State \Return $t_{high}$
\label{lin:cirank_return_bound}
\EndFunction
\label{lin:cirank_rankbound_end}
\For{$u \gets 1, 2, \ldots, k$}
    \label{lin:cirank_for_start}
    \State $\rho, \beta \gets \rhoRank / k, (\gamma+1)/2$
    \label{lin:cirank_rho}
    \label{lin:cirank_beta}
    \State {\footnotesize $\I_{u} \gets (\Call{RankBound}{p_{u}, 0.1\rho, \beta, -1}, \Call{RankBound}{p_{u}, 0.9\rho, \beta, +1})$}
    \label{lin:cirank_ci} 
\EndFor
\label{lin:cirank_for_end}
\State \Return $\I_1, \I_2, \ldots, \I_k$
\end{algorithmic}
\label{alg:find_rank_ci}
\end{algorithm}

%% file: appendix-algorithm.tex
\section{Algorithm Variants}

\subsection{General User Question}
\label{sec:ext_general_question}

In this section, we introduce a general form of user question through weighted sum, such that more groups can be involved in the question and the comparison between groups can be more flexible. This covers the cases of the original questions, since a single group difference can also be treated as a weighted sum between two groups.
We also discuss how the explanation framework should be adapted to this general form. Finally, we give a use case for privately explaining a general user question.

\begin{definition}[General User Question]
\label{def:general_user_question}
Given a database $D$ an aggregate query $q$, a DP mechanism $\mechanism$, and  noisy group aggregation releases $\hat{o}_{i_1}, \hat{o}_{i_2}, \ldots, \hat{o}_{i_m}$ of the groups $\alpha_{i_1}, \alpha_{i_2}, \ldots, \alpha_{i_m}$ from the query $q$, 
a general user question $\userquery$ is represented by $m$ weights and a constant $c$: $(w_{i_1}, w_{i_2}, \ldots, w_{i_m}, c)$. Intuitively, the question is interpreted as ``Why $\sum_{j=i_1}^{i_m} w_j \hat{o}_j \geq c$''. 
\end{definition}

\Cref{def:general_user_question} allows more interesting questions, such as "Why the total salary of group A and B is larger than the total salary of group C and D?" or "Why the average salary of group A is 10 times larger than the one of group B?".
Next we illustrate how the algorithms for each problem related to our framework should be adapted in the case of general user question. 

\paratitle{Private Confidence Interval of Question}
Given a general user question $(w_{i_1}, w_{i_2}, \ldots, w_{i_m}, c)$, we discuss how to derive the confidence interval of $\sum_{j=i_1}^{i_m} w_j o_j - c$. Comparing to the case of a simple user question $(\alpha_i, >, \alpha_j)$, where the target of confidence interval is $o_i - o_j$, here we have a weighted sum of multiple group results. Therefore, when $agg$ is $CNT$ or $SUM$, the noisy weighted sum follows the Gaussian distribution with scale $\sqrt{\sum_{j=i_1}^{i_m}w_j^2} \sigma$, where $\sigma$ is the noise scale used in query answering. When $agg$ is $AVG$, the noisy weighted sum can also be viewed as a combination of multiple Gaussian variables. In conclusion, we consider the adaptaions as follows:
\begin{enumerate}
    \item For $agg = CNT$ or $agg = SUM$, update the margin $\sqrt{2}(\sqrt{2}\sigma)\erf^{-1}(\gamma)$ as $\sqrt{2}(\sqrt{\sum_{j=i_1}^{i_m}w_j^2}\sigma)\erf^{-1}(\gamma)$.
    \item For $agg = AVG$, update the sub confidence level $\beta$ to be $(\gamma-1)/(2m)+1$ 
    , 
    and the image of sub confidence intervals to be $\sum_{j=i_1}^{i_m} \I^S_{j} / \I^C_{j} - c$  
    .
\end{enumerate}

\paratitle{Private Top-k Explanation Predicates}

Since the user question has a new form, the influence function and its corresponding score function should also be adapted. We consider their natural extensions as follows:

\begin{definition}[General Influence Function]
\label{def:general_impact_function}
Given a database $D$ and a general user question $\userquery = (w_{i_1}, w_{i_2}, \ldots, w_{i_m}, c)$ with respect to the query \texttt{SELECT $\Agb$, agg($\Aagg$) FROM R WHERE $\phi$ GROUP BY $\Agb$}, the influence of an explanation predicate $p$ is defined follows:
\begin{small}
\begin{align*}
    \imp(p; Q, D) = &
    \Bigg( 
    \sum_{j=i_1}^{i_m} w_j q(g_j(D))
    -
    \sum_{j=i_1}^{i_m} w_j q(g_j(\neg p (D)))
    \Bigg)
    \\ & \times
    \begin{cases}
        \frac{
            \min\limits_{t \in \{i_1, i_2, \ldots, i_m\}} \abs{g_t(\neg p (D))}
        }{
            \max\limits_{t \in \{i_1, i_2, \ldots, i_m\}} \abs{g_t(D)} + 1
        }
        &, agg \in \{COUNT, SUM\} \\
        \min\limits_{t \in \{i_1, i_2, \ldots, i_m\}} \abs{g_t(\neg p (D))}
        &, agg = AVG
    \end{cases}
\end{align*}
\end{small}
\end{definition}


We can plug-in the new influence function into \cref{alg:noisy_topk} to find the noisy top-k explanation predicates. The corresponding sensitivity of the new influence function is given as follows:

\begin{restatable}{theorem}{generalInfluSens}[General Influence Function Sensitivity]
\label{thm:general_score_sens}
Given an explanation predicate $p$ and a general user question $\userquery = (w_{i_1}, w_{i_2}, \ldots, w_{i_m}, c)$ with respect to a group-by query with aggregation $agg$, the following holds:
\begin{enumerate}
    \item If $agg = CNT$, the sensitivity of $\imp(p; Q, D)$ is $2 \sum_{j=i_1}^{i_m} \abs{w_j}$.
    \item If $agg = SUM$, the sensitivity of $\imp(p; Q, D)$ is $2 \sum_{j=i_1}^{i_m} \abs{w_j} \dommax{\Aagg}  $. 
    \item If $agg = AVG$, the sensitivity of $\imp(p; Q, D)$ is $8 \sum_{j=i_1}^{i_m} \abs{w_j} \dommax{\Aagg} $.
\end{enumerate}
\end{restatable}


\begin{proof}
It is a weighted version of \Cref{thm:influ_sens}.
\end{proof}

We also allow explanation predicates to include disjunction and allow the framework to specify a specific set of explanation predicates by enumeration. 

\paratitle{Private Confidence Interval of Influence}
We can plug-in the new influence function and their sensitivities into the original algorithm to find the confidence interval of influence.


\paratitle{Private Confidence Interval of Rank}
We can plug-in the new influence function and their sensitivities into \cref{alg:find_rank_ci} to find the confidence interval of rank. 

\paratitle{Use Case: Taxi-Imbalance}
We consider the New York City taxi trips dataset \cite{nyctaxi} in January and February, 2019, as a use case. We preprocessed the dataset such that it includes 4 columns: {\tt PU\_Zone}, {\tt PU\_Borough}, {\tt DO\_Zone}, {\tt DO\_Borough}. In this case we analyze the traffic volume between boroughs.
With privacy budget $\rhoQuery = 0.1$, the framework answers the user query as
``SELECT PU\_Borough, DO\_Borough, CNT(*) FROM R GROUP BY PU\_Borough, DO\_Borough''. There are in total 49 groups, and among the query answers we have \textit{(Brooklyn, Queens): 11,431} and \textit{(Queens, Brooklyn): 121,934}. 
User then asks ``Why Queens to Brooklyn has more than 10 times the number of trips from Brooklyn to Queens?''
This corresponds to the question "why $q_1 - 10 q_2 \geq 0$", or in the form of weights $(1, -10, 0)$.
 The confidence interval of the question is $(7580, 7668)$, which validates the question. 
To explain the question, 
we consider a predicate space of the form "PU\_Zone = <zone> $\lor$ DO\_Zone = <zone>" with in total 127 different zones.
With $\rhoTopk = 0.025$, $\rhoInflu = 0.025$, and $\rhoRank = 0.95$, we have the explanation table as shown in \cref{fig:taxi_imbalance_exptable}. The relative influence is relative to the noisy difference $\hat{o}_1 - 10 \hat{o}_2 = 7624$.  From this table, we can find that two airports, JFK and LaGuardia airports that are located in Queens, are the major reasons for why there are more traffic volume from Queens to Brooklyn since there are more incoming taxi traffic to the airports instead of outgoing taxi traffic.

\begin{figure}[h]
    \centering
\begin{tcolorbox}[colback=white,left=1.5pt,right=1.5pt,top=0pt,bottom=0pt]
    \small
    \begin{tabular}{@{} >{\raggedright}p{3.8cm} c c c c @{}}
        \toprule
        \multirow{2}{*}{explanation predicate} &
        \multicolumn{2}{c}{Rel Influ 95\%-CI} &
        \multicolumn{2}{c}{Rank 95\%-CI} \\
        \cmidrule(lr){2-3}
        \cmidrule(lr){4-5}
        & L & U & L & U \\
        \midrule
zone = "JFK Airport"                            &  55.21\% & 72.18\% &   1 &   1  \\
zone = "LaGuardia Airport"                      &  28.75\% & 45.72\% &   1 &   3  \\
zone = "Bay Ridge"                              &  -6.64\% & 23.60\% &   3 & 127  \\
zone = "Queensboro Hill"                        & -10.75\% &  6.22\% &   3 & 127  \\
zone = "Flushing"                               & -12.52\% &  4.25\% &   3 & 127  \\
        \bottomrule
    \end{tabular}
\end{tcolorbox}    
    \caption{Top-5 explanations for Taxi-Imbalance.}
    \label{fig:taxi_imbalance_exptable}
\end{figure}

\subsection{Finding Top-k by arbitrary influence function}



In the noisy binary search of \cref{alg:find_rank_ci}, we use the difference between $\imp(p; D)$ and $\imp(\rank^{-1}(t; D, \mathcal{P}, \imp); D)$ as an indicator for each branch. The utility of this algorithm depends on the global sensitivity of the influence function. When we extend the entire framework to support more queries and questions, the influence function can be more complex and sensitive.
For example, given a question such as why $q_1(D)/q_2(D)$ is higher than expected, for some query $q_1$ and $q_2$ in the first phase, one influence function could be $\imp(p; D) = (1-\abs{p(D)}/\abs{D})(q_1(D)/q_2(D) - q_1(\neg p(D))/q_2(\neg p(D)))$. In this case, one can always find $p$ and $D$ and $D'$ such that the absolute difference between $\imp(p; D)$ and $\imp(p; D')$ is arbitrary high. A typical work around is to bound the ranges of basic queries; however, it introduces bias and may destroy the ranking order. Moreover, the bound needs to be chosen without looking the data, which makes it even more impossible.

On the other hand, the difference between $\imp(p; D)$ and $\imp(\rank^{-1}(t; D, \mathcal{P}, \imp); D)$ is not the only choice of branch indicator. Denote $S$ as a function of the form $S(p, t; D, \mathcal{P}, \imp)$. In general, if this function satisfies three properties as listed in the theorem below, which are also the only properties that the proof of \Cref{thm:rank_ci} requires, using this function as the branch indicator in \cref{alg:find_rank_ci} still allows this algorithm to satisfy $\rho$-zCDP and the guarantee of confidence interval of rank.

\begin{theorem}
\label{thm:condition_of_confidence_rank_bound_algorithm}
Substituting $\imp(p) - \imp(\rank^{-1}(t))$ by $S(p, t; D, \mathcal{P}, \imp)$ and $2 \Delta_{\imp}$ by $\Delta_S$ for \cref{alg:find_rank_ci}, the new algorithm satisfies $\rho$-zCDP
and outputs correct confidence intervals of rank
if the following holds for $S$ and $\Delta_S$:
\begin{itemize}
    \item Center Zero. $S(p, \rank(p; D, \mathcal{P}, \imp); D, \mathcal{P}, \imp) = 0$
    \item Non-Decreasing. For any $i < j$, $S(p, i; D, \mathcal{P}, \imp) \leq S(p, j; D, \mathcal{P}, \imp)$.
    \item Stable. For any two neighboring datasets $D \neighbor D'$, $\abs{S(p, t; D, \mathcal{P}, \imp) -  S(p, t; D', \mathcal{P}, \imp)} \leq \Delta_S$. 
\end{itemize}
\end{theorem}
 
A natural choice of $S$ is to define $S(p, t; D, \mathcal{P}, I) = \imp(p; D) - \imp(\rank^{-1}(t; D, \mathcal{P}, I); D)$, the difference between the influence of $p$ and the $t$-th largest influence. With the "Center Zero" and "Non-Decreasing" properties, the indicator function $S$ can tell that a number $t$ is a rank bound of $\rank(p; D, \mathcal{P}, \imp)$ if $S(p, t; D, \mathcal{P}, \imp) > 0$. If $i$ and $j$ are both rank bound of $\rank(p; D, \mathcal{P}, \imp)$ and $i$ is closer to the target rank than $j$, $S(p, i; D, \mathcal{P}, \imp)$ is also closer to 0 than $S(p, j; D, \mathcal{P}, \imp)$.  However, for the natural choice of $S$, sensitivity $\Delta_S = 2\Delta_{\imp}$ and $\Delta_{\imp}$ could be unbounded for some $\imp$, which results in poor utility. Instead, we can define $S$ in a way such that it still reflects the difference between the influence of $p$ and the $t$-th largest influence, but has low sensitivity. 

Inspired by inverse sensitivity and other techniques that share the same spirit \cite{asi2020near, DBLP:conf/nips/AsiD20, DBLP:journals/pvldb/FariasBFMMS20, smith2011privacy}, we present a stable branch indicator function $S(p, t; D, \mathcal{P}, \imp)$ such that it is approximately the least number of tuples that need to be changed to move the rank of $p$ beyond $t$. Specially, when $t \leq \rank(p; D, \mathcal{P}, \imp)$, $S(p, t; D, \mathcal{P}, \imp) = 0$.

Denote $D \triangle D'$ as the symmetric difference between two datasets $D$ and $D'$. Denote influence lower bound $ILB(p, d; D, \imp) = \inf\{\imp(p; D') \mid |D' \triangle D| \leq d \}$ the least influence of $p$  and influence upper bound $IUB(p, d ; D, \imp) = \sup\{\imp(p; D') \mid |D' \triangle D| \leq d \}$ the largest influence of $p$ within distance $d$ to $D$. Given two predicates $p$ and $\tilde{p}$, if $IUB(\tilde{p}, d; D, \imp) < ILB(p, d; D, \imp)$, it indicates that there is no dataset $D'$ within distance $d$ to $D$ such that the influence of $\tilde{p}$ is higher than or equal to the one of $p$. 

Denote the complementary size of such predicate $\tilde{P}$ in $\mathcal{P}$ as $B(p, d; D, \mathcal{P}, \imp) = \abs{\mathcal{P}} -  \lvert\{\tilde{p} \in \mathcal{P} \mid IUB(\tilde{p}, d; D, \imp) < ILB(p, d; D, \imp)\} \rvert$. This gives a rank bound of $\rank(p; D', \mathcal{P}, \imp)$ for any dataset $D'$ such that $\abs{D \Delta D'} \leq d$. 

\eat{
\begin{example}
Given a predicate $p$, a dataset $D$ and an influence function $I(p; D) = (1-\abs{p(D)}/\abs{D})(q_1(D)/q_2(D) - q_1(\neg p(D))/q_2(\neg p(D)))$ where $q_1(D)$ and $q_2(D)$ are predicate counting queries with sensitivity 1. Suppose $\abs{p(D)} = 50, \abs{D} = 100, q_1(D) = 10, q_2(D) = 20, q_1(\neg p(D)) = 5, q_2(\neg p(D)) = 15$. One possible influence lower bound within distance 1 is $ILB(I, p, D, 1) = (1 - (50+1)/(100-1))((10-1)/(20+1)-(5+1)/(15-1)) = 0$. One possible influence upper bound within distance 1 is $ILB(I, p, D, 1) = (1 - (50-1)/(100+1))((10+1)/(20-1)-(5-1)/(15+1)) \approx 0.169$. 
\end{example}
}

\eat{
\begin{lemma}
Given a predicate $p$, a distance $d$, a dataset $D$, a set of predicates $\mathcal{P}$ and an influence function $I$, $B(p, d; D, \mathcal{P}, I)$ is a rank bound at distance $d$ for $p$ such that of $\rank(p; D, \mathcal{P}, I)$.
\end{lemma}
}

\begin{example}
Suppose $\mathcal{P}$ has 5 predicates $p_1$, $p_2$, $p_3$, $p_4$ and $p_5$. Now we show $B(p_3, 2)$. Suppose at distance 2, $IUB(p_1, 2) = 1$, $IUB(p_2, 2) = 3$, $IUB(p_3, 2) = 6$, $IUB(p_4, 2) = 7$, $IUB(p_5, 2) = 10$, and $ILB(p_3, 2) = 4$. In this case, $B(p_3, 2) = 5 - 2 = 3$ since predicate $p_1$ and $p_2$ have lower $IUB$ than the $ILB$ of $p_3$. This indicates, by adding or removing $2$ tuples from $D$, the rank of $p_3$ cannot be beyond $3$.  
\end{example}

\begin{lemma}
Given a predicate $p$, a dataset $D$, a set of predicates $\mathcal{P}$ and an influence function $\imp$, for any dataset $D'$ such that $\abs{D \Delta D'} \leq 1$ and any distance $d$, we have:
\begin{align}
    B(p, d; D, \mathcal{P}, \imp) \leq B(p, d+1; D', \mathcal{P}, \imp)
\end{align}
\end{lemma}

\begin{proof}
Denote $\mathcal{D}_1 = \{D'' \mid \abs{D'' \Delta D} = d\}$ and $\mathcal{D}_2 = \{D'' \mid \abs{D'' \Delta D'} = d+1\}$. Notice that $B(p, d; D, \mathcal{P}, \imp)$ (or $B(p, d+1; D', \mathcal{P}, \imp)$) is counting the complementary size of predicate $\tilde{p}$ in $\mathcal{P}$ such that no dataset $D''$ in $\mathcal{D}_1$ (or $\mathcal{D}_2$) satisfies $\imp(\tilde{p}; D'') \geq \imp(p; D'')$.  Since $\abs{D \Delta D'} \leq 1$, we have $\mathcal{D}_1 \subseteq \mathcal{D}_2$, therefore $B(p, d; D, \mathcal{P}, \imp) \leq B(p, d+1; D', \mathcal{P}, \imp)$.
\end{proof}

If $IUB$ is a loose influence upper bound and $ILB$ is a loose influence lower bound, the lemma above still holds. We show an example of the function $B$ on two neighboring datasets as follows in \Cref{tbl:example_of_bd}.

\begin{table}[h!]
\caption{Example of $B$}
\centering
\begin{tabular}{|l|l|l|l|l|l|} 
\hline
d         & 0 & 1 & 2 & 3 & 4   \\ 
\hline
$B(p, d; D)$   & 2 & 2 & 4 & 6 & 10  \\ 
\hline
$B(p, d; D')$ & 2 & 3 & 5 & 7 & 8   \\
\hline
\end{tabular}
\label{tbl:example_of_bd}
\end{table}

\begin{definition}
Given a predicate $p$, a dataset $D$, a set of predicates $\mathcal{P}$ and an influence function $\imp$, $\omega$ is a stable branch indicator function as
\begin{align*}
    \omega(p, t; D, \mathcal{P}, \imp) = \min\{d \geq 0 \mid B(p, d; D, \mathcal{P}, \imp) \geq t\}
\end{align*}
\end{definition}

Below, we show an example of a stable branch indicator in \Cref{tbl:example_of_omegat}.

\begin{table}[h!]
\caption{Example of $\omega$}
\centering
\begin{tabular}{|l|l|l|l|l|l|l|l|l|} 
\hline
t        & 1 & 2 & 3 & 4 & 5 & 6 & 7 & \ldots  \\ 
\hline
$\omega(p, t)$ & 0 & 0 & 2 & 2 & 3 & 3 & 10 & \ldots  \\
\hline
\end{tabular}
\label{tbl:example_of_omegat}
\end{table}

\begin{theorem}
Given a predicate $p$, a dataset $D$, a set of predicates $\mathcal{P}$ and an influence function $\imp$, $\omega$, the three conditions of \cref{thm:condition_of_confidence_rank_bound_algorithm} is satisfied if function $S = \omega$ and sensitivity $\Delta_S = 1$.
\end{theorem}

\begin{proof}~

\textbf{Center Zero.} $B(p, 0; D, \mathcal{P}, I) = \rank(p; D, \mathcal{P}, I)$.

\textbf{Non-Decreasing.} Since $B$ is non-decreasing in terms of $d$ given $D, \mathcal{P}, I$, $\omega_t(D)$ is also non-decreasing.

\textbf{Stable.} 
Drop $\mathcal{P}, \imp$ for simplicity.
Suppose t is fixed. Denote $d^* = \omega(p, t; D)$. By definition, we have $B(p, d^*; D) \geq t$. For any neighboring dataset $D' \sim D$, since $B(p, d^*; D) \leq B(p, d^*+1; D')$, it indicates $B(p, d^*+1; D') \geq t$ and thus we have $\omega(p, t; D') \leq d^*+1$.

When $d^* < 2$, which means $\omega(p, t; D) < 2$, it is impossible to have $\omega(p, t; D') - \omega(p, t; D) < -1$ since $\omega(p, t; D') \geq 0$ is always true. When $d^* \geq 1$, we show that it is impossible to have $B(p, d^*-2; D') \geq t$. If $B(p, d^*-2; D') \geq t$, we have $B(p, d^*-1; D) \geq B(p, d^*-2; D') \geq t$, which indicates $\omega(p, t; D) \leq d^*-1$ and leads to a contradiction. Therefore, we have $B(p, d^*-2; D') < t$, which indicates the impossibility of $\omega(p, t; D') \leq d^*-2$. Therefore, we have $\omega(p, t; D') \geq d^*-1$.

Since $d^*-1 \leq \omega(p, t; D') \leq d^*+1$, we have $|\omega(p, t; D) - \omega(p, t; D')| \leq 1$.
\end{proof}

The branch indicator function $\omega$ finds the minimum $d$ such that $B(p, d; D, \mathcal{P}, I) \geq t$. If we add a constraint $d \geq C$ with some constant $C$, the theorem above still holds.

\subsection{Large Domain Private Top-k Selection}


\Cref{alg:report_top_k} gives a practical version for find top-k elements given an score function from a large domain. It assumes that the domain $\mathcal{P}$ is partitioned into an active domain $\mathcal{P}_{act}$ and an idle domain $\mathcal{P}_{idle}$, such that the elements in the idle domain all have the same score $C$. We assume a random draw from the idle domain $\mathcal{P}_{idle}$ could be done in $O(1)$, so the runtime of the algorithm only depends on the size of the active domain $\mathcal{P}_{act}$ as $O(k \abs{\mathcal{P}_{act}})$. This algorithm satisfies $k \epsilon^2/8$-zCDP or $k\epsilon$-DP. 

\begin{algorithm}[h]
\caption{Report Noisy Top-k Elements from a Large Domain}
\begin{algorithmic}[1]
\Require A private dataset $D$, an active domain $\mathcal{P}_{act}$ of predicate class $\mathcal{P}$, an idle domain $\mathcal{P}_{idle}$ of predicate class $\mathcal{P}$,  a score function $u$, global sensitivity of $u$ as $\Delta_u$, a constant $C$ as the score for any element from the idle domain $\mathcal{P}_{idle}$, and a privacy parameter $\epsilon$.
\Ensure Top-k predicates ordered by scores.
\State Compute $u(p,D)$ for every $p \in \mathcal{P}_{act}$ without releasing the results.
\For {$i \gets 1 \ldots k$}
    \State $s \gets - \infty$
    \For {$p \gets $ iterate the space of $\mathcal{P}_{act}$}
        \State $s' \gets u(p, D) + Gumbel(2\Delta_u/\epsilon) $
        \If{$s' > s$}
            \State $p_{r_i} \gets p$
            \State $s \gets s'$
        \EndIf
    \EndFor
    \State $s' \gets \frac{2\Delta_u}{\epsilon} \ln(|\mathcal{P}_{idle}) + C + Gumbel(2\Delta_u/\epsilon)$
    \If{$s' > s$}
        \State $\hat{p}_i \gets $ a random draw from $\mathcal{P}_{idle}$
    \EndIf
    \If{$\hat{p}_i \in \mathcal{P}_{act}$}
        \State $\mathcal{P}_{act} \gets \mathcal{P}_{act} \setminus \{\hat{p}_i\}$
    \Else
        \State $\mathcal{P}_{idle} \gets \mathcal{P}_{idle} \setminus \{\hat{p}_i\}$
    \EndIf
\EndFor
\State \Return $(\hat{p}_1, \hat{p}_2, \ldots, \hat{p}_k)$.
\end{algorithmic}
\label{alg:report_top_k}
\end{algorithm}

\subsection{Computing Confidence Interval of General Arithmetic Combinations}
\label{sec:general_ci}








Formally, a query $q$ can be expressed as an arithmetic combination of queries if it can be expressed as $q(D) = f(q_1(D), q_2(D), \ldots, q_\ell(D))$ where function $f$ includes the operators in $\{+, -, *, /, \exp, \log\}$ and for each sub-query $q_i$, a noisy answer $\hat{\subo}_i =  N(q_i(D), \sigma_i^2)$ is released under $\rho$-zCDP \footnote{Although user doesn't see the intermediate differentially private query result, we assume the framework stores them.}, where $\sigma_i = \Delta_{q_i} / \sqrt{2\rho/\ell}$ and $\Delta_{q_i}$ is the sensitivity for sub-query $q_i$. The rest of this sub section discusses how to derive the confidence interval for $q(D)$ based on the noisy releases $\hat{\subo}_1, \hat{\subo}_2, \ldots, \hat{\subo}_\ell$ and function $f$.

Given the noisy releases $\hat{\subo}_1$, $\hat{\subo}_2$, $\ldots$, $\hat{\subo}_\ell$ through the Gaussian mechanism,  the confidence intervals of $q_1(D),$ $q_2(D),$ $\ldots,$ $q_\ell(D)$ can be derived by Gaussian confidence interval 
, and there is a clear connection between these queries to $q(D)$ through function $f$. Therefore, we can compute the image of these confidence intervals through the function $f$, which is also a valid confidence interval for $q_f(D)$.
Given a function $f: X \rightarrow Y$, denote by $f[A]$ the image of $f$ under $A\subseteq X$ i.e. $f[A] = \{f(a): a \in A\}$.

\begin{restatable}{theorem}{generalci}
\label{thm:general_ci}
Given a database $D$ and a query $q$ that can be expressed as $q_f(D)$ $=$ $f(q_1(D),$ $q_2(D),$ $\ldots,$ $q_\ell(D))$, 
where $f$ includes the operators in $\{+, -, *, /, \exp, \log\}$, and confidence intervals at confidence level $\beta$ for $q_1(D), q_2(D), \ldots, q_\ell(D)$ as $\I_1, \I_2, \ldots, \I_\ell$. Let $\I=f[\I_1 \times \I_2 \times \ldots \times \I_\ell]$ be the image of $\I_1 \times \I_2 \times \ldots \times \I_\ell$ under $f$, i.e., the set of numbers composed of each mapping of $f$ for a combination of values from $\I_1 \times \I_2 \times \ldots \times \I_\ell$. Also assume that $f$ is defined for each vector in $\I_1 \times \I_2 \times \ldots \times \I_\ell$. 
Let $\I^L = \inf I$ and $\I^U = \sup I$.  We have
\begin{align*}
    Pr[\I^L \leq q_f(D) \leq \I^U] \geq \ell(\beta-1)+1
\end{align*}
\end{restatable}

\begin{proof}
Since the event $\bigwedge_{i=1}^{\ell} q_i(D) \in \I_i$ implies $q_f(D) \in \I$ due to $\I$ is the image of $\I_1 \times \I_2 \times \ldots \times \I_\ell$ under $f$, we have $Pr[q_f(D) \in \I] \geq Pr[\bigwedge_{i=1}^{\ell} (q_i(D) \in \I_i)]$. Secondly, by \Cref{lem:event_intersection} and by definition about $Pr[q_i(D) \in \I_i] \geq \beta$ for $\forall i$, we have
$Pr[\bigwedge_{i=1}^{\ell} (q_i(D) \in \I_i)] \geq \sum_{i=1}^{\ell}{Pr[q_i(D) \in \I_i]} - (\ell-1) \geq \ell(\beta-1)+1$. Thirdly, since $\I^L = \inf I$ and $\I^U = \sup I$, we have $I \subseteq [\I^L, \I^U]$, and therefore $Pr[\I^L \leq q_f(D) \leq \I^U] \geq Pr[q_f(D) \in \I]$. Together, we have $Pr[\I^L \leq q_f(D) \leq \I^U] \geq Pr[q_f(D) \in \I] \geq Pr[\bigwedge_{i=1}^{\ell} (q_i(D) \in \I_i)] \geq \ell(\beta-1)+1$.

\end{proof}


Although it might not be obvious to find the analytical form of the image, we can use numerical methods to find the approximations of the supremum and infimum of the
image. The width of such a interval is not determined.

Given a query $q$ decomposed by function $f$ and the noisy answers, algorithm \ref{alg:image_ci} summarizes the approach for deriving the confidence interval for $q(D)$. We first derive the confidence interval of each sub-query (line \ref{lin:gen_ci_for_start} to \ref{lin:gen_ci_for_end}) with confidence level $\beta = 1 - \frac{1 - \gamma}{\ell}$ (\cref{lin:gen_ci_set_beta}) and finally compute the confidence interval for $q(D)$ (line \ref{lin:gen_ci_l} and \ref{lin:gen_ci_u}). 

\begin{algorithm}
\caption{Image-based Confidence Interval}
\label{alg:image_ci}
\begin{algorithmic}[1]
\Require A query $q$ such that $q(D) = f(q_1(D), q_2(D), \ldots, q_\ell(D))$, noisy answers $\hat{\subo}_1, \hat{\subo}_2, \ldots, \hat{\subo}_\ell$ of  queries $q_1, q_2, \ldots, q_\ell$ using Gaussian mechanisms with scales $\sigma_1$, $\sigma_2$, $\ldots,$ $\sigma_\ell$, confidence level $\gamma$.
\State $\beta \gets 1 - (1 - \gamma) / \ell$
\label{lin:gen_ci_set_beta}
\For{$i \in 1 \ldots \ell$}
\label{lin:gen_ci_for_start}
    \State $\I_i = (\hat{\subo}_i - \sigma_i\sqrt{2}\erf^{-1}(\beta), \hat{\subo}_i + \sigma_i\sqrt{2}\erf^{-1}(\beta))$
\EndFor
\label{lin:gen_ci_for_end}
\State $\I^L \gets \inf_{x_i \in \I_i \forall x_i} f(x_1, x_2, \ldots, x_\ell)$
\label{lin:gen_ci_l}
\State $\I^U \gets \sup_{x_i \in \I_i \forall x_i} f(x_1, x_2, \ldots, x_\ell)$
\label{lin:gen_ci_u}
\State $\I_q = (\I^L, \I^U)$
\State \Return $\I_q$.
\end{algorithmic}
\end{algorithm}

\subsection{Bootstrap Confidence Interval}
\label{sec:bootstrap}

\begin{algorithm}
\caption{Bootstrap Confidence Interval}
\begin{algorithmic}[1]
\Require A query $q$ such that $q(D) = f(q_1(D), q_2(D), \ldots, q_\ell(D))$, noisy answers $\hat{o}_1, \hat{o}_2, \ldots, \hat{o}_\ell$ of  queries $q_1, q_2, \ldots, q_\ell$ using Gaussian mechanisms with scales $\sigma_1$, $\sigma_2$, $\ldots,$ $\sigma_\ell$, confidence level $\gamma$, and a bootstrap step size $B$.
\Ensure A confidence interval for $q_f(D)$ at confidence level $\gamma$.
\For{$b \gets 1 \ldots B$}
    \For{$i \gets 1 \ldots \ell$}
        \State $o^*_i \gets \hat{o}_i + N(0, \sigma_i^2)$
    \EndFor
    \State $\theta^*_b \gets f(o^*_1, o^*_2, \ldots, o^*_\ell)$
\EndFor
\State $\hat{\theta} \gets f(\hat{o}_1, \hat{o}_2, \ldots, \hat{o}_\ell)$
\State $z_0 = \Phi^{-1}(\frac{1}{B}\sum_{b=1}^{B}\mathbbm{1}_{\theta^*_b < \hat{\theta}})$
\State $\I^L \gets \min \{s | \frac{1}{B}\sum_{b=1}^{B}\mathbbm{1}_{\theta^*_b < s} \geq \Phi(2 z_0 + \Phi^{-1}(\frac{1-\gamma}{2})\}$
\State $\I^U \gets \max \{s | \frac{1}{B}\sum_{b=1}^{B}\mathbbm{1}_{\theta^*_b < s} \leq \Phi(2 z_0 + \Phi^{-1}(\frac{1+\gamma}{2})\}$
\State $\I_q = (\I^L, \I^U)$
\State \Return $\I_q$.
\end{algorithmic}
\label{alg:bootstrap_CI}
\end{algorithm}


Bootstrap is an old yet powerful technique started by Bradley Efron in late 70's \cite{efron1982jackknife} which can be used for computing the confidence interval of an unknown statistic. Although it is not an exact confidence interval, it enjoys a theoretical guarantee on the correctness of the approximation. Traditional bootstrap assumes there are multiple samples that is samples from a unknown distribution with the parameters of interest. Here the parameters of interest is $q_1(D)$ to $q_\ell(D)$, the distribution is a multivariate Gaussian distribution, and we only observe one sample from it. Therefore, we apply the method introduced from \cite{efron1985bootstrap} to construct an confidence interval for $q(D)$, which consider the similar problem of finding the confidence interval of $q(D)$ with one observation of noisy $q_1(D)$ to $q_\ell(D)$.

In section 5 of \cite{efron1985bootstrap}, it describes a parametric bootstrap. The main idea is to assume the observation is from a parametric distribution, use the observation to infer the parameters using maximum likelihood estimate, resample from the estimated distribution, and use bias-corrected percentile method to construct a confidence interval. \cref{alg:bootstrap_CI} illustrates an application of parametric bootstrap confidence interval for $q(D)$.

The most related work to this private confidence interval problem that are also based on bootstrap are \cite{ferrando2020general, covington2021unbiased}, which construct the CI that encodes the randomness from both sampling and noise, while we only consider the randomness from noise. Traditional CI is closely related to sampling, which assumes some population parameters and data is sampled according to those parameters, therefore the population parameter can be inferred from the sampled data. However, for differential privacy, the setting is totally different. Data is considered as fixed and the statistics of interest is only based on the fixed data. Therefore, there is no randomness in sampling a dataset.

%% file: appendix-experiment.tex
\section{Supplementary Experiment}

\subsection{Another example of \Cref{fig:phase3}}
\label{sec:another_example_phase3}

See \Cref{fig:another_phase3}.

\begin{figure}[h]
    \begin{tcolorbox}[colback=white,left=1.5pt,right=1.5pt,top=0pt,bottom=0pt]
        {\bf \small Answer-Phase-3:}
        \normalsize
        \setlength\tabcolsep{1.5pt}
        \begin{center}
        {\footnotesize
        \begin{tabular}{@{} >{\raggedright}p{4.2cm} c c P{0.8cm} P{0.8cm} @{}}
            \toprule
            \multirow{2}{*}{explanation predicate} &
            \multicolumn{2}{c}{Rel Influ 95\%-CI} &
            \multicolumn{2}{c}{Rank 95\%-CI} \\
            \cmidrule(lr){2-3}
            \cmidrule(lr){4-5}
            & L & U & L & U \\
            \midrule
\tt        education = "Bachelors" &  4.51\% & 11.38\% & 1 &  5 \\
\tt occupation = "Exec-managerial" &  3.04\% & 9.91\% & 1 &  8 \\
\tt               age = "(40, 50]" &  1.98\% & 8.85\% & 1 & 14 \\
\tt     relationship = "Own-child" & -1.53\% & 5.34\% & 1 & 51 \\
\tt     workclass = "Self-emp-inc" & -2.34\% & 4.53\% & 1 & 87 \\
            \bottomrule
        \end{tabular} 
        }
        \end{center}
    \end{tcolorbox}  
    \caption{Another example (in a random run) of \Cref{fig:phase3} for Phase-3 of \oursys.}
    \label{fig:another_phase3}
\end{figure}

\subsection{Confidence Level}
\label{sec:confidence_level}

\begin{figure}[h]
    \centering
    \includegraphics[width=\linewidth]{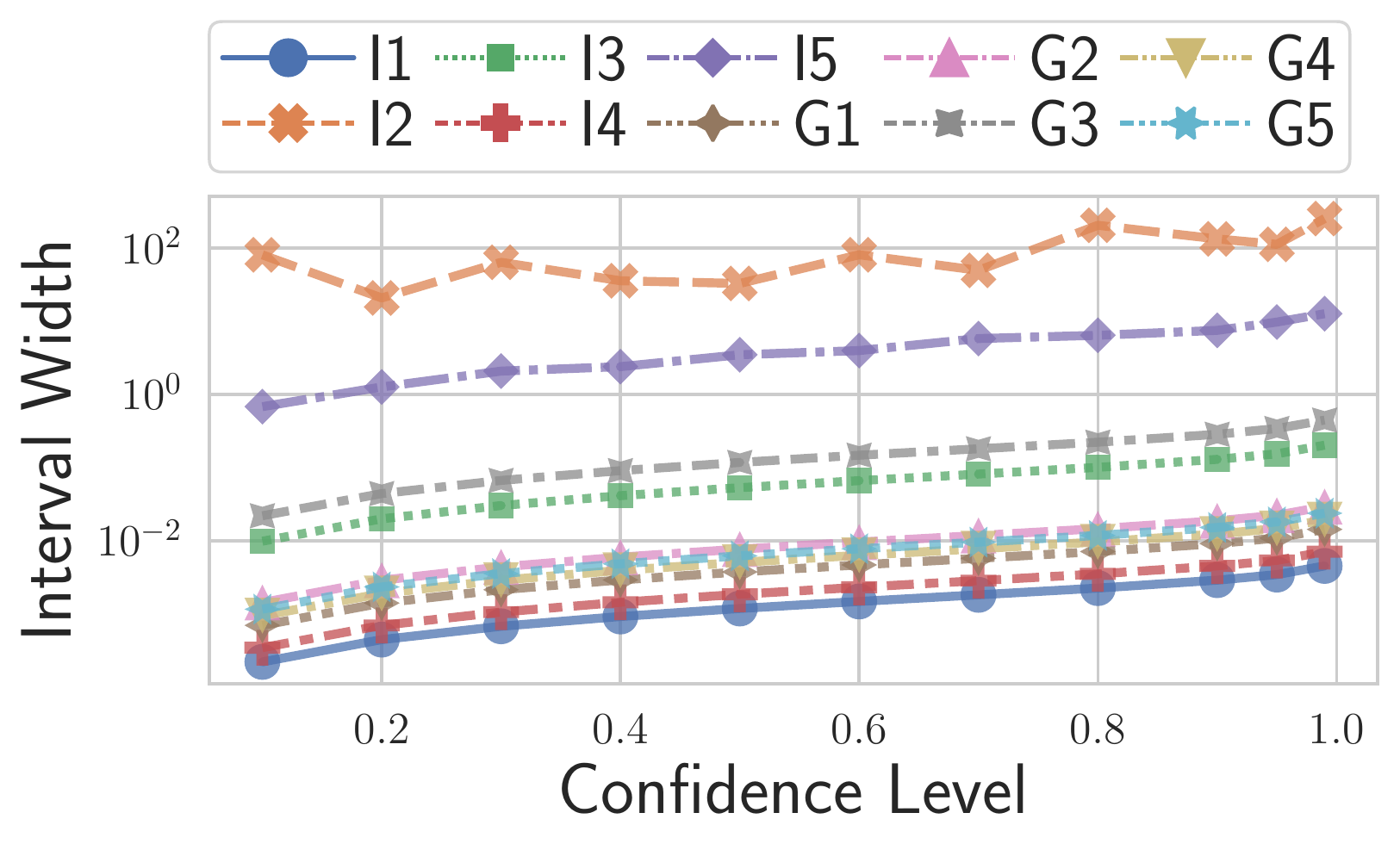}
    \caption{The width of confidence interval of relative influence versus the confidence level.}
    \label{fig:influeci_gamma}
\end{figure}

\Cref{fig:influeci_gamma} shows the relationship between the average interval width of the confidence interval of relative influence and the confidence level.

\begin{figure}[h]
    \centering
    \includegraphics[width=\linewidth]{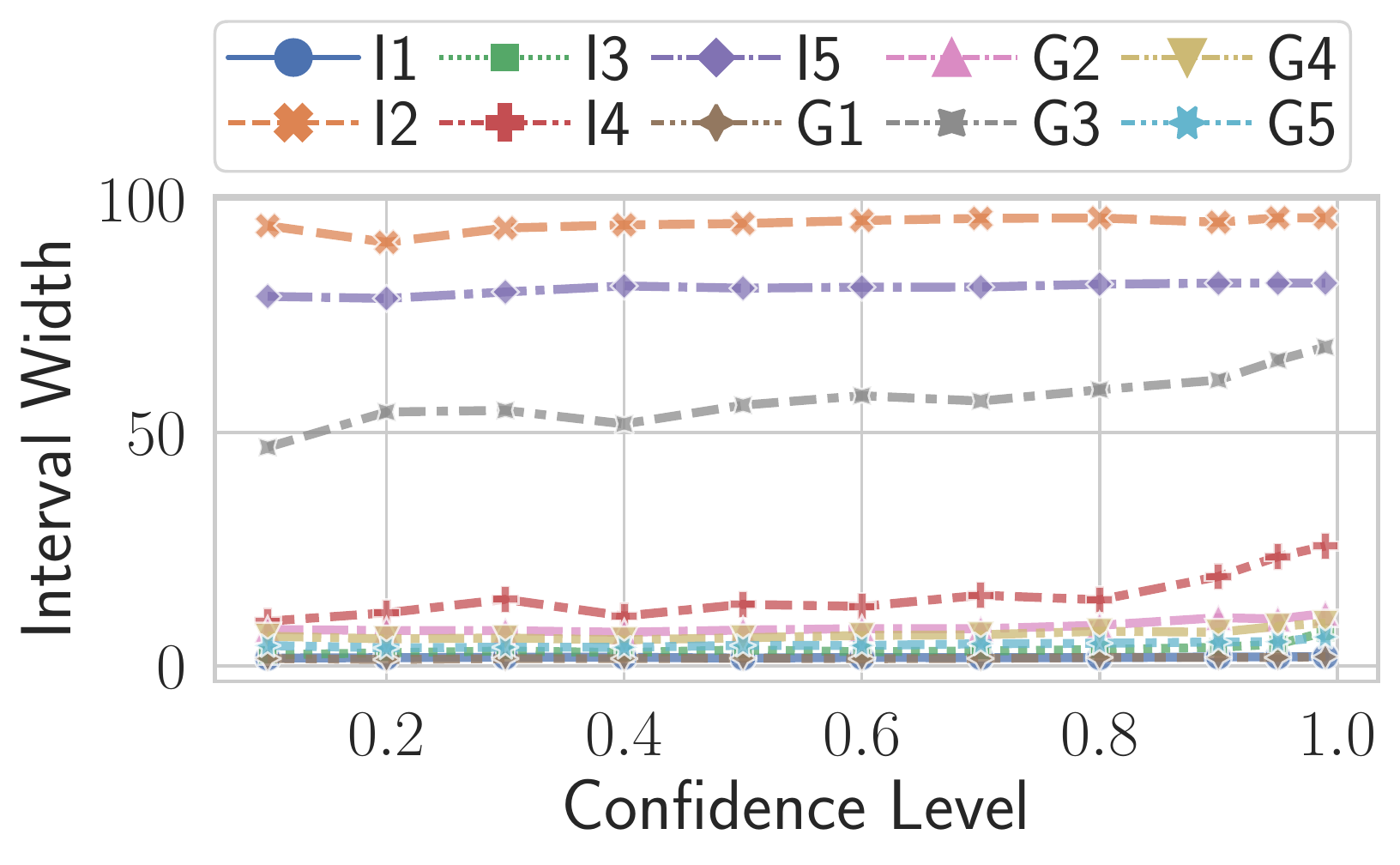}
    \caption{The width of confidence interval of rank versus the confidence level.}
    \label{fig:rankci_gamma}
\end{figure}

\Cref{fig:rankci_gamma} shows the relationship between the average interval width of the confidence interval of rank and the confidence level. 

\subsection{Full Ranking}
\label{sec:full_ranking}

\begin{figure}[h]
    \centering
    \includegraphics[width=\linewidth]{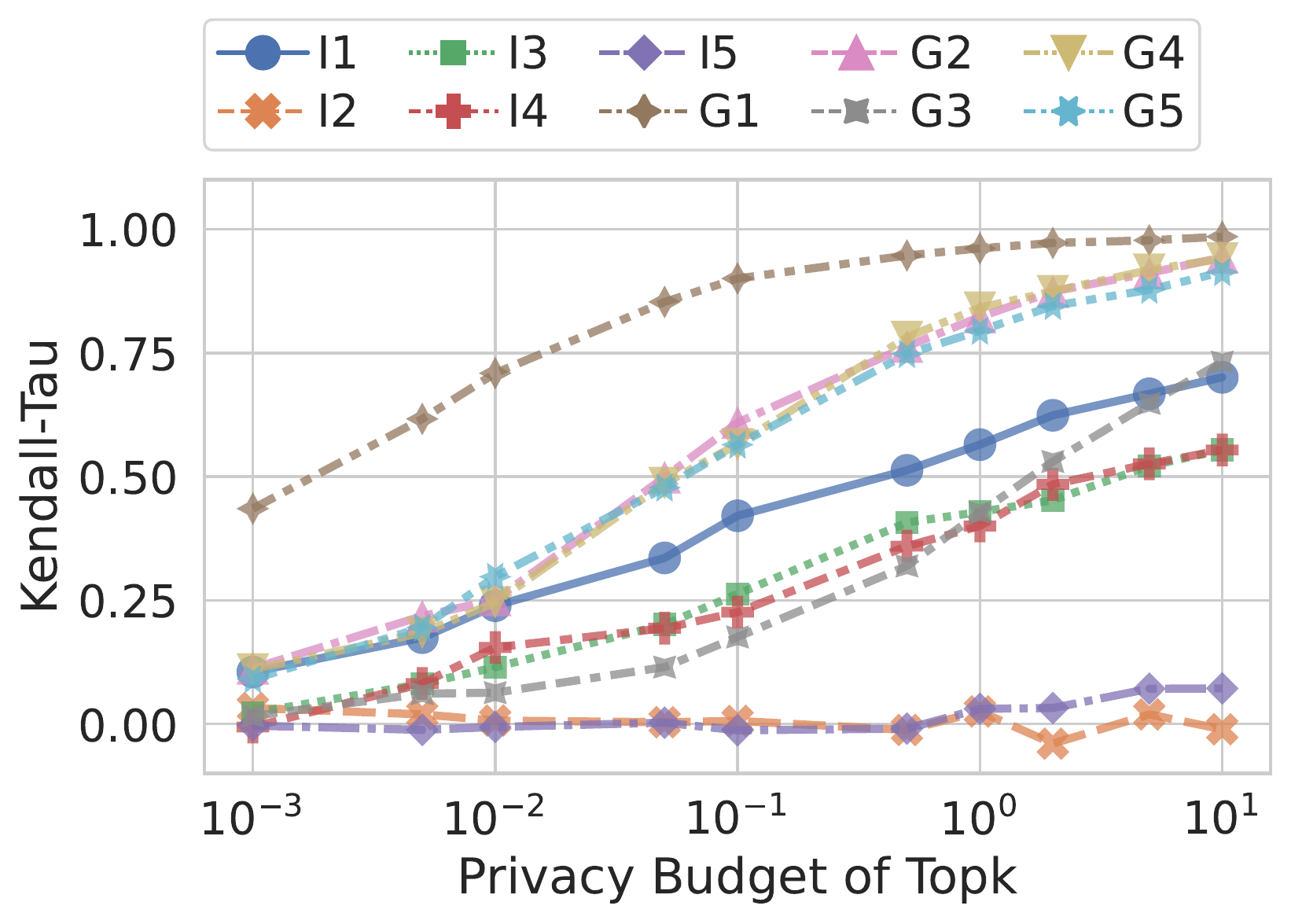}
    \caption{Kendall-Tau of top-k selection by \oursys. }
    \label{fig:exp-kendall}
\end{figure}

To further understand the performance of the top-k selection in \oursys, 
we set $k$ to be the maximum size to have a full ranking of all the explanation predicates and stops \oursys~ at the step of top-k selection. 
We measure the quality of the full ranking by Kendall-Tau \cite{kendall1938new}. From \Cref{fig:exp-kendall}, we find that for question G1 its Kendall-Tau is always above 0.4 for privacy budget of topk $\rhoTopk \geq 0.001$, while for question I1 its Kendall-Tau starts to be above 0.4 when $\rhoTopk \geq 0.1$. 
Though the interpretation of Kendall-Tau is not unified, a correlation coefficient above 0.4 indicates a moderate rank association to the true ranking and above 0.7 indicates a strong rank association \cite{akoglu2018user}. However, when we increase the number of conjuncts $l$ from 1 to 2 to 3, the correlation coefficient drops significantly:
for I1, it drops from 0.513 to 0.029 to 0.001, and for G1, it drops from 0.947 to 0.466 to 0.060 
. This is because increasing the number of conjuncts $l$ will exponentially increase the number of explanation predicates and thus increase the difficulty of a full ranking.  

